\documentclass[italian,openany,12pt,a4paper,titlepage]{book}%
\usepackage{amsfonts}
\usepackage{amsmath}
\usepackage[italian]{babel}
\usepackage{amssymb}
\usepackage{graphicx}
\newtheorem{theorem}{Teorema}

\newtheorem{definition}[theorem]{Definizione}

\newtheorem{problem}[theorem]{Problem}
\newtheorem{proposition}[theorem]{Proposizione}

\begin{document}

\frontmatter
\tableofcontents

\chapter{Introduzione}

Lo scopo di questa tesi fornire una panoramica dell' ottica geometrica
affrontando alcuni aspetti salienti della disciplina. In particolare l'
attenzione viene rivolta allo studio dei sistemi ottici centrati che
costituiscono l' oggetto principale d' indagine di questo lavoro.

Nel primo capitolo, seguendo l' impostazione ormai tipica della fisica teorica
moderna, si pensi alla teoria dei campi quantizzati, le leggi che regolano
l'ottica geometrica sono fatte discendere da un principio variazionale, il
principio di Fermat. Il principio \`{e} formulato nel caso di superfici di
discontinuit\`{a} per l' indice di rifrazione $N\left(  \mathbf{x}\right)  .$
Viene trattato il problema dei dati al contorno, e vengono presentate le
analogie con la meccanica classica. Viene data anche la chiave di lettura dei
principi variazionali della fisica classica, attraverso una delle pi\`{u}
belle teorie concepite dalla fisica del novecento, gli integrali sui cammini
di Feynman.

Il secondo capitolo presenta il passaggio dal formalismo di Lagrange a quello
di Hamilton, il pi\`{u} adatto a trattare questo tipo di problemi; si
introducono le funzioni principali di Hamilton che costituiscono la base della
teoria delle aberrazioni che verr\`{a} presentata nei successivi capitoli. Si
indagano le conseguenze della simmetria assiale dei sistemi ottici, viene
presentato l' invariante ottico di Lagrange, e nel caso notevole in cui le
superfici ottiche presentino simmetria sferica, si ricava l' espressione di
due nuovi invarianti ottici, sconosciuti in letteratura. Sono presentati i
cosidetti invarianti di rotazione, e lo studio delle loro trasformazioni in un
sistema ottico centrato.

Il capitolo tre \`{e} dedicato allo sviluppo dell' ottica Gaussiana attraverso
l' uso delle funzioni principali di Hamilton ed alla definizioni delle
grandezze che caratterizzano un sistema ottico a quest' ordine di
approssimazione. Viene presentata un' analisi delle aberrazioni del primo ordine.

Il capitolo quarto, dopo una prima analisi che permette di comprendere le
prime propriet\`{a} delle aberrazioni del terzo ordine, presenta la teoria
cos\`{\i} come sviluppata da Petzval, von Seidel e Schwarzschild, che conduce
alle formule che esprimono i cinque coefficienti di aberrazione. Viene esposto
il significato fisico dei coefficienti di aberrazione del terzo e del quinto ordine.

L' uso delle funzioni principali \`{e} abbandonato nel quinto capitolo a
favore delle pi\`{u} semplici equazioni di Hamilton. Seguendo la strada
percorsa da R. K. Luneburg, in modo semplice ed intuitivo, vengono ricostruite
l' ottica del primo ordine ed \`{e} dato un approccio alternativo per il
calcolo dei coefficienti del terzo ordine. Il procedimento consiste nel
supporre l' indice di rifrazione continuo, calcolare i coefficienti di
aberrazione per questa situazione fisica e nel passare poi al limite per i
consueti sistemi che presentano indice di rifrazione costante a tratti. Le
formule di Luneburg vengono ricavate in una situazione generalizzata in cui
possono essere presenti superfici asferiche o riflettenti. Vengono anche
proposte le formule integrali punto di partenza per il calcolo delle
aberrazioni del quinto ordine.

Nel sesto capitolo si verificano le relazioni ricavate in precedenza per l'
analisi di alcuni sistemi telescopici che presentano superfici asferiche e
riflettenti : il telescopio Newtoniano e la camera di Schmidt. I risultati
ottenuti concordano con quelli presenti in letteratura, e con programmi di
simulazione numerica.\newpage\vspace*{13cm}
\ \ \ \ \ \ \ \ \ \ \ \ \ \ \ \ \ \ \ \ \ \ \ \ \ \ \ \ \ \ \ \ \ \ \ \ \ \ \ \ \ \ \ \ \ \ \ \ \ \ \ \ \ \ \ \ \ \ \ \ \ \ \ \ \ \ \ \ \ \ \ \ \ \ \textbf{Ringraziamenti}%

Colgo questa occasione per ringraziare tutti coloro che in questi anni mi
hanno supportato e, soprattutto, sopportato. Un abbraccio va a mio fratello
Alessandro e a Tiziana per l' ospitalit\`{a} che mi hanno concesso nei periodi
bui; a mia madre Dora per avermi insegnato a cucinare la pasta alla siciliana;
a Manuela per avermi fatto fare l' esame di Esp.II; a\ Ciro perch\'{e} rimane
interista; ai ragazzi dell' auletta perch\`{e} sono ottimi compagni di
tresette; a Biagio per avermi spronato quando non riuscivo pi\`{u} a studiare;
al SPSF folle speranza della mia terra. Un ringraziamento particolare va al
mio relatore il Prof. Antonio Romano per essere riuscito a condurmi in porto
nonostante remassi contro. Un bacio lungo un anno a Floriana.

\mainmatter

\chapter{Il principio di Fermat}

\section{Il principio di Fermat}

Si consideri la propagazione di un' onda elettromagetica monocromatica in un
mezzo isotropo e non omogeneo $\Sigma$. Se la lunghezza d' onda della luce
\`{e} trascurabile rispetto alle dimensioni della regione in cui si propaga e
degli ostacoli interposti, la propagazione della luce pu\`{o} essere descritta
da \textbf{raggi luminosi}.

\begin{definition}
Quando la luce si propaga in un mezzo isotropo, le traiettorie ortogonali ai
fronti d'onda vengono dette \textbf{raggi luminosi.}
\end{definition}

In seguito si supporr\`{a} che l' indice di rifrazione $N\left(
\mathbf{x}\right)  $ sia una funzione continua e sufficientemente regolare del
vettore posizione $\mathbf{x}\in\Sigma$, fatta eccezione per un numero finito
di superfici $S_{n}$ $n=1,\ldots,l$ sulle quali $N\left(  \mathbf{x}\right)  $
ha discontinuit\`{a} finite.

\begin{definition}
Si definisce la \textbf{lunghezza del cammino ottico }lungo una curva
$\gamma_{0}$ \textbf{\ }%
\begin{equation}
OPL\left(  \gamma_{0}\right)  =\int_{\gamma_{0}}N\left(  \mathbf{x}\right)  ds
\end{equation}

\end{definition}

Se $\gamma_{0}$ \`{e} un raggio luminoso, indicando con $v$ la velocit\`{a}
della luce nel mezzo $\Sigma$, avremo
\begin{equation}
Nds=\frac{c}{v}ds=cdt
\end{equation}
di conseguenza la lunghezza del cammino ottico \emph{lungo un raggio }\`{e}
proporzionale al tempo di propagazione lungo $\gamma_{0}$.

Formuliamo ora il seguente

\textbf{Principio di Fermat : }\emph{lungo un raggio luminoso dal punto
}$\overline{\mathbf{x}}$ \emph{al punto} $\overline{\mathbf{x}}^{\prime}%
$\emph{, la lunghezza del cammino ottico \`{e} un estremale rispetto a tutte
le lunghezze dei cammini ottici corrispondenti a qualunque altro cammino tra i
punti in esame.}

Si pongono immediatamente alcune fondamentali domande:

\begin{enumerate}
\item Come determinare il raggio $\gamma_{0}$ dopo che siano stati assegnati i
due estremi e l' indice di rifrazione?

\item Come trattare le discontinuit\`{a} nel principio di Fermat?
\end{enumerate}

Il principo di Fermat costituisce uno dei primi principi variazionali della
fisica, una categoria di problemi che caratterizza l' incognita del problema
(il raggio luminoso in questo caso) richiedendo che un' espressione integrale
abbia come estremale proprio la nostra incognita. Si arriva cos\`{\i} ad un
problema differente da un consueto problema di Cauchy, per il quale esiste il
ben noto teorema di esistenza ed unicit\`{a} della soluzione.

Si inizi col caratterizzare gli estremali del nostro funzionale. In primo
luogo ogni cammino tra $\overline{\mathbf{x}}$ ed\textbf{\ }$\overline
{\mathbf{x}}^{\prime}$ \`{e} una curva $\gamma$ che ha questi punti come
estremi. Introdotto un sistema di riferimento cartesiano di coordinate
$\left(  x_{i}\right)  $ $i=1,2,3,$ una curva $\gamma$\ \`{e} rappresentata da
una terna di equazioni $x_{i}=f_{i}\left(  t\right)  $ $t\in\left[
a,a^{\prime}\right]  $, i punti $\overline{\mathbf{x}}$ e $\overline
{\mathbf{x}}^{\prime}$ saranno gli estremi di $\gamma$ imponendo che si abbia%
\begin{equation}
f_{i}\left(  a\right)  =\overline{x}_{i}~~f_{i}\left(  a^{\prime}\right)
=\overline{x}_{i}^{\prime} \label{estr}%
\end{equation}
In particolare si indicheranno le equazioni parametriche del \emph{raggio
}$\gamma_{0}$ tra $\overline{\mathbf{x}}$ e $\overline{\mathbf{x}}^{\prime}$,
ossia del cammino ottico \emph{effettivo }con le funzioni $x_{i}\left(
t\right)  ,i=1,2,3,$ che verificano le condizioni \ref{estr}. Il passo
successivo sar\`{a} quello di cercare un' espressione di $OPL\left(
\gamma\right)  $ che soddisfi le \ref{estr} per ogni cammino e che ci permetta
di individuare i suoi estremali. Si pu\`{o} scrivere
\begin{equation}
ds=\sqrt{\sum_{i=1}^{3}\dot{f}_{i}^{2}}dt
\end{equation}
dove $\dot{f}=\frac{df}{dt}$ . Si pu\`{o} ora esprimere $OPL\left(
\gamma\right)  $ per un cammino qualunque tra gli estremi $\overline
{\mathbf{x}}$ e $\overline{\mathbf{x}}^{\prime}$:%
\begin{equation}
OPL\left(  \gamma\right)  =\int_{\gamma}N\left(  \mathbf{x}\right)  \sqrt
{\sum_{i=1}^{3}\dot{f}_{i}^{2}}dt \label{oplgammag}%
\end{equation}
Si deve ora far variare la curva $f_{i}$, consideriamo la famiglia di curve ad
un un parametro $\Gamma$ con estremi fissati $\overline{\mathbf{x}}$ e
$\overline{\mathbf{x}}^{\prime}$ e cui appartiene $\gamma_{0}$
\begin{equation}
x_{i}=f_{i}\left(  t,\varepsilon\right)  \text{~~}\forall t\in\left[
a,a^{\prime}\right]  i=1,2,3,
\end{equation}
dove%
\begin{equation}
f_{i}\left(  a,\varepsilon\right)  =\overline{x}_{i},~~f_{i}\left(  a^{\prime
},\varepsilon\right)  =\overline{x}_{i}^{\prime},~~\forall\varepsilon
\in-\left(  \delta,\delta\right)  \text{~~}f_{i}\left(  t,0\right)
=x_{i}\left(  t\right)  \label{def fam}%
\end{equation}
e $\left(  -\delta,\delta\right)  $ \`{e} un intorno del valore $\varepsilon
=0$, il parametro $t$ ci far\`{a} muovere lungo la curva, mentre $\varepsilon$
seleziona una delle curve di $\Gamma$. $OPL$ dipende quindi dal valore di
$\varepsilon$, come si vede dalla seguente formula che \`{e} ottenuta dalla
\ref{oplgammag} quando si impone $\gamma\in\Gamma$
\begin{equation}
OPL\left(  \varepsilon\right)  =\int_{a}^{a^{\prime}}N\left(  \mathbf{x}%
\right)  \sqrt{\sum_{i=1}^{3}\dot{f}_{i}^{2}\left(  t,\varepsilon\right)  }dt.
\end{equation}
Si pu\`{o} riformulare il principio di Fermat nel modo seguente: determinare
sotto quali condizioni la funzione $OPL\left(  \varepsilon\right)  $\ ha un
estremo per\emph{ }$\varepsilon=0$ per una qualunque scelta della famiglia di
cammini che verificano le \ref{estr}. Una condizione necessaria \`{e}%
\begin{equation}
OPL\left(  \varepsilon\right)  =0
\end{equation}
per una qualunque scelta di $\Gamma$. Dovremo quindi:

\begin{enumerate}
\item calcolare la derivata prima di $OPL\left(  \varepsilon\right)  $ in
$\varepsilon=0$;

\item determinare sotto quali condizioni si annulla per ogni scelta di
$\Gamma$;

\item verificare che queste condizioni determinano $\gamma_{0}$;

\item controllare che $\varepsilon=0$ sia un punto di estremo per $OPL\left(
\varepsilon\right)  $.
\end{enumerate}

Conviene considerare una situazione pi\`{u} generale in vista di futuri
sviluppi in modo da non ripetere calcoli dello stesso tipo. Pi\`{u}
precisamente, conviene considerare la derivata della seguente funzione
\begin{equation}
I\left(  \varepsilon\right)  =\int_{t_{1\left(  \varepsilon\right)  }}%
^{t_{2}\left(  \varepsilon\right)  }N\left(  \mathbf{x}\right)  \sqrt
{\sum_{i=1}^{3}\dot{f}_{i}^{2}\left(  t,\varepsilon\right)  }dt
\end{equation}
e si definisce la famiglia di curve $\Delta$\ verifica le seguenti
propriet\`{a}
\begin{subequations}
\label{curve delta}%
\begin{gather}
x_{i}=f_{i}\left(  t,\varepsilon\right)  ~~\forall t\in\left[  t_{1}\left(
\varepsilon\right)  ,t_{2}\left(  \varepsilon\right)  \right]  ~~i=1,2,3,\\
f_{i}\left(  t_{1}\left(  0\right)  ,0\right)  =\overline{x}_{i}%
,~~f_{i}\left(  t_{2}\left(  0\right)  ,0\right)  =\overline{x}_{i}^{\prime
},~~\forall\in\left(  -\delta,\delta\right)  ~~
\end{gather}
\ $f_{i}\left(  t,0\right)  =x_{i}\left(  t\right)  $dove $\left(
-\delta,\delta\right)  $ \`{e} un intorno di zero, a differenza di $\Gamma$
dove gli estremi delle curve erano fissati, si \`{e} preferito considerare
estremi variabili al variare di $\varepsilon$, $\Delta$ si riduce a $\Gamma$,
e $I\left(  \varepsilon\right)  $ si riduce a $OPL\left(  \varepsilon\right)
$, se
\end{subequations}
\begin{equation}
t_{1}\left(  \varepsilon\right)  =a~~t_{2}\left(  \varepsilon\right)
=a^{\prime}\text{~~}\forall\varepsilon\in\left(  -\delta,\delta\right)
\end{equation}
prima di determininare $I^{\prime}\left(  \varepsilon\right)  $ e valutarla
per $\varepsilon=0$, \`{e} importante osservare che le coordinate degli
estremi delle curve $\gamma\in\Delta$ sono
\[
f_{i}\left(  t_{1}\left(  \varepsilon\right)  ,\varepsilon\right)
,~~f_{i}\left(  t_{2}\left(  \varepsilon\right)  ,\varepsilon\right)
\]
cosicch\'{e}, andando dal raggio $\gamma_{0}$, corrispondente al valore
$\varepsilon=0,$ ad una curva vicina $\gamma$ corrispondente al valore
$d\varepsilon$ di $\varepsilon$, le coordinate degli estremi variano secondo
le relazioni
\begin{equation}
d\overline{x}_{i}=\left(  \dot{x}_{i}\frac{dt_{1}}{d\varepsilon}%
+\frac{\partial f_{i}}{\partial\varepsilon}\right)  _{\varepsilon
=0}d\varepsilon,\text{~~}d\overline{x}_{i}^{\prime}=\left(  \dot{x}_{i}%
\frac{dt_{2}}{d\varepsilon}+\frac{\partial f_{i}}{\partial\varepsilon}\right)
_{\varepsilon=0}d\varepsilon.
\end{equation}
L' indice di rifrazione ha discontinuit\`{a} finite sulle superfici $S_{i},$
$i=1,\ldots,n,$ che sono le superfici del sistema ottico in considerazione.
Per semplicit\`{a} nel seguito si supporr\`{a} che vi sia una sola superficie
di discontinuit\`{a}, di conseguenza dobbiamo ipotizzare che anche le funzioni
$f_{i}\left(  t,\varepsilon\right)  $ possano avere derivate prime
discontinue, ossia che il vettore tangente alle curve possa cambiare
bruscamente direzione attraversando $S$. Sia $F\left(  \mathbf{x}\right)  =0$
l' equazione implicita della superficie $S$, la discontinuit\`{a} deve essere
localizzata nel punto di intersezione della curva di $\Delta$ con $S$. Per
determinare il luogo $\varphi$ di questi punti, si deve verificare che l'
equazione
\begin{equation}
F\left(  f_{i}\left(  t,\varepsilon\right)  \right)  =0 \label{intersezione}%
\end{equation}
definisca implicitamente una funzione $t=\psi\left(  \varepsilon\right)  $ che
da il valore di $t$ corrispondente al punto di intersezione $P$ tra $\gamma$
ed $S$. Infatti la condizione del Dini
\begin{equation}
\frac{\partial}{\partial t}F\left(  f_{i}\left(  t,\varepsilon\right)
\right)  \not =0,
\end{equation}
si scrive
\begin{equation}
\frac{\partial F}{\partial x_{i}}\frac{\partial f}{\partial t}\not =0
\end{equation}
o, equivalentemente
\begin{equation}
\mathbf{n\cdot t}\ \not =0
\end{equation}
dove $\mathbf{n=}\nabla F/\left\vert \nabla F\right\vert $ \`{e} il vettore
unitario normale interno ad $S$ e $\mathbf{t}$ il vettore tangente unitario a
$\gamma$ in $P$. Allora, se $\gamma$ non \`{e} tangente ad $S$ in $P$, la
\ref{intersezione} definisce implicitamente una funzione $t=\psi\left(
\varepsilon\right)  $ e il luogo dei punti $\varphi$ \`{e} una curva e la cui
equazione parametrica risulta essere
\begin{equation}
\phi_{i}\left(  \varepsilon\right)  =f_{i}\left(  \psi\left(  \varepsilon
\right)  ,\varepsilon\right)  , \label{phi di e}%
\end{equation}
dove
\begin{equation}
F\left(  \phi_{i}\left(  \varepsilon\right)  \right)  =F\left(  f_{i}\left(
\psi\left(  \varepsilon\right)  ,\varepsilon\right)  \right)  =0.
\label{intersezione2}%
\end{equation}
Rispettando la continuit\`{a} delle curve di $\Delta$ nell'attraversare $S$,
$\phi\left(  \varepsilon\right)  $ \`{e} supposta essere una curva regolare.
Differenziando la \ref{intersezione2} rispetto ad $\varepsilon$ e tenendo
conto della \ref{phi di e} otteniamo le seguenti due condizioni
\begin{gather}
n_{i}\left(  \frac{\partial f_{i}}{\partial t}\frac{\partial\psi}%
{\partial\varepsilon}+\frac{\partial f_{i}}{\partial\varepsilon}\right)
^{-}\equiv n_{i}\xi_{i}^{-}\label{limite destro sinistro}\\
n_{i}\left(  \frac{\partial f_{i}}{\partial t}\frac{\partial\psi}%
{\partial\varepsilon}+\frac{\partial f_{i}}{\partial\varepsilon}\right)
^{+}\equiv n_{i}\xi_{i}^{+}%
\end{gather}
dove la somma sull' indice $i$ \`{e} sottintesa e con la notazione
$(\ldots)^{+}~$e $(\ldots)^{-}$ si sono indicati i limiti destro e sinistro.
Inoltre la regolarit\`{a} di $\phi\left(  \varepsilon\right)  $ implica
\begin{equation}
\phi^{\prime}\left(  \varepsilon\right)  =\left(  \frac{\partial f_{i}%
}{\partial t}\frac{\partial\psi}{\partial\varepsilon}+\frac{\partial f_{i}%
}{\partial\varepsilon}\right)  ^{-}=\left(  \frac{\partial f_{i}}{\partial
t}\frac{\partial\psi}{\partial\varepsilon}+\frac{\partial f_{i}}%
{\partial\varepsilon}\right)  ^{+} \label{limite destro sinistro2}%
\end{equation}
possiamo quindi scrivere
\begin{equation}
\xi_{i}^{-}=\xi_{i}^{+}.
\end{equation}
Nel seguito indicheremo il salto di ogni grandezza sulla superficie $S$ con
\begin{equation}
\left[  \left[  a\right]  \right]  =a^{+}-a^{-}%
\end{equation}
Si \`{e} ora in grado di valutare la derivata $I^{\prime}\left(  0\right)  .$
Si ha
\begin{equation}
I^{\prime}\left(  \varepsilon\right)  =\int_{t_{1\left(  \varepsilon\right)
}}^{\psi_{\left(  \varepsilon\right)  }}L\left(  f_{i}\left(  t,\varepsilon
\right)  ,\dot{f}\left(  t,\varepsilon\right)  \right)  dt~+\int
_{\psi_{\left(  \varepsilon\right)  }}^{t_{2\left(  \varepsilon\right)  }%
}L\left(  f_{i}\left(  t,\varepsilon\right)  ,\dot{f}\left(  t,\varepsilon
\right)  \right)  dt,
\end{equation}
dove si \`{e} posto
\begin{equation}
L=N\left(  \mathbf{x}\right)  \sqrt{\sum_{i=1}^{3}\left(  \dot{f}\right)
^{2}}.
\end{equation}
Applicando una ben nota formula di derivazione di un integrale, la precedente
derivata si scrive
\begin{gather}
\frac{dI}{d\varepsilon}\left(  0\right)  =L\left(  \bar{x}_{i}^{\prime}%
,\dot{x}_{i}\left(  a^{\prime}\right)  \right)  t_{2}^{\prime}\left(
0\right)  -L\left(  \bar{x}_{i},\dot{x}_{i}\left(  a\right)  \right)
t_{1}^{\prime}\left(  0\right)  +\nonumber\\
+\int_{a}^{\psi\left(  0\right)  }\left[  \frac{\partial L}{\partial x_{i}%
}\frac{\partial f_{i}}{\partial\varepsilon}+\frac{\partial L}{\partial\dot
{x}_{i}}\frac{d}{dt}\left(  \frac{\partial f_{i}}{\partial\varepsilon}\right)
\right]  dt+\int_{\psi\left(  0\right)  }^{a^{\prime}}\left[  \frac{\partial
L}{\partial x_{i}}\frac{\partial f_{i}}{\partial\varepsilon}+\frac{\partial
L}{\partial\dot{x}_{i}}\frac{d}{dt}\left(  \frac{\partial f_{i}}%
{\partial\varepsilon}\right)  \right]  dt~-\left[  \left[  L\right]  \right]
\psi^{\prime}\left(  0\right)  ,
\end{gather}
dove tutte le quantit\`{a} del secondo membro sono valutate ad $\varepsilon
=0$. Integrando per parti il secondo addendo in ogni integrale, si ottiene
\begin{gather}
\frac{dI}{d\varepsilon}\left(  0\right)  =L\left(  \bar{x}_{i}^{\prime}%
,\dot{x}_{i}\left(  a^{\prime}\right)  \right)  t_{2}^{\prime}\left(
0\right)  -L\left(  \bar{x}_{i},\dot{x}_{i}\left(  a\right)  \right)
t_{1}^{\prime}\left(  0\right)  ~+\nonumber\\
\int_{a}^{a^{\prime}}\left[  \frac{\partial L}{\partial x_{i}}-\frac{d}%
{dt}\frac{\partial L}{\partial\dot{x}_{i}}\right]  \frac{\partial f_{i}%
}{\partial\varepsilon}dt+\left[  \frac{\partial L}{\partial\dot{x}_{i}}%
\frac{\partial f_{i}}{\partial\varepsilon}\right]  _{a}^{\psi\left(  0\right)
}+\left[  \frac{\partial L}{\partial\dot{x}_{i}}\frac{\partial f_{i}}%
{\partial\varepsilon}\right]  _{\psi\left(  0\right)  }^{a^{\prime}}-\left[
\left[  L\right]  \right]  \psi^{\prime}\left(  0\right)  ,
\end{gather}
possiamo quindi scrivere
\begin{gather}
\frac{dI}{d\varepsilon}=-\left(  \frac{\partial L}{\partial\dot{x}_{i}%
}-L\right)  _{a^{\prime}}t_{2}^{\prime}\left(  0\right)  +\left(
\frac{\partial L}{\partial\dot{x}_{i}}\dot{x}_{i}-L\right)  _{a}t_{1}^{\prime
}\left(  0\right)  +\int_{a}^{a^{\prime}}\left(  \frac{\partial L}{\partial
x_{i}}-\frac{d}{dt}\frac{\partial L}{\partial\dot{x}_{i}}\right)
\frac{\partial f_{i}}{\partial\varepsilon}dt+\\
\left(  \frac{\partial L}{\partial\dot{x}_{i}}\left(  \frac{\partial f_{i}%
}{\partial\varepsilon}+\dot{x}_{i}t_{2}^{\prime}\right)  \right)  _{a^{\prime
}}-\left(  \frac{\partial L}{\partial\dot{x}_{i}}\left(  \frac{\partial f_{i}%
}{\partial\varepsilon}+\dot{x}_{i}t_{1}^{\prime}\right)  \right)  _{a}-\left[
\left[  \frac{\partial L}{\partial\dot{x}_{i}}\frac{\partial f_{i}}%
{\partial\varepsilon}\right]  \right]  _{\psi\left(  0\right)  }-\left[
\left[  L\right]  \right]  \psi^{\prime}\left(  0\right)  .\nonumber
\end{gather}
Infine ricordando le \ref{limite destro sinistro}%
,\ref{limite destro sinistro2} si ottiene%
\begin{multline}
\frac{dI}{d\varepsilon}=-\left(  \frac{\partial L}{\partial\dot{x}_{i}%
}-L\right)  _{a^{\prime}}t_{2}^{\prime}\left(  0\right)  +\left(
\frac{\partial L}{\partial\dot{x}_{i}}\dot{x}_{i}-L\right)  _{a}t_{1}^{\prime
}\left(  0\right)  +\nonumber\\
\int_{a}^{a^{\prime}}\left(  \frac{\partial L}{\partial x_{i}}-\frac{d}%
{dt}\frac{\partial L}{\partial\dot{x}_{i}}\right)  \frac{\partial f_{i}%
}{\partial\varepsilon}dt+\left(  \frac{\partial L}{\partial\dot{x}_{i}%
}\right)  _{a^{\prime}}\frac{d\bar{x}_{i}^{\prime}}{d\varepsilon}-\left(
\frac{\partial L}{\partial\dot{x}_{i}}\right)  _{a}\frac{d\bar{x}_{i}%
}{d\varepsilon}-\\
\left[  \left[  \frac{\partial L}{\partial\dot{x}_{i}}\left(  \frac{\partial
x_{i}}{\partial\varepsilon}+\dot{x}_{i}\psi^{\prime}\left(  0\right)  \right)
\right]  \right]  _{\psi\left(  0\right)  }-\left[  \left[  L-\frac{\partial
L}{\partial\dot{x}_{i}}\dot{x}_{i}\right]  \right]  _{\psi\left(  0\right)
}\psi^{\prime}\left(  0\right)  .\nonumber
\end{multline}
Poich\'{e} $L$ \`{e} una funzione omgenea del primo grado nella variabile
$\dot{x}_{i},$ si ha
\begin{equation}
L-\frac{\partial L}{\partial\dot{x}_{i}}\dot{x}_{i}=0,
\end{equation}
e l'espressione precedente si riscrive come
\begin{subequations}
\label{di deps def}%
\begin{multline}
\frac{dI}{d\varepsilon}=\int_{a}^{a^{\prime}}\left(  \frac{\partial
L}{\partial x_{i}}-\frac{d}{dt}\frac{\partial L}{\partial\dot{x}_{i}}\right)
\frac{\partial f_{i}}{\partial\varepsilon}dt+\left(  \frac{\partial
L}{\partial\dot{x}_{i}}\right)  _{a^{\prime}}\frac{d\bar{x}_{i}^{\prime}%
}{d\varepsilon}-\left(  \frac{\partial L}{\partial\dot{x}_{i}}\right)
_{a}\frac{d\bar{x}_{i}}{d\varepsilon}\nonumber\\
-\left[  \left[  \frac{\partial L}{\partial\dot{x}_{i}}\left(  \frac{\partial
x_{i}}{\partial\varepsilon}+\dot{x}_{i}\psi^{\prime}\left(  0\right)  \right)
\right]  \right]  _{\psi\left(  0\right)  }%
\end{multline}
Per applicare il principio di Fermat a quest'ultima espressione deve essere
imposto che tutte le curve abbiano gli stessi estremi, applicando le
\ref{limite destro sinistro}, \ref{limite destro sinistro2} per $\varepsilon
=0,$ si ha la seguente equazione
\end{subequations}
\begin{equation}
\frac{dI}{d\varepsilon}=\int_{a}^{a^{\prime}}\left(  \frac{\partial
L}{\partial x_{i}}-\frac{d}{dt}\frac{\partial L}{\partial\dot{x}_{i}}\right)
\frac{\partial f_{i}}{\partial\varepsilon}dt+\left[  \left[  \frac{\partial
L}{\partial\dot{x}_{i}}\right]  \right]  _{\psi\left(  0\right)  }\xi_{i}
\label{Int lagrange}%
\end{equation}
con%
\begin{equation}
\xi_{i}=\left(  \frac{\partial x_{i}}{\partial\varepsilon}+\dot{x}_{i}%
\psi^{\prime}\left(  0\right)  \right)  _{\psi\left(  0\right)  }.
\label{csi i}%
\end{equation}
Il principio di Fermat richiede che la derivata di $OPL$ si annulli per
qualunque scelta delle $\left(  \partial x_{i}/\partial\varepsilon\right)
_{\psi\left(  o\right)  }$ e $\xi_{i}$,\ tali che
\begin{equation}
\xi_{i}n_{i}=0
\end{equation}
possiamo riscrivere le \ref{Int lagrange},%
\begin{align}
\frac{\partial L}{\partial x_{i}}-\frac{d}{dt}\frac{\partial L}{\partial
\dot{x}_{i}}  &  =0\label{lagrange}\\
\left[  \left[  \frac{\partial L}{\partial\dot{x}_{i}}\right]  \right]
_{\psi\left(  0\right)  }  &  =\lambda n_{i}~~i=1,2,3
\end{align}
dove $\lambda$ \`{e} un moltiplicatore di Lagrange incognito.

Le equazioni \ref{lagrange} sono le equazioni di Eulero-Lagrange e la
condizione di salto di Weierstrass-Erdman per il funzionale
\begin{equation}
\int_{a}^{a^{\prime}}Ldt=\int_{a}^{a^{\prime}}\sqrt{\sum_{i=1}^{3}\left(
\dot{x}_{i}\right)  ^{2}}dt.
\end{equation}
Si possono riscrivere le equazioni di Lagrange in una forma pi\`{u}
espressiva, introdotto il il versore tangente al raggio $\gamma_{0}$ che va
dal punto $\bar{x}$ al punto $\bar{x}^{\prime}$
\begin{equation}
\mathbf{t=}\frac{\dot{x}_{i}}{\sqrt{\sum_{i=1}^{3}\left(  \dot{x}_{i}\right)
^{2}}},
\end{equation}
e ricordando che%
\begin{equation}
\frac{ds}{dt}=\sqrt{\sum_{i=1}^{3}\left(  \dot{x}_{i}\right)  ^{2}}%
\end{equation}
\`{e} possibile porle nella forma%
\begin{equation}
\frac{d}{ds}\left(  N\mathbf{t}\right)  =\nabla N.
\end{equation}
Si osservi che queste equazioni implicano che la luce si propaga in linea
retta in tutte le regioni che hanno indice di rifrazione $N$ costante.
Riscrivendo anche la condizione sul salto con le stesse notazioni e
\begin{equation}
N^{\prime}\mathbf{t}^{\prime}-N\mathbf{t=}\lambda\mathbf{n}.
\end{equation}
si evince che il raggio incidente, quello rifratto ed $\mathbf{n,}$ il versore
normale ad $S$ ed orientato nel verso della propagazione della luce, sono
complanari al punto di incidenza. Inoltre moltiplicando vettorialmente la
precedente relazione per $\mathbf{n}$, si ha
\begin{equation}
N^{\prime}\mathbf{t}^{\prime}\times\mathbf{n=}N~\mathbf{t}\times\mathbf{n}%
\end{equation}
da cui segue
\begin{equation}
N^{\prime}\sin i^{\prime}=N\sin i \label{snell}%
\end{equation}
dove $i,~i^{\prime}$ sono rispettivamente gli angoli incidente e rifratto
ossia gli angoli che $\mathbf{t}$ e $\mathbf{t}^{\prime}$ formano con
$\mathbf{n}$. La formula \ref{snell} insieme alla propriet\`{a} che i raggi
incidenti e rifratti e la normale alla superficie sono complanari
rappresentano \textbf{la legge di rifrazione }o \textbf{legge di Snell
}dell'ottica geometrica. Valutiamo ora il moltiplicatore di Lagrange $\lambda
$. Moltiplicando scalarmente per $\mathbf{n}$ si verifica
\begin{equation}
\lambda=N^{\prime}\cos i^{\prime}-N\cos i=N^{\prime}\sqrt{1-\sin^{2}i^{\prime
}}-N\cos i^{\prime}%
\end{equation}
che per la \ref{snell} si pu\`{o} riscrivere come
\begin{equation}
\lambda=\sqrt{N^{\prime2}-N^{2}\sin^{2}i}-N\cos i
\end{equation}
E' cos\`{\i} dimostrato che il principio di Fermat include tutte le
propriet\`{a} dei raggi luminosi, la legge di riflessione, citata all' inizio
di questo paragrafo si ottiene dalla legge di Snell sostituendo $N^{\prime
}=-N.$

\section{Il problema dei dati al contorno}

Si osservi che la determinazione del raggio luminoso tra due punti
$\mathbf{x~}$e $\mathbf{x}^{\prime}$per mezzo del principio di Fermat \`{e}
stata ridotta alla ricerca della soluzione $\left(  x_{i}\left(  t\right)
\right)  ,~i=1,2,3,$ del seguente sistema di equazioni differenziali
\begin{subequations}
\begin{gather}
\frac{\partial L}{\partial x_{i}}-\frac{d}{dt}\frac{\partial L}{\partial
\dot{x}_{i}}=0,\label{lagrange sist}\\
x_{i}\left(  a\right)  =x_{a}^{i}~~~~~~x_{i}\left(  b\right)  =x_{b}^{i}\\
\left[  \left[  \frac{\partial L}{\partial\dot{x}_{i}}\right]  \right]
_{\psi\left(  0\right)  }=\lambda n_{i}%
\end{gather}
che, come \`{e} gi\`{a} stato sottilineato, non ammette un teorema di
esistenza ed unicit\`{a} per la soluzione. Si pu\`{o} verificare tramite
esempi che il problema pu\`{o} ammettere infinite soluzioni.

Verrano ora ricavate altre utili informazioni dal problema cos\`{\i} com'
\`{e} posto. Ricordardando che una curva di $R^{3}$ \`{e} definita come un
applicazione
\end{subequations}
\[
\gamma:t\in\left[  a,b\right]  \rightarrow\left(  x_{1}\left(  t\right)
,x_{2}\left(  t\right)  ,x_{3}\left(  t\right)  \right)  .
\]
Di consegueza un cambio del parametro $t=t\left(  \tau\right)  $ genera una
nuova curva. In altre parole una curva \`{e} una carta $t\rightarrow R^{3},$
non un luogo di punti di $R^{3}.$ D' altro canto \`{e} chiaro che il
funzionale che compare nel principio di Fermat \`{e} indipendente dalla
\ scelta del parametro $t$, poich\'{e} il principio di Fermat definisce un
raggio come un luogo di punti. Pertanto il problema con dati al contorno a cui
ci stiamo interessando \`{e} definito a meno di un cambio di parametro,
cosicch\'{e} le equazioni \ref{lagrange sist} non sono indipendenti. Infatti
l'identit\`{a} (la somma su $i$ \`{e} sottointesa)%
\[
L-\dot{x}_{i}\frac{\partial L}{\partial\dot{x}_{i}}=0,
\]
implica%
\begin{equation}
\frac{\partial L}{\partial x_{h}}=\dot{x}_{i}\frac{\partial^{2}L}{\partial
\dot{x}_{i}\partial x_{h}},~~\frac{\partial L}{\partial\dot{x}_{h}}%
=\frac{\partial L}{\partial\dot{x}_{h}}+\dot{x}_{i}\frac{\partial^{2}%
L}{\partial\dot{x}_{i}\partial x_{h}}%
\end{equation}
e quindi%
\begin{equation}
\dot{x}_{i}\frac{\partial^{2}L}{\partial\dot{x}_{i}\partial x_{h}}=0.
\end{equation}
Tenendo presenti le due ultime relazioni si verifica immediatamente che vale
la seguente%
\[
\dot{x}_{i}\left(  \frac{\partial L}{\partial x_{i}}-\frac{d}{dt}%
\frac{\partial L}{\partial\dot{x}_{i}}\right)  =0,
\]
di consenguenza una delle tre equazioni \`{e} verificata come conseguenza
delle altre e pu\`{o} essere eliminata. Poich\'{e} le soluzioni sono
indipendenti dal parametro $t$, supponiamo che sia possibile introdurre una
superficie di coordinate in cui tutti i raggi a cui siamo interessati pu\`{o}
essere rappresentata dalle seguenti equazioni parametriche%
\begin{equation}
x_{\alpha}=x_{\alpha}\left(  x_{3}\right)  ,~~\alpha=1,2.
\end{equation}
Sotto queste ipotesi, abbiamo%
\begin{equation}
L=N\left(  x_{i}\right)  \sqrt{1+\dot{x}_{1}^{2}+\dot{x}_{2}^{2}},
\end{equation}
d'ora in avanti il punto indicher\`{a} la derivazione rispetto ad x$_{3}$ e e
le equazioni indipendenti sono
\begin{subequations}
\begin{gather}
\frac{\partial L}{\partial x_{\alpha}}-\frac{d}{dt}\frac{\partial L}%
{\partial\dot{x}_{\alpha}}=0,\\
x_{\alpha}\left(  a\right)  =x_{a}^{\alpha}~~x_{\alpha}\left(  b\right)
=x_{b}^{\alpha}\\
\left[  \left[  \frac{\partial L}{\partial\dot{x}_{i}}\right]  \right]
_{\psi\left(  0\right)  }=\lambda n_{\alpha},~~\alpha=1,2.
\end{gather}

\section{Analogie con la meccanica classica}

In questo paragrafo verrano messe in evidenza alcune analogie tra l' ottica
Hamiltoniana e la meccanica classica. Sia $K$ un sistema meccanico con vincoli
olonomi, fissi e lisci, sul quale agiscono forze conservative derivanti dall'
energia potenziale U. Si supponga che il il sistema K abbia $n$ gradi di
libert\`{a} e indichiamo con $V_{n}$ il suo spazio delle configurazioni.
Introducendo su $V_{n}$ un sistema di coordinate lagrangiane $\mathbf{q=}%
\left(  q_{1},\ldots,q_{n}\right)  $ il moto di $K$ sar\`{a} descritto
geometricamente da una curva $\mathbf{q}\left(  t\right)  \mathbf{=}\left(
q_{1}\left(  t\right)  ,\ldots,q_{n}\left(  t\right)  \right)  $ su $V_{n}$.
Le curve di questo tipo sono definite a partire da un altro principio
varazionale. Introducendo l' energia cinetica lagrangiana
\end{subequations}
\begin{equation}
T=\frac{1}{2}a_{hk}\left(  \mathbf{q}\right)  \dot{q}_{h}\dot{q}_{k},
\end{equation}
consideriamo la classe dei moti aventi un valore fissato dell' energia E, e la
variet\`{a} di Riemann ottenuta dotando $V_{n}$ della metrica%
\begin{equation}
d\sigma^{2}=2\left(  E-U\left(  \mathbf{q}\right)  \right)  a_{hk}\left(
\mathbf{q}\right)  q_{h}q_{k}, \label{metrica}%
\end{equation}
possiamo formulare il \textbf{Principio di Maupertuis}:

\emph{Sia K un sistema meccanico con vincoli olonomi, fissi e lisci, sul quale
agiscono forze conservative derivanti dall' energia potenziale U. Le curve di
}$V_{n}$\emph{\ che descrivono i moti di K con energia E tra due
configurazioni q}$_{1}$\emph{\ e q}$_{2}$\emph{, sono geodediche per la
metrica di Riemann \ref{metrica} ovvero sono estremali per il funzionale}%
\begin{equation}
\int_{\mathbf{q}_{1}}^{\mathbf{q}_{2}}d\sigma.
\end{equation}
\emph{Inoltre il punto, che rappresenta K in V}$_{n}$\emph{, si muove lungo
queste curve secondo la legge}%
\begin{equation}
\frac{d\sigma}{dt}=2\left(  E-U\right)  .
\end{equation}
In particolare, se considerando un punto materiale di massa m e introduciamo
coordinate cartesiane ortogonali, abbiamo%
\[
T=\frac{1}{2}\sum_{i=1}^{3}\left(  \dot{x}_{i}\right)  ^{2},
\]
da cui la metrica ed il funzionale assumono la forma
\begin{subequations}
\label{punto m}%
\begin{gather}
d\sigma^{2}=2\left(  E-U\right)  \sum_{i=1}^{3}dx_{i}^{2},\\
\int_{\mathbf{x}_{1}}^{\mathbf{x}_{2}}\sqrt{2\left(  E-U\right)  \sum
_{i=1}^{3}dx_{i}^{2}},
\end{gather}
dove $\mathbf{x}_{1},~\mathbf{x}_{2}$ sono le posizioni iniziale e finale di P.

Un confronto tra le \ref{punto m} e le equazioni corrispondenti per l' ottica
geometrica, mostra chiaramente che fissate le configurazioni iniziale e
finale, il punto P, soggetto a forze che discendono dall' energia potenziale
U, descrive la stessa traiettoria di un raggio di luce con gli stessi estremi
a condizione che sia soddisfatta la seguente uguaglianza%
\end{subequations}
\begin{equation}
\sqrt{2\left(  E-U\right)  }=N.
\end{equation}

\section{Da Fermat a Feynman}

Lo scopo di questo paragrafo \`{e} presentare una breve analisi storica e
concettuale dei principi variazionali. E' noto che le equazioni fondamentali
della fisica, nei suoi campi pi\`{u} disparati, possono essere ricavati dal
calcolo degli estremali di un funzionale che assume la forma generale%
\begin{equation}
S\left(  \psi,\dot{\psi}\right)  =\int d^{4}x\mathcal{L}\left(  \psi^{i}%
,\dot{\psi}^{j},x^{\mu}\right)  . \label{azione}%
\end{equation}
Dove $\psi$ indica la grandezza incognita, $\dot{\psi}$ le sue derivate
rispetto alle $x^{\mu}$ che sono le variabili indipendenti del sistema ed
$\mathcal{L}$ \`{e} una funzione di tutte queste grandezze detta
\emph{densit\`{a} lagrangiana}. Grazie a questo procedimento si arriva alle
note equazioni di Lagrange che costituiscono il punto di partenza per la
risoluzione di numerosi problemi. Storicamente la circostanza che la natura
evolva soddisfando un cos\`{\i} elegante principio ha generato le pi\`{u}
svariate interpretazioni, tra cui, ovviamente, la prova di un' origine divina
di tali principi e ha condotto all' attribuire un significato metafisico a
tali principi. Una corretta interpretazione dei principi variazionali tramite
il loro inquadramento nell' ambito della teoria quantistica dei campi si \`{e}
avuta intorno al 1950 ad opera di Feynman. L' idea consiste nel caratterizzare
la probabilit\`{a} di transizione da uno stato iniziale $|i>$ ad uno stato
finale $|f>$ attraverso alla funzione $S$ definita nella \ref{azione}, pi\`{u}
precisamente
\begin{equation}
\left\langle i\left\vert \exp-iHt/\hbar\right\vert f\right\rangle
=\int\mathcal{D}\psi\exp\left\{  \frac{i}{\hbar}\mathcal{L}\left(  \psi
,\dot{\psi}\right)  \right\}  .
\end{equation}
Questa formula si propone come un'alternativa alle regole di quantizzazione
canonica in quanto permette di definire il propagatore a partire dalla
Lagrangiana classica. Si noti l' eleganza di questa idea che rende molto
intuitivo il legame tra la meccanica classica e quantistica: ogni cammino
virtuale contribuisce all' ampiezza totale con un ampiezza di modulo uno e
dunque nessun cammino \`{e} pi\`{u} probabile degli altri. Tuttavia, nel
limite in cui la costante $\hbar$ sia piccola rispetto alle azioni in gioco
nel sistema fisico, il principio della fase stazionaria ci dice che i
contributi di tutti i cammini si cancellano per interferenza, tranne quelli
per cui la fase $S\left(  \psi,\dot{\psi}\right)  $ risulti stazionaria
rispetto a piccole variazioni del cammino, il che costituisce il principio di
Maupertuis, che determina le equazioni del moto classiche. Ecco svelato il
segreto di tutti i principi variazionali della meccanica classica: si tratta
semplicemente di manifestazioni della fase stazionaria applicata alla
formulazione quantistica in termini di somma sui cammini!

\chapter{Ottica Hamiltoniana}

\section{Formulazione Hamiltoniana dell' ottica geometrica}

Risulta particolarmente conveniente il passaggio dal formalismo Lagrangiano a
quello Hamiltoniano.

\begin{definition}
Si ponga com' \`{e} d' uso per i momenti cinetici
\begin{equation}
p_{\alpha}=\frac{\partial L}{\partial\dot{x}_{\alpha}}=N\frac{\dot{x}_{\alpha
}}{\sqrt{1+\dot{x}_{1}^{2}+\dot{x}_{2}^{2}}}. \label{defp}%
\end{equation}

\end{definition}

Per scrivere l' Hamiltoniana in funzione delle coordinate canoniche si devono
esprimere le variabili $\dot{x}_{\alpha}$ in termini dei momenti cinetici. Si
osservi che
\begin{equation}
p_{3}^{2}=N^{2}-p_{1}^{2}-p_{2}^{2}=\frac{N^{2}}{1+\dot{x}_{1}^{2}+\dot{x}%
_{2}^{2}}%
\end{equation}
quindi
\begin{equation}
\sqrt{N^{2}-p_{1}^{2}-p_{2}^{2}}=\frac{N}{\sqrt{1+\dot{x}_{1}^{2}+\dot{x}%
_{2}^{2}}}%
\end{equation}
usando la \ref{defp} si ottiene
\begin{equation}
\dot{x}_{\alpha}=\frac{p_{\alpha}}{\sqrt{N^{2}-p_{1}^{2}-p_{2}^{2}}}.
\end{equation}
Si pu\`{o} ora scrivere%
\begin{equation}
H(x_{\alpha},p_{\alpha},x_{3})=\sum_{\alpha=1}^{2}p_{\alpha}\dot{x}_{\alpha
}-L=-\sqrt{N^{2}-p_{1}^{2}-p_{2}^{2}}. \label{hamiltoniana}%
\end{equation}
Le equazioni di Hamilton sono%
\begin{align}
\dot{x}_{\alpha}  &  =\frac{\partial H}{\partial p_{\alpha}},\\
\dot{p}_{\alpha}  &  =-\frac{\partial H}{\partial x_{\alpha}},
\end{align}
quindi le condizioni di salto diventano%
\begin{equation}
\lbrack\lbrack p_{\alpha}]]=\lambda n_{\alpha}. \label{p0}%
\end{equation}

\section{Funzioni principali di Hamilton}

Sebbene le equazioni di Hamilton siano facilmente integrabili, al loro uso si
preferisce quello delle funzioni principali di Hamilton. La conoscenza di una
di queste funzioni permetterebbe una analisi completa delle aberrazioni di un
sistema ottico, purtroppo non \`{e} possibile determinarle in maniera esatta.
La loro importanza \`{e} comunque grande, dato che a partire dalle loro
espressioni approssimate ad ordini successivi si ricavano versioni della
teoria via via pi\`{u} accurate. Dalla prima approssimazione si ottiene l'
ottica gaussiana, dalla seconda la teoria delle aberrazioni del terzo ordine,
dalla successiva la teoria del quinto ordine e cos\`{\i} via. Ovviamente i
calcoli per ottenere approssimazioni migliori diventano sempre pi\`{u}
complicati man mano che cresce il livello di precisione che si vuole ottenere,
cosicch\'{e} solo le prime due sono realmente utili.

D' ora in avanti, con la dizione spazio oggetto si denoter\`{a} quella regione
che contiene l' oggetto di cui il sistema ottico in esame deve formare l'
immagine. La luce proviene questa regione, che si assume essere situata alla
sinistra del sistema ottico. Analogamente, lo spazio immagine \`{e} la regione
contenente l' immagine formata dal sistema ottico che pu\`{o} coincidere o
meno con la prima regione come nel caso di sistemi riflettenti. A seconda dei
casi sar\`{a} conveniente riferirsi ai due spazi usando il medesimo sistema di
coordinate oppure utilizzando due sistemi distinti.

Al fin di introdurre la prima funzione principale, supponiamo che:

\textbf{per una qualunque coppia di punti }$\mathbf{x}$\textbf{ e
}$\mathbf{x\prime}$\textbf{, il primo appartenente allo spazio oggetto il
secondo allo spazio immagine, esiste uno ed un solo raggio }$\mathbf{\gamma}$
\textbf{con estremi }$\mathbf{x}$\textbf{ ed }$\mathbf{x\prime}$\textbf{.}

Questa ipotesi implica che la lunghezza del cammino ottico lungo un raggio
dipenda solo dalla scelta degli estremi poich\'{e} il raggio $\gamma$ che li
collega \`{e} univocamente determinato.

La funzione principale di punto o puntuale \`{e} definita come
\begin{equation}
V\left(  \mathbf{x,x}^{\prime}\right)  =OPL\left(  \gamma\right)  ,
\end{equation}
dove $OPL\left(  \gamma\right)  $ \`{e} valutata lungo l' unico raggio che li collega.

Per apprezzare la grande importanza della funzione principale $V\left(
\mathbf{x,x}^{\prime}\right)  $, bisogna valutarne il differenziale su una
coppia di punti $\mathbf{x,x\prime}$. Di conseguenza basta considerare la
famiglia $\Delta$ di curve ad un parametro con estremi variabili (si veda
\ref{curve delta}) e calcolare il differenziale della funzione $I\left(
\varepsilon\right)  $ per $\varepsilon=0$; in altre parole usiamo la
\ref{di deps def} lungo una famiglia di raggi. Dalla \ref{di deps def} una
volta che si prenda in considerazione anche le equazioni di Lagrange si ha%
\begin{equation}
dV=\frac{\partial V}{\partial x_{i}}dx_{i}+\frac{\partial V}{\partial
x_{i}^{^{\prime}}}dx_{i}^{^{\prime}}=-\frac{\partial L}{\partial\dot{x}_{i}%
}dx_{i}+\frac{\partial L}{\partial\dot{x}_{i}^{^{\prime}}}dx_{i}^{^{\prime}}
\label{dv}%
\end{equation}
essendo le variazioni $dx$ ed $dx^{\prime}$ arbitrarie possiamo scrivere le
seguenti uguaglianze%
\begin{gather}
\frac{\partial V}{\partial x_{i}}=-\frac{\partial L}{\partial\dot{x}_{i}};\\
\frac{\partial V}{\partial x_{i}^{^{\prime}}}=\frac{\partial L}{\partial
\dot{x}_{i}^{^{\prime}}}%
\end{gather}
ovvero
\begin{subequations}
\label{ppp}%
\begin{gather}
p_{i}=-\frac{\partial V}{\partial x_{i}}=p_{i}\left(  \mathbf{x,x}^{\prime
}\right)  ,\\
p_{i}^{\prime}=\frac{\partial V}{\partial x_{i}^{\prime}}=p_{i}^{\prime
}\left(  \mathbf{x,x}^{\prime}\right)  .
\end{gather}
Le \ref{ppp} associano ad ogni coppia di punti, uno nello spazio oggetto e l'
altro nello spazio immagine, le corrispondenti direzioni $\mathbf{p}$ e
$\mathbf{p\prime}$\textbf{ }dell' unico raggio che li connette. Va
sottolineato che le \ref{ppp} non sono indipendenti poich\'{e} valgono le
seguenti identit\`{a}
\end{subequations}
\begin{equation}
\mathbf{p\cdot p=}N^{2}~~\mathbf{p}^{\prime}\mathbf{\cdot p}^{\prime
}\mathbf{=}N^{^{\prime}2}, \label{ppN2}%
\end{equation}
che possono anche essere scritte come
\begin{equation}
\sum_{i=1}^{3}\left(  \frac{\partial V}{\partial x_{i}}\right)  =N^{2}%
,~\ \sum_{i=1}^{3}\left(  \frac{\partial V}{\partial x_{i}^{\prime}}\right)
=N^{\prime2}.
\end{equation}
Se ne conclude che V \`{e} una soluzione dell'equazione dell' iconale nello
spazio oggetto ed in quello immagine. Supponendo che quattro delle \ref{ppp}
formino un sistema di equazioni indipendenti, senza perdita di generalit\`{a}
scegliamo le prime due equazioni sia per della \ref{ppp}a che per la
\ref{ppp}b,
\begin{subequations}
\label{pppp}%
\begin{gather}
p_{\alpha}=-\frac{\partial V}{\partial x_{\alpha}}=p_{\alpha}\left(
x_{\alpha},x_{\alpha}^{\prime},x_{3},x_{3}^{\prime}\right)  ,\\
p_{\alpha}^{\prime}=\frac{\partial V}{\partial x_{\alpha}^{\prime}}=p_{\alpha
}^{\prime}\left(  x_{\alpha},x_{\alpha}^{\prime},x_{3},x_{3}^{\prime}\right)
.
\end{gather}
con $\alpha,\beta=1,2$ vale la condizione
\end{subequations}
\begin{equation}
\det\left(  \frac{\partial^{2}V}{\partial x_{\alpha}\partial x_{\beta}%
^{\prime}}\right)  \neq0.
\end{equation}
Con queste ipotesi \`{e} possibile invertire le prime due equazioni e porre il
sistema nella forma
\begin{subequations}
\label{ppp2}%
\begin{gather}
x_{\alpha}=x_{\alpha}^{\prime}\left(  x_{\alpha},x_{\alpha}^{\prime}%
,x_{3},x_{3}^{\prime}\right)  ,\\
p_{\alpha}^{\prime}=p_{\alpha}^{\prime}\left(  x_{\alpha},x_{\alpha}^{\prime
},x_{3},x_{3}^{\prime}\right)  .
\end{gather}
Assegnando due piani nello spazio oggetto e nello spazio immagine,
rispettivamente $x_{3}=\bar{x}_{3},~x_{3}^{\prime}=\bar{x}_{3}^{\prime}$. D'
ora in avanti questi piani saranno indicati rispettivamente come piano base
anteriore e piano base posteriore. Possiamo dire che
\end{subequations}
\begin{proposition}
le \ref{ppp2} definiscono una corrispondenza che ad ogni punto $x_{\alpha}$
del piano base anteriore $x_{3}=\bar{x}_{3}$ e ad ogni direzione del raggio
uscente da questo punto associa uno ed un solo punto nel piano base posteriore
$x_{3}^{\prime}=\bar{x}_{3}^{\prime}$ ed una sola direzione del raggio.
\end{proposition}

Oltre alla funzione principale di punto V, possono essere introdotte altre
funzioni principali che si rivelano utili in diverse situazioni. Infatti V non
pu\`{o} essere utilizzata se i piani $x_{3}=\bar{x}_{3},~x_{3}^{\prime}%
=\bar{x}_{3}^{\prime}$ contengono punti stigmatici \footnote{
\begin{definition}
Due punti $\mathbf{x,}$ $\mathbf{x}^{\prime}$si dicono stigmatici se tutti i
raggi uscenti da $\mathbf{x}$ si intersecano in $\mathbf{x}^{\prime}$
\end{definition}
}.

Si definisca la funzione principale angolare, o pi\`{u} brevemente funzione
angolare
\begin{equation}
T=V+\mathbf{p\cdot x-p}^{\prime}\cdot\mathbf{x}^{\prime}.
\end{equation}
Dalla \ref{ppp} e dalla definizione di T si ottiene
\begin{equation}
dT=\mathbf{x\cdot}d\mathbf{p-x}^{\prime}\mathbf{\cdot}d\mathbf{p}^{\prime},
\label{dT}%
\end{equation}
dove le variazioni $d\mathbf{p}$ e $d\mathbf{p}^{\prime}$ non sono arbitrarie
perch\'{e} dalla \ref{ppN2} si ha
\begin{equation}
\mathbf{p\cdot}d\mathbf{p}=0~~\mathbf{p}^{\prime}\mathbf{\cdot}d\mathbf{p}%
^{\prime}=0.
\end{equation}
Di conseguenza
\begin{equation}
dp_{3}=\frac{1}{p_{3}}\sum_{\alpha=1}^{2}p_{\alpha}dp_{\alpha},~\ \ \ ~~dp_{3}%
^{\prime}=\frac{1}{p_{3}^{\prime}}\sum_{\alpha=1}^{2}p_{\alpha}^{\prime
}dp_{\alpha}^{\prime}, \label{dp3}%
\end{equation}
cosicch\'{e}, sostituendo queste espressioni nella \ref{dT} otteniamo le
seguenti relazioni%
\begin{equation}
dT=\sum_{\alpha=1}^{2}\left[  \left(  x_{a}-\frac{p_{\alpha}}{p_{3}}\right)
dp_{\alpha}-\left(  x_{a}^{\prime}-\frac{p_{\alpha}^{\prime}}{p_{3}^{\prime}%
}\right)  dp_{\alpha}^{\prime}\right]
\end{equation}
dove le variazioni $dp_{\alpha},~dp_{\alpha}^{\prime}$ sono ora arbitrarie. Si
ottiene pertanto
\begin{subequations}
\label{eq hamjac}%
\begin{align}
x_{a}  &  =\frac{p_{\alpha}}{p_{3}}x_{3}+\frac{\partial T}{\partial p_{\alpha
}},\\
x_{a}^{\prime}  &  =\frac{p_{\alpha}^{\prime}}{p_{3}^{\prime}}x_{3}^{\prime
}-\frac{\partial T}{\partial p_{\alpha}^{\prime}}.
\end{align}
Concludendo, se la funzione angolare $T\left(  p_{\alpha},p_{\alpha}^{\prime
}\right)  $ \`{e} nota e sono assegnati i piani base principali, le precedenti
relazioni associano ad una coppia di direzioni $\mathbf{p,p\prime}$ i punti
$\mathbf{x,x\prime}$\textbf{ }in cui i raggi considerati intersezionano i
piani base.

Si possono definire le funzioni principali miste o funzioni miste in questo
modo%
\end{subequations}
\begin{align}
W_{1}  &  =V+\mathbf{p\cdot x}\\
W_{2}  &  =V-\mathbf{p}^{\prime}\mathbf{\cdot x}^{\prime}.
\end{align}
Dalla (\ref{dv}) e dalle precedenti definizioni si ha
\begin{align}
dW_{1}  &  =\mathbf{-x}\cdot d\mathbf{p+p}^{\prime}\cdot d\mathbf{x}^{\prime
},\\
dW_{2}  &  =\mathbf{-p}\cdot dx\mathbf{+x}^{\prime}\cdot d\mathbf{p}^{\prime}.
\end{align}
Tenendo conto delle \ref{dp3} e fissato il piano base posteriore, otteniamo le
seguenti relazioni
\begin{align}
x_{\alpha}  &  =\frac{p_{\alpha}}{p_{3}}x_{3}-\frac{\partial W_{1}}{\partial
p_{\alpha}},\\
p_{\alpha}^{\prime}  &  =\frac{\partial W_{1}}{\partial x_{\alpha}^{\prime}},
\end{align}%
\begin{align}
x_{\alpha}^{\prime}  &  =\frac{p_{\alpha}^{\prime}}{p_{3}^{\prime}}%
x_{3}^{\prime}-\frac{\partial W_{2}}{\partial p_{\alpha}^{\prime}},\\
p_{\alpha}^{\prime}  &  =-\frac{\partial W_{2}}{\partial x_{\alpha}}.
\end{align}
Si possono ripetere le stesse considerazioni fatte fin qui per il sistema
ottenuto dalla funzione angolare. Anche per le funzioni $T,W_{1},W_{2}$
cos\`{\i} come per la funzione $V$ possono esistere situazioni in cui esse non
sono utilizzabili. Ad esempio, la funzione $T$ non pu\`{o} essere utilizzata
nei casi in cui, per $\mathbf{p,p\prime}$\textbf{ }fissati, esistano infinite
coppie di punti $\mathbf{x}$ e$\mathbf{\ x\prime}$ che soddisfano la
\ref{eq hamjac}, (sistemi telescopici). La funzione $W_{1}$ non pu\`{o} essere
usata se a $\mathbf{x,p\prime}$\textbf{ }fissati vengono associate infinite
coppie $\mathbf{p,x\prime}$\textbf{ }(punti focali nel piano oggetto). Una
condizione simile vale per $W_{2}$ (punti focali nel piano immagine).

\section{Simmetria assiale di un sistema ottico}

D' ora in avanti verranno considerati sistemi ottici aventi un asse di
simmetria $s$ , che sar\`{a} chiamato asse ottico del sistema. Analizzeremo in
particolare le conseguenze sulle funzioni principali.

Per un sistema con simmetria assiale, l' asse s \`{e} anche un raggio per il
sistema ottico. Nel seguito l' asse $Ox_{3}$ del sistema di riferimento
$Ox_{1}x_{2}x_{3}$\ sar\`{a} scelto coincidente con l' asse di simmetria, la
stessa convenzione sar\`{a} adottata per il sistema di riferimento nello
spazio immagine $Ox_{1}^{\prime}x_{2}^{\prime}x_{3}^{\prime}$\ e gli altri due
assi paralleli con quelli dello spazio oggetto. Di conseguenza l' indice di
rifrazione $N\left(  \mathbf{x}\right)  $ verifica la condizione%
\begin{equation}
N\left(  \mathbf{x}\right)  =N\left(  U,x_{3}\right)  ,~~U=x_{1}^{2}+x_{2}%
^{2}.
\end{equation}
Inoltre, nel caso di un sistema ottico con simmetria assiale, la funzione
principale $V\left(  \mathbf{x,x}^{\prime}\right)  $ pu\`{o} essere posta
nella forma%
\begin{equation}
V\left(  \mathbf{x,x}^{\prime}\right)  =V\left(  U,U^{\prime},x_{3}%
,x_{3}^{\prime}\right)  ,
\end{equation}
dove
\begin{subequations}
\begin{align}
U  &  =x_{1}^{2}+x_{2}^{2},\\
U^{\prime}  &  =x_{1}^{\prime2}+x_{2}^{\prime2},\\
\omega &  =x_{1}x_{1}^{\prime}+x_{2}x_{2}^{\prime}.
\end{align}
Considerazioni simili valgono anche per le altre funzioni principali, ad
esempio per la funzione angolare
\end{subequations}
\begin{equation}
T\left(  \mathbf{p,p}^{\prime}\right)  =T\left(  \xi,\xi,\eta\right)  ,
\end{equation}
con
\begin{subequations}
\begin{align}
\xi &  =p_{1}^{2}+p_{2}^{2},\\
\xi^{\prime}  &  =p_{1}^{\prime2}+p_{2}^{\prime2},\\
\eta &  =p_{1}p_{1}^{\prime}+p_{2}p_{2}^{\prime}.
\end{align}
Si osservi che le componenti $p_{3},~p_{3}^{^{\prime}}$ non compaiono
esplicitamente in quanto possono essere scritte come
\end{subequations}
\begin{align*}
p_{3}^{2}  &  =N^{2}-\xi,\\
p_{3}^{\prime2}  &  =N^{\prime2}-\xi^{\prime}.
\end{align*}

\section{L' invariante ottico di Lagrange}

Diretta conseguenza dell' invarianza per rotazioni attorno all' asse s
(coincidente con $Ox_{3}$) sar\`{a} la presenza di un integrale primo, $L_{3}%
$. Baster\`{a} applicare il teorema di Noether nel formalismo lagrangiano per
averne un' espressione esplicita. Si \`{e} preferito l' uso del formalismo
Lagrangiano perch\'{e} a differenza di quello Hamiltoniano d\`{a} direttamente
l' espressione dell' integrale primo, mentre il secondo fornisce un sistema di
equazioni dal quale esplicitarlo. Sia $S$ un sistema ottico centrato, $s$ l'
asse di simmetria e si scelga l' asse di simmetria coincidente con con il
terzo asse del riferimento $Ox_{1}x_{2}x_{3}$. Adottando coordinate polari
$\left(  \rho,\theta\right)  $
\begin{align}
x_{1}  &  =\rho\cos\theta,\\
x_{2}  &  =\rho\sin\theta,
\end{align}
dove
\begin{align}
\rho &  =\sqrt{x_{1}^{2}+x_{2}^{2}},\\
\theta &  =\arctan\frac{x_{2}}{x_{1}}.
\end{align}
La Lagrangiana nelle nuove coordinate si scrive%
\begin{equation}
L=N\sqrt{1+\dot{\rho}^{2}+\rho^{2}\dot{\theta}^{2}}.
\end{equation}
Dato che $L$ non dipende da $\theta$ la quantit\`{a}
\begin{equation}
L_{3}=\frac{\partial L}{\partial\dot{\theta}}=N\frac{\rho^{2}\dot{\theta}^{2}%
}{\sqrt{1+\dot{\rho}^{2}+\rho^{2}\dot{\theta}^{2}}},
\end{equation}
\`{e} indipendente da $x_{3}$. Ritornando alle coordinate cartesiane ed
esprimendo $L_{3}$ in termini delle $p_{\alpha}$ usando le \ref{defp} si ha
\begin{equation}
L_{3}=p_{2}x_{1}-p_{1}x_{2}.
\end{equation}
Ossia $L_{3}$ \`{e} costante in ogni regione in cui l' indice di rifrazione
\`{e} costante. Verr\`{a} dimostrato ora che questa quantit\`{a} si conserva
anche nel passaggio tra mezzi con indice di rifrazione diversi. Senza
perdit\`{a} di generalit\`{a} ci limiteremo a considerare il passaggio
attraverso una sola superficie. Le quantit\`{a} della prima regione verranno
distinte da quelle della seconda indicando quest' ultime con il medesimo
simbolo con un apice. Sia
\begin{align}
x_{1}\left(  z\right)   &  =\frac{p_{1}}{r}\left(  z-z_{0}\right)  +x_{0},\\
x_{2}\left(  z\right)   &  =\frac{p_{2}}{r}\left(  z-z_{0}\right)  +y_{0}.
\end{align}
l' espressione del raggio nella prima regione. Nella seconda regione avremo%
\begin{align}
x_{1}^{\prime}\left(  z\right)   &  =\frac{p_{1}^{\prime}}{r^{\prime}}\left(
z-z_{0}^{\prime}\right)  +x_{0}^{\prime},\\
x_{2}^{\prime}\left(  z\right)   &  =\frac{p_{2}^{\prime}}{r^{\prime}}\left(
z-z_{0}^{\prime}\right)  +y_{0}^{\prime}.
\end{align}
Si osservi che $\left(  x_{0}^{\prime},y_{0}^{\prime},z_{0}^{\prime}\right)  $
\`{e} il punto d' intersezione della superficie con la superficie ottica. Per
continuit\`{a} si deve avere
\begin{equation}
x_{\alpha}\left(  z_{0}^{\prime}\right)  =x_{\alpha}^{\prime}\left(
z_{0}^{\prime}\right)  ,
\end{equation}
quindi
\begin{subequations}
\label{x0}%
\begin{align}
x_{0}^{\prime}  &  =\frac{p_{1}}{r}\left(  z_{0}^{\prime}-z_{0}\right)
+x_{0},\\
y_{0}^{\prime}  &  =\frac{p_{2}}{r}\left(  z_{0}^{\prime}-z_{0}\right)
+y_{0}.
\end{align}
$p_{\alpha}^{\prime}$ sar\`{a} dato dalla \ref{p0}. Sia $F\left(  x^{2}%
+y^{2},z\right)  =0$ l'espressione implicita della superficie, le prime due
componenti del versore normale nel punto di intersezione sono
\end{subequations}
\begin{align}
n_{1}  &  =\frac{1}{\left\vert \nabla F\right\vert }\frac{\partial F}{\partial
x_{1}}=\frac{1}{\left\vert \nabla F\right\vert }\frac{\partial F}{\partial
U}\frac{\partial U}{\partial x_{1}}=2x_{0}^{\prime}\bar{n},\\
n_{2}  &  =\frac{1}{\left\vert \nabla F\right\vert }\frac{\partial F}{\partial
x_{2}}=\frac{1}{\left\vert \nabla F\right\vert }\frac{\partial F}{\partial
U}\frac{\partial U}{\partial x_{2}}=2x_{0}^{\prime}\bar{n},
\end{align}
dove si \`{e} posto $\bar{n}=\frac{1}{\left\vert \nabla F\right\vert }%
\frac{\partial F}{\partial U}$ ed $U_{2}=x^{2}+y^{2}$. Sostituendo le
relazioni test\`{e} trovate nell' espressione di $L_{3}^{\prime}$ dopo calcoli
elementari si ottiene%
\begin{equation}
L_{3}=L_{3}^{\prime},
\end{equation}
ovvero la quantit\`{a} $L_{3}$ si conserva passando attraverso una superficie ottica.

\section{Superfici sferiche, nuovi invarianti ottici}

Viene ora preso in considerazione il caso in cui ci sia una sola superficie
rifrangente sferica. Si fissi il sistema di riferimento con l'origine
coincidente col centro della sfera. Sia
\[
x^{2}+y^{2}+z^{2}=R^{2}.
\]
l' equazione della superficie sferica. Siano%
\begin{align*}
x\left(  z_{0}\right)   &  =x_{0}~~p\left(  z_{0}\right)  =p_{0}\\
y\left(  z_{0}\right)   &  =y_{0}~~q\left(  z_{0}\right)  =q_{0}%
\end{align*}
i dati iniziali ed $N$ ed $N^{\prime}$ gli indici di rifrazione. Risulta
conveniente scrivere il raggio in forma parametrica%
\[
\mathbf{x=p}t+\mathbf{x}_{0}%
\]
con ovvio significato della notazione usata e
\begin{equation}
t=\frac{\left(  z-z_{0}\right)  }{p_{3}}.
\end{equation}
Il valore di $t$ per il quale il raggio incontra la superficie \`{e}%
\begin{equation}
t_{int}=\frac{-\left(  \mathbf{p\cdot x}_{0}\right)  -\sqrt{N^{2}%
R^{2}-\mathbf{L}^{2}}}{N^{2}},
\end{equation}
dove $\mathbf{L}$ \`{e} il vettore di componenti $L_{k}\mathbf{=}%
\varepsilon_{ijk}x_{i}p_{j},$ dove $\varepsilon_{ijk}$ \`{e} il tensore di
Levi Civita. Si dimostrer\`{a} in seguito che $\mathbf{L}$ \`{e} un invariante
ottico nel caso in cui le superfici ottiche siano a simmetria sferica. Si
pu\`{o} ora calcolare il valore di $\lambda$. Essendo
\[
\cos i=\frac{\mathbf{p\cdot x}\left(  t_{int}\right)  }{NR}=\sqrt{N^{2}%
R^{2}-\mathbf{L}^{2}},
\]
si ottiene%
\begin{equation}
\lambda=\frac{\sqrt{N^{\prime2}R^{2}-\mathbf{L}^{2}}-\sqrt{N^{2}%
R^{2}-\mathbf{L}^{2}}}{R}.
\end{equation}
Il versore $\mathbf{n}$ assume la seguente forma
\begin{equation}
\mathbf{n}=\frac{\mathbf{p}t_{int}+\mathbf{x}_{0}}{R}.
\end{equation}
L' espressione del raggio nella regione a destra della superficie \`{e} quindi%
\[
\mathbf{x}\left(  z\right)  =(\mathbf{p}_{1}+\lambda_{1}\mathbf{n}_{1})\left(
t-t_{1}\right)  +\mathbf{x}\left(  t_{1}\right)  .
\]
Si pu\`{o} ora tornare alla consueta forma dell'espressione del raggio%
\begin{align}
x_{\alpha}\left(  z\right)   &  =(p_{\alpha}+\lambda n_{\alpha})\frac{\left(
z-z_{i}\right)  }{p_{3}^{\prime}}+p_{\alpha}\left(  \frac{z_{i-}z_{0}}{p_{3}%
}\right)  +x_{0\alpha}=\label{aberr}\\
&  (p_{\alpha}+\lambda n_{\alpha})\left(  t-t_{i}\right)  +p_{\alpha}\left(
t_{i}-t_{0}\right)  +x_{0\alpha}\nonumber\\
x_{\alpha}\left(  z\right)   &  =p_{\alpha}\left[  \left(  t-t_{i}\right)
+\frac{\lambda}{R}\left(  t_{i}-t_{0}\right)  \left(  t-t_{i}\right)  +\left(
t_{i}-t_{0}\right)  \right]  +x_{\alpha0}\left[  \frac{\lambda}{R}\left(
t_{i}-t_{0}\right)  +1\right]  .
\end{align}
Verr\`{a} dimostrato che le quantit\`{a}%
\begin{align}
L_{1}  &  =qz-ry\\
L_{2}  &  =rx-pz
\end{align}
sono invarianti ottici. Iniziamo con l'osservare che
\begin{equation}
\frac{d}{dz}L_{1}=q-r\frac{q}{r}=0.
\end{equation}
Nella seconda regione avremo%
\begin{gather}
L_{1}^{^{\prime}}=\left(  q+\lambda n_{2}\right)  [\left(  r+\lambda
n_{3}\right)  \left(  t-t_{i}\right)  +r\left(  t_{i}-t_{0}\right)
+z_{1}]-\nonumber\\
\left(  r+\lambda n_{3}\right)  \left[  \left(  q+\lambda n_{2}\right)
\left(  t-t_{i}\right)  +q\left(  t_{i}-t_{0}\right)  +y_{1}\right]  =\\
L_{1}+\frac{\lambda}{R}\left[  q\left(  t_{i}-t_{0}\right)  +y_{1}\right]
\left[  r\left(  t_{i}-t_{0}\right)  +z_{1}\right]  -\frac{\lambda}{R}\left[
q\left(  t_{i}-t_{0}\right)  +y_{1}\right]  \left[  r\left(  t_{i}%
-t_{0}\right)  +z_{1}\right]  =L_{1}.\nonumber
\end{gather}
Lo stesso risultato vale per $L_{2}$. Le due grandezze sono dunque invarianti ottici.

\section{Trasformazioni degli invarianti di rotazione\ attraverso \newline un
sistema ottico centrato}

Sia $S$ un sistema ottico centrato composto da $M$ superfici ottiche $S_{m}$
$\left(  m=1\ldots M\right)  $ con indice di rifrazione $N_{m}$. Le regioni
tra le varie superfici ottiche sono numerate con lo stesso indice della
superficie che ne delimita il lato destro. Tutte le quantit\`{a} che verranno
introdotte con indice $m$ saranno quindi riferite alla regione od alla
superficie m-sima. Un raggio che attraversa il sistema \`{e} una spezzata, in
ogni regione che compone il sistema ottico ha espressione%
\begin{align*}
X_{m}  &  =p_{m}\left(  \frac{z-z_{m}^{int}}{r_{m}}\right)  +x_{m},\\
Y_{m}  &  =q_{m}\left(  \frac{z-z_{m}^{int}}{r_{m}}\right)  +y_{m}.
\end{align*}
Una volta calcolato il punto d' intersezione $z_{m}^{int}$del raggio con la
superfice ottica $S_{m}$ che lo rifrange si pu\`{o} dunque caratterizzare ogni
raggio grazie alle quattro costanti $\left(  x_{m},y_{m},p_{m},q_{m}\right)
,$ che, scritte in funzione delle grandezze che caratterizzano il raggio nella
regione precedente, danno%
\begin{align}
x_{m}  &  =p_{m-1}\left(  \frac{z_{m}^{int}-z_{m-1}^{int}}{r_{m-1}}\right)
+x_{m-1},\\
y_{m}  &  =q_{m-1}\left(  \frac{z_{m}^{int}-z_{m-1}^{int}}{r_{m-1}}\right)
+y_{m-1},\\
p_{m}  &  =p_{m-1}+\lambda_{m}n_{x},\\
q_{m}  &  =q_{m-1}+\lambda_{m}n_{y}.
\end{align}
Queste relazioni altro non sono che la condizione di continuit\`{a} del raggio
sulla superficie e la condizione di salto di Weierstrass-Erdaman. Si pu\`{o}
esprimere $\lambda_{m}$ in funzione del dato iniziale invece che dell' angolo
d' incidenza sulla superficie. Sia $F_{m}\left(  x^{2}+y^{2},z\right)
=F_{m}\left(  u,z\right)  $ l'equazione implicita della superficie, di
conseguenza il versore normale ha equazione%
\begin{equation}
\mathbf{n}_{m}\mathbf{=}\frac{\nabla F_{m}}{\left\vert \nabla F_{m}\right\vert
}=\frac{1}{\left\vert \nabla F_{m}\right\vert }\left(  \frac{\partial F_{m}%
}{\partial u}\frac{\partial u}{\partial x},\frac{\partial F_{m}}{\partial
u}\frac{\partial u}{\partial y},\frac{\partial F_{m}}{\partial z}\right)  ,
\end{equation}
posto
\begin{align}
R_{m}  &  =\left\vert \nabla F_{m}\right\vert \\
n_{u}  &  =2\frac{\partial F_{m}}{\partial u}%
\end{align}
si ha
\begin{equation}
\mathbf{n}_{m}\mathbf{=}\frac{1}{R_{m}}\left(  n_{u}x,n_{u}y,\frac{\partial
F_{m}}{\partial z}\right)
\end{equation}
Si pu\`{o} ricavare il coseno dell'angolo d'incidenza osservando che
\begin{equation}
\mathbf{p\cdot n=}\left\vert \mathbf{p}\right\vert \left\vert \mathbf{n}%
\right\vert \cos i_{m}=N\cos i_{m}%
\end{equation}
allora
\begin{equation}
\lambda_{m}=\sqrt{N_{m}-\left(  N^{2}-\left(  \frac{\mathbf{p\cdot n}}{N^{2}%
}\right)  \right)  }-\mathbf{p\cdot n.}%
\end{equation}
L'espressione dei nuovi numeri direttori \`{e} dunque%
\begin{align}
p_{m+1}  &  =p_{m}\left(  1+\frac{n_{u}\lambda_{m}}{R_{m}}\frac{\left(
z_{m+1}-z_{m}\right)  }{r_{m}}\right)  +\frac{\lambda_{m}n_{u}}{R_{m}}x_{m}\\
q_{m+1}  &  =q_{m}\left(  1+\frac{n_{u}\lambda_{m}}{R_{m}}\frac{\left(
z_{m+1}-z_{m}\right)  }{r_{m}}\right)  +\frac{\lambda_{m}n_{u}}{R_{m}}y_{m}.
\end{align}
Se si scrivono gli invarianti di rotazione in funzione di questi dati si
arriva alla seguente caratterizzazione%
\begin{equation}
\mathbf{u}_{m}=T\mathbf{u}_{m-1}%
\end{equation}
dove si \`{e} posto $\mathbf{u}_{m}=\left(  u_{m},v_{m},w_{m}\right)  $ e $T$
\`{e} una opportuna matrice che dipende dalle caratteristiche fisiche del
sistema in esame e dai dati iniziali del raggio. Applicando la relazione
precedente a tutte le superfici che compongono il sistema ottico si ha per gli
invarianti ottici dello spazio immagine%
\begin{equation}
\mathbf{u}_{M}=\mathbf{Tu}_{1} \label{utum}%
\end{equation}
dove
\begin{equation}
\mathbf{T=}\prod\limits_{i=1}^{M}\mathbf{T}_{i}.
\end{equation}

\chapter{Ottica Gaussiana}

\section{Introduzione}

Sia $K$ un sistema ottico centrato e sia $s$ il suo asse di simmetria. Si
indicher\`{a} con $\pi$ e $\pi^{\prime}$ due piani contenuti rispettivamente
nello spazio oggetto e nello spazio immagine ed ortogonali ad $s$. Il piano
$\pi$ sar\`{a} identificato col piano base anteriore, mentre il piano
$\pi^{\prime}$ con quello posteriore. Sebbene tutte le derivazioni saranno
effettuate a partire dalla funzione angolare $T$, possono essere ripetute per
una qualunque delle funzioni principali. Risulta decisamente conveniente
cambiare la notazione adottata con la seguente%
\begin{subequations}
\begin{align}
x_{1}  &  =x,~~x_{2}=y,~~x_{3}=z,\\
x_{1}^{\prime}  &  =x^{\prime},~~x_{2}^{\prime}=y^{\prime},~~x_{3}^{\prime
}=z^{\prime},\\
p_{1}  &  =p,~\ p_{2}=q,~\ p_{3}=r,\\
p_{1}^{\prime}  &  =p^{\prime},~\ p_{2}^{\prime}=q^{\prime},~\ p_{3}^{\prime
}=r^{\prime}.
\end{align}
Si introduca nello spazio oggetto un sistema di coordinate $Oxyz$ tale che
$Oz$ coincida con l' asse s e $O\in\pi$; analoga scelta verr\`{a} fatta nello
spazio immagine, con un sistema di coordinate $O^{\prime}x^{\prime}y^{\prime
}z^{\prime}$ con $O^{\prime}z^{\prime}$ coincidente con l' asse $s$ e
$O^{\prime}\in\pi^{\prime}$. La prima conseguenza di queste scelte \`{e} che
le equazioni dei piani base sono%
\end{subequations}
\begin{equation}
z=0,~~z^{\prime}=0.
\end{equation}
Dalle \ref{eq hamjac}, nella nuova notazione si ha
\begin{subequations}
\label{eqhamjac2}%
\begin{align}
x  &  =\frac{\partial T}{\partial p},~\ x^{\prime}=-\frac{\partial T}{\partial
p^{\prime}},\\
y  &  =\frac{\partial T}{\partial q},~~y^{\prime}=-\frac{\partial T}{\partial
q^{\prime}}.
\end{align}
Va osservato che la funzione angolare dipende esplicitamente dalla scelta dei
piani base $\pi,$ $\pi^{\prime}$ ossia dalla distanza $a$ della prima
superficie di $K$ da $\pi$ e dalla distanza $a^{\prime}$ dell' ultima
superficie di $K$ da $\pi^{\prime}$. Quando i due piani base coincidono con i
piani tangenti ai vertici delle ultime superfici la funzione angolare si dice
intrinseca. Per determinare come $T$ dipende da questa scelta, si considerino
due piani base $\pi,~\pi_{1}$ e due riferimenti associati a questi piani
$Oxyz$ e $O_{1}x_{1}y_{1}z_{1}$ nello spazio oggetto ed altri due piani nello
spazio immagine $\pi^{\prime}$, $\pi^{\prime}$. Si denotino rispettivamente
con $d$ e $d^{\prime}$ le distanze tra le due coppie di piani. Sia $\gamma$ un
raggio di equazione $\mathbf{x=p}t+\mathbf{x}_{0}$, dove $t$ \`{e} un
parametro, $\mathbf{p}$ la direzione del raggio e $\mathbf{x}_{0}$ le
coordinate del raggio per $t=0$ quando interseca il piano $\pi$. Il raggio
intersecher\`{a} il piano $\pi_{1}$ per $t=t_{1}=\frac{d}{r}$. In maniera
assolutamente identica un raggio $\gamma^{\prime}$ intersecher\`{a} il piano
$\pi^{\prime}$ per $t^{\prime}=t_{1}^{\prime}=\frac{d^{\prime}}{r}$. Se ne
conclude che che le corrispondenze prodotte dai raggi tra le coppie di piani
$\pi$, $\pi_{1}$ e $\pi^{\prime},\pi_{1}^{\prime}$ sono espresse dalle
seguenti relazioni%
\end{subequations}
\begin{gather}
x_{1}=x+p\frac{d}{r},~~y_{1}=y+q\frac{d}{r},\\
x_{1}^{\prime}=x^{\prime}+p^{\prime}\frac{d^{\prime}}{r^{\prime}}%
,~~y_{1}^{\prime}=y^{\prime}+q^{\prime}\frac{d^{\prime}}{r^{\prime}}.
\end{gather}
Utilizzando le \ref{eqhamjac2} nelle relazioni precedenti seguono
\begin{subequations}
\label{1234}%
\begin{gather}
x_{1}=\frac{\partial T}{\partial p}+\frac{d}{r}p,~~y_{1}=\frac{\partial
T}{\partial q}+\frac{d}{r}q,\\
x_{1}^{\prime}=-\frac{\partial T}{\partial p^{\prime}}+\frac{d^{\prime}%
}{r^{\prime}}p^{\prime},~~y_{1}^{\prime}=-\frac{\partial T}{\partial
q^{\prime}}+\frac{d^{\prime}}{r^{\prime}}q^{\prime}.
\end{gather}
Tenendo conto delle \ref{eqhamjac2} e delle \ref{1234} si ottiene facilmente
\end{subequations}
\begin{subequations}
\label{123}%
\begin{gather}
\frac{\partial T_{1}}{\partial p}=\frac{\partial T}{\partial p}+\frac{d}%
{r}p,~~\frac{\partial T_{1}}{\partial q}=\frac{\partial T}{\partial q}%
+\frac{d}{r}q,\\
\frac{\partial T_{1}}{\partial p^{\prime}}=\frac{\partial T}{\partial
p^{\prime}}-\frac{d^{\prime}}{r^{\prime}}p^{\prime},~~\frac{\partial T_{1}%
}{\partial q^{\prime}}=\frac{\partial T}{\partial q^{\prime}}-\frac{d^{\prime
}}{r^{\prime}}q^{\prime},
\end{gather}
dove $T_{1}$ indica la funzione angolare relativa ai piani $\pi_{1}$ e
$\pi_{1}^{\prime}$.

\section{Approssimazione parassiale per i sistemi ottici centrati}

Possiamo ora introdurre la pi\`{u} semplice descrizione di un sistema ottico.
Nel seguito si supporranno verificate le seguenti condizioni:
\end{subequations}
\begin{enumerate}
\item Gli angoli tra tutti i raggi che attraversano $K$ e l' asse ottico sono
"piccoli", cosicch\'{e} possiamo considerare le componenti di $\mathbf{p}%
$\textbf{ }e\textbf{ }$\mathbf{p\prime}$\textbf{ }come quantit\`{a} del primo ordine.

\item Le componenti dei punti $\mathbf{x},\mathbf{x}$\textbf{' }in cui i raggi
intersecano i piani base e le superfici di $K$ sono anch' esse quantit\`{a}
del primo ordine.
\end{enumerate}

Sotto queste ipotesi, nelle trasformazioni tra i due piani base, tutti i
termini che contengono potenze maggiori di uno nelle quantit\`{a}
$x,x\prime,y,y\prime,p,p\prime,q,q\prime$ possono essere trascurate.
Equivalentemente ricordando che tali trasformazioni sono state ricavate dalla
derivazione di $T$, tale funzione deve essere approssimata fino a termini del
secondo ordine nelle $p,p\prime$,$q,q^{\prime}$\ Le condizioni precedenti
costituiscono l' approssimazione parassiale o Gaussiana e la teoria che ne
deriva \`{e} detta ottica parassiale o Gaussiana. La funzione angolare assume
dunque la forma
\begin{equation}
T=T_{0}+A_{\xi}\xi+A_{\xi^{\prime}}\xi^{\prime}+A_{\eta}\eta,
\end{equation}
essendo $T_{0},A_{\xi},A_{\xi^{\prime}},A_{\eta}$ costanti che dipendono dalle
caratteristiche del sistema ottico e dalla posizione dei piani base rispetto a
$K$. Inoltre si \`{e} interssati solo alle derivate di $T$, di conseguenza si
pu\`{o} trascurare il termine $T_{0}$. Una volta che sia nota $T$ per una
coppia di piani $\pi,\pi^{\prime}$ pu\`{o} essere calcolata la sua omologa per
un' altra coppia di piani $\pi_{1},\pi_{1}^{\prime}$. Infatti, poich\'{e}
\begin{equation}
r=\sqrt{N^{2}-\xi^{2}},~~r^{\prime}=\sqrt{N^{^{\prime}2}-\xi^{\prime2}}%
\end{equation}
possiamo scrivere nell' approssimazione Gaussiana
\begin{equation}
r\simeq N-\frac{\xi}{2N},~~r^{\prime}\simeq N^{\prime}-\frac{\xi^{\prime}%
}{2N^{\prime}}%
\end{equation}
e dalle \ref{123} otteniamo
\begin{equation}
A_{1\xi}=A_{\xi}+\frac{d}{2N},~~A_{1\xi^{\prime}}=A_{\xi^{\prime}}%
-\frac{d^{\prime}}{2N^{\prime}},~~A_{1\eta}=A_{\eta}.
\end{equation}
Se indichiamo con con $\pi^{\ast},\pi^{\ast\prime}$ indichiamo i piani
tangenti, ortogonali ad $s$, alla prima e all' ultima superficie del sistema
ottico $K$, la funzione angolare $T^{\ast}$, relativa a quest' ultima coppia
di piani assume la forma%
\begin{equation}
T^{\ast}=A_{\xi}^{\ast}\xi+A_{\xi^{\prime}}^{\ast\prime}\xi^{\prime}+A_{\eta
}^{\ast}\eta.
\end{equation}
Dalle ultime due relazioni possiamo ricavare l' espressione di $T$ in funzione
dei coefficienti della funzione angolare intrinseca%
\begin{equation}
T=T_{0}+\left(  A_{\xi}-\frac{d}{2N}\right)  \xi+\left(  A_{\xi^{\prime}%
}+\frac{d^{\prime}}{2N^{\prime}}\right)  \xi^{\prime}+A^{\ast}\eta
\end{equation}

\section{Trasformazioni tra i piani base nell' ottica Gaussiana}

Consideriamo le \ref{eq hamjac} e scriviamo le
equazioni corrispondenti
\begin{subequations}
\label{eqhamjac3}%
\begin{align}
x  &  =2\left(  A_{\xi}^{\ast}-\frac{a}{2N}\right)  p+A_{\eta}^{\ast}%
p^{\prime},\\
y  &  =2\left(  A_{\xi}^{\ast}-\frac{a}{2N}\right)  p+A_{\eta}^{\ast}%
q^{\prime},\\
x^{\prime}  &  =-2\left(  A_{\xi^{\prime}}+\frac{a^{\prime}}{2N^{\prime}%
}\right)  p^{\prime}-A_{\eta}^{\ast}p,\\
y^{\prime}  &  =-2\left(  A_{\xi^{\prime}}+\frac{a^{\prime}}{2N^{\prime}%
}\right)  q^{\prime}-A_{\eta}^{\ast}q.
\end{align}
Se \`{e} soddisfatta la condizione
\end{subequations}
\begin{equation}
A_{\eta}^{\ast}\not =0,
\end{equation}
\`{e} possibile risolvere le prime due equazioni rispetto a $p^{\prime
},q^{\prime}$. Sostituendo queste relazioni nelle altre due equazioni si
ottiene la corrispondenza tra i piani base $\pi$ e $\pi^{\prime}$
\begin{subequations}
\label{corrispondenza}%
\begin{align}
x^{\prime}  &  =-\frac{2}{A_{\eta}^{\ast}}\left(  A_{\xi^{\prime}}^{\ast
}+\frac{a^{\prime}}{2N^{\prime}}\right)  x+\frac{1}{A_{\eta}}\left[  4\left(
A_{\xi}^{\ast}+\frac{a}{2N}\right)  \left(  A_{\xi}^{\ast}-\frac{a}%
{2N}\right)  -A_{\eta}^{\ast2}\right]  p,\\
y^{\prime}  &  =-\frac{2}{A_{\eta}^{\ast}}\left(  A_{\xi^{\prime}}^{\ast
}+\frac{a^{\prime}}{2N^{\prime}}\right)  y+\frac{1}{A_{\eta}}\left[  4\left(
A_{\xi}^{\ast}+\frac{a}{2N}\right)  \left(  A_{\xi}^{\ast}-\frac{a}%
{2N}\right)  -A_{\eta}^{\ast2}\right]  q,\\
p^{\prime}  &  =\frac{1}{A_{\eta}^{\ast}}x-\frac{1}{A_{\eta}^{\ast}}\left(
A_{\xi}^{\ast}-\frac{a}{2N}\right)  p,\\
q^{\prime}  &  =\frac{1}{A_{\eta}^{\ast}}x-\frac{1}{A_{\eta}^{\ast}}\left(
A_{\xi}^{\ast}-\frac{a}{2N}\right)  q,
\end{align}
che associa ad ogni raggio di numeri direttori $\left(  p,q\right)  $ uscente
dal punto $\left(  x,y\right)  $ del piano $\pi$ appartenente allo spazio
oggetto, sia la direzione $\left(  p^{\prime},q^{\prime}\right)  $ del raggio
emergente dal sistema ottico, sia il punto $\left(  x^{\prime},y^{\prime
}\right)  $ in cui interseca il piano $\pi^{\prime}$ nello spazio immagine. E'
possibile derivare una propriet\`{a} fondamentale dal sistema
\ref{corrispondenza}:
\end{subequations}
\begin{theorem}
Per ogni fissato piano $\pi$ nello spazio oggetto, esiste un opportuno piano
$\pi^{\prime}$ nello spazio immagine tale che tutti i raggi con numeri
direttori $\left(  p,q\right)  $ uscenti da un punto fissato $\left(
x,y\right)  \in\pi$ intersecano il piano $\pi^{\prime}$ nello stesso punto
$\left(  x^{\prime},y^{\prime}\right)  $. Inoltre tale propriet\`{a} vale per
ogni $\left(  x,y\right)  \in\pi$.
\end{theorem}

E' immediato ricavare questa notevole propriet\`{a} dalle \ref{corrispondenza}%
, basta osservare che le coordinate del punto $\left(  x^{\prime},y^{\prime
}\right)  $ sono indipendenti dalla raggio del raggio entrante in $K$ se e
solo se \`{e} soddisfatta la seguente condizione%
\begin{equation}
4\left(  A_{\xi}^{\ast}+\frac{a}{2N}\right)  \left(  A_{\xi}^{\ast}-\frac
{a}{2N}\right)  -A_{\eta}^{\ast2}=0. \label{cond pcon}%
\end{equation}
I piani base che verificano questa propriet\`{a} sono \textbf{piani
coniugati}, in particolare il piano base anteriore \`{e} detto \textbf{piano
oggetto} ed il piano base posteriore \`{e} detto \textbf{piano immagine};
inoltre l'immagine $\left(  x^{\prime},y^{\prime}\right)  \in\pi^{\prime}$ di
$\left(  x,y\right)  $ \`{e} detta \textbf{immagine parassiale }o
\textbf{Gaussiana }di $\left(  x,y\right)  $. Per una generica coppia di piani
coniugati le \ref{corrispondenza} risultano semplificate ed assumono la forma
\begin{subequations}
\label{corrispondenza 2}%
\begin{align}
x^{\prime}  &  =-\frac{2}{A_{\eta}^{\ast}}\left(  A_{\xi^{\prime}}^{\ast
}+\frac{a^{\prime}}{2N^{\prime}}\right)  x,\\
y^{\prime}  &  =-\frac{2}{A_{\eta}^{\ast}}\left(  A_{\xi^{\prime}}^{\ast
}+\frac{a^{\prime}}{2N^{\prime}}\right)  y,\\
p^{\prime}  &  =\frac{1}{A_{\eta}^{\ast}}x-\frac{1}{A_{\eta}^{\ast}}\left(
A_{\xi}^{\ast}-\frac{a}{2N}\right)  p,\\
q^{\prime}  &  =\frac{1}{A_{\eta}^{\ast}}x-\frac{1}{A_{\eta}^{\ast}}\left(
A_{\xi}^{\ast}-\frac{a}{2N}\right)  q.
\end{align}
Va ricordato che anche se le \ref{corrispondenza}, e le \ref{corrispondenza 2}
sono formalmente valide per ogni scelta del punto $\left(  x,y\right)  $ e
delle direzioni $\left(  p,q\right)  $ le coordinate, in realt\`{a} forniscono
risultati fisicamente accettabili solo se il punto \`{e} in un intorno dell'
asse ottico, ed i raggi uscenti formano angoli piccoli sempre con $s$. Inoltre
il rapporto%
\end{subequations}
\begin{equation}
M=\frac{x^{\prime}}{x}=\frac{y^{\prime}}{y}=-\frac{2}{A_{\eta}^{\ast}}\left(
A_{\xi^{\prime}}^{\ast}+\frac{a^{\prime}}{2N^{\prime}}\right)
\label{ingrandimento}%
\end{equation}
\`{e} indipendente dal punto $\left(  x,y\right)  $ cosicch\'{e} l' immagine
\`{e} ingrandita \emph{proporzionalmente }se $\left\vert M\right\vert >1$ o
rimpiciolita se $\left\vert M\right\vert <1$. $M$ \`{e} detto ingrandimento
relativo alla coppia di piani coniugati. In particolare l' immagine risulta
invertita se $M<0$.

\section{Fuochi e punti nodali di un sistema ottico}

A partire dalle \ref{corrispondenza 2} \`{e} possibile definire alcune
fondamentali grandezze parassiali di un sistema ottico $K$. I due piani
coniugati $\pi_{p}$, $\pi_{p}^{\prime}$ per i quali risulta
\begin{equation}
x=x^{\prime},~~y=y^{\prime},
\end{equation}
sono detti rispettivamente piano principale anteriore e piano principale
posteriore. Per determinare la distanza di questi piani dalle superfici
anteriore e posteriore, nelle \ref{cond pcon}, \ref{ingrandimento} va imposto
\begin{align}
2\left(  A_{\xi^{\prime}}^{\ast}+\frac{a_{p}^{\prime}}{2N^{\prime}}\right)
+A_{\eta}^{\ast}  &  =0,\\
2\left(  A_{\xi}^{\ast}+\frac{a_{p}}{2N}\right)  -A_{\eta}^{\ast}  &  =0,
\end{align}
che forniscono
\begin{equation}
a_{p}=N\left(  A_{\eta}^{\ast}+2A_{\xi}^{\ast}\right)  ,~~a_{p}=-N^{\prime
}\left(  A_{\eta}^{\ast}+2A_{\xi^{\prime}}^{\ast}\right)  . \label{fuochi}%
\end{equation}
I punti P e P' in cui i piani principali incontrano l' asse ottico sono detti
rispettivamente \textbf{punto principale anteriore }e\textbf{ posteriore}.
Introdurremo ora altre definizioni utili quando si ha a che fare con un
sistema ottico nell' approssimazione parassiale. Calcoliamo la posizione del
piano principale posteriore quando il piano principale anteriore \`{e} posto
all' infinito, dalla \ref{cond pcon} si ottiene
\[
a_{f}^{\prime}=-2N^{\prime}A_{\xi^{\prime}}^{\ast}.
\]
Il punto $F^{\prime}$ sull' asse ottico posto a distanza $a_{f}^{\prime}$ dal
piano principale posteriore del sistema \`{e} detto fuoco posteriore. La
differenza
\begin{equation}
f^{\prime}=a_{f}^{\prime}-a_{p}^{\prime}=N^{\prime}A_{\eta}^{\ast},
\end{equation}
\`{e} detta distanza focale posteriore. Analogamente, si definisce la distanza
focale anteriore $f$
\begin{equation}
f=a_{f}-a_{p}=-NA_{\eta}^{\ast}%
\end{equation}
ed il fuoco anteriore $F$.

Per concludere questa sezione chiameremo punti nodali di un sistema ottico
quei punti coniugati sull' asse ottico $Q,Q^{\prime}$ tali che a un qualunque
raggio uscente da Q ne corrisponda uno da Q' parallelo al primo. Dalla
condizione di parallelismo
\begin{equation}
\frac{p}{N}=\frac{p^{\prime}}{N^{\prime}},~~\frac{q}{N}=\frac{q^{\prime}%
}{N^{\prime}},
\end{equation}
tenuto conto della \ref{corrispondenza} in cui poniamo $x=0,~y=0$, segue
\begin{equation}
2NA_{\xi}+N^{\prime}A_{\eta}=0.
\end{equation}
Questa relazione, usando la \ref{cond pcon}, diviene%
\begin{equation}
2N^{\prime}A_{\xi^{\prime}}+NA_{\eta}=0. \label{nodi}%
\end{equation}
Confrontando la \ref{nodi} con la \ref{fuochi} si pu\`{o} concludere che i
punti nodali coincidono con i fuochi quando $N=N^{\prime}$. Tutti i risultati
prececenti possono essere riassunti come segue:

\emph{Nell' approssimazione parassiale, un sistema ottico fornisce un'
immagine nitida di qualunque piano oggetto }$\pi$\emph{ nel corrispondente
piano immagine }$\pi^{\prime}$\emph{. Nelle trasformazioni dal piano oggetto
al piano immagine gli angoli non sono modificati mentre le lunghezze variano
di un fattore costante }$M.$

\section{La relazione di Huygens}

Spesso risulta conveniente esprimere le relazioni \ref{corrispondenza 2} in
termini delle distanze $d,d\prime$ dei piani principali dai piani oggetto ed
immagine. poich\'{e}
\begin{equation}
d=a-a_{p},~~d^{\prime}=a^{\prime}-a_{p}^{\prime},
\end{equation}
le \ref{corrispondenza 2} assumono la forma%
\begin{equation}
\left(  2A_{\xi^{\prime}}^{\ast}+\frac{d^{\prime}}{N^{\prime}}+\frac
{a_{p}^{\prime}}{N^{\prime}}\right)  \left(  2A_{\xi}^{\ast}+\frac{d}{N}%
+\frac{a_{p}}{N}\right)  -A_{\eta}^{\ast2}=0
\end{equation}
che con un po' di algebra pu\`{o} essere posta nella forma%
\begin{equation}
\left(  \frac{d^{\prime}}{N^{\prime}}-A_{\eta}^{\ast}\right)  \left(  A_{\eta
}^{\ast}+\frac{d}{N}\right)  +A_{\eta}^{\ast2}=0,
\end{equation}
che implica
\begin{equation}
\frac{d^{\prime}}{N^{\prime}}\frac{d}{N}+\frac{d^{\prime}}{N^{\prime}}A_{\eta
}^{\ast}-\frac{d}{N}A_{\eta}^{\ast}=0.
\end{equation}
Dopo un ultimo passaggio si arriva a%
\begin{equation}
\frac{N^{\prime}}{d^{\prime}}-\frac{N}{d}=\frac{1}{A_{\eta}^{\ast}}%
=\frac{N^{\prime}}{f^{\prime}}=-\frac{N}{f}%
\end{equation}
che \`{e} la nota relazione di Huygens.

\section{La funzione angolare di una superficie nel limite dell' ottica
Gaussiana}

Dalla definizione della funzione angolare $T$, si pu\`{o} derivare la sua
espressione.\ Siano $\mathbf{p}$ e $\mathbf{p\prime}$ le direzioni dei raggi
uscenti, $\mathbf{r}$ ed $\mathbf{r}^{\prime}$ i vettori uscenti dalle origini
degli spazi oggetto ed immagine ed aventi come coordinate quelle del punto
$P$, intersezione tra il raggio e la superficie, allora la funzione angolare
assume la forma%
\begin{equation}
T=\mathbf{pr-p}^{\prime}\mathbf{r}^{\prime}.
\end{equation}
Per la convenzione adottata nella scelta dei riferimenti si ha%
\[
x^{\prime}=x,~~y^{\prime}=y,~~z^{\prime}=z-\left(  a-a^{\prime}\right)  ,
\]
e si riscrive $T$ come%
\begin{equation}
T=\left(  p-p^{\prime}\right)  x+\left(  q-q^{\prime}\right)  y+\left(
r-r^{\prime}\right)  z+r^{\prime}\left(  a-a^{\prime}\right)  .
\end{equation}
D' altro canto, le coordinate del punto $P,$ in cui il raggio interseca la
superficie $S$, devono minimizzare la lunghezza del cammino ottico, ovvero la
funzione
\[
T\left(  x,y,z\right)  +\lambda F\left(  x,y,z\right)  ,
\]
dove $\lambda$ \`{e} un moltiplicatore di Lagrange e $F\left(  x,y,z\right)
=z-z\left(  x,y\right)  =0$ \`{e} l'equazione di $S$. La condizione necessaria
per un minimo si scrive come%
\begin{equation}
\nabla\left(  T+\lambda F\right)  =0
\end{equation}
o equivalentemente,%
\begin{equation}
p-p^{\prime}=\lambda\frac{\partial z}{\partial x},~~q-q^{\prime}=\lambda
\frac{\partial z}{\partial y},~~r-r^{\prime}=-\lambda.
\end{equation}
Sostituendo il valore di $\lambda$ ricavato dalla terza equazione nelle altre,
si ha
\begin{equation}
\frac{p-p^{\prime}}{r-r^{\prime}}=\lambda\frac{\partial z}{\partial x}%
,~~\frac{q-q^{\prime}}{r-r^{\prime}}=\lambda\frac{\partial z}{\partial
y},~~r-r^{\prime}=-\lambda. \label{eq hamjac4}%
\end{equation}
Fino ad ora sono state utilizzate relazioni esatte. Si \`{e} interessati a
relazioni per l' ottica parassiale, si inizi con lo scrivere l' espressione
approssimata della superficie $S$ che fino al secondo ordine \`{e}%
\begin{equation}
z=\frac{\left(  x^{2}+y^{2}\right)  }{2R}+O\left(  4\right)  ,
\end{equation}
dove $R$ \`{e} il raggio di curvatura della superficie. Le espressioni di $r$
ed $r^{\prime}$ al medesimo ordine di approssimazione sono%
\begin{equation}
r=N-\frac{1}{2N}\left(  p^{2}+q^{2}\right)  +O\left(  4\right)  ,\text{~~}%
r^{\prime}=N^{\prime}-\frac{1}{2N^{\prime}}\left(  p^{\prime2}+q^{\prime
2}\right)  +O\left(  4\right)  .
\end{equation}
Sostituendo nelle \ref{eq hamjac4}, dopo alcuni calcoli, si ha infine l'
espressione della funzione angolare per una superficie nel limite di Gauss%
\begin{equation}
T=aN-a^{\prime}N^{\prime}+A_{\xi}\xi+A_{\xi^{\prime}}\xi^{\prime}+A_{\eta}%
\eta, \label{T ottica gauss}%
\end{equation}
dove
\begin{subequations}
\label{def coeff T ottica gauss}%
\begin{gather}
A_{\xi}=-\frac{a}{2N}+A_{\xi}^{\ast},\\
A_{\xi}^{\prime}=-\frac{a^{\prime}}{2N^{\prime}}+A_{\xi}^{\prime\ast},\\
A_{\eta}=A_{\eta}^{\ast},\\
2A_{\xi}^{\ast}=2A_{\xi^{\prime}}^{\ast}=-A_{\eta}^{\ast}=\frac{R}{N^{\prime
}-N}%
\end{gather}

\section{La funzione angolare di un sistema ottico composto}

Sia $K$ un sistema ottico composto da $n$ sottosistemi $K_{1},\ldots,K_{n}$ di
cui sono note le singole funzioni angolari $T_{1},\ldots,T_{2}$. Dalla
definizione di lunghezza di un cammino ottico si ha
\end{subequations}
\begin{equation}
T=\sum_{i=1}^{n}T_{i}. \label{T sistema composto}%
\end{equation}
Si osservi che ogni termine della sommatoria dipende dalle $p_{i}%
,p_{i}^{\prime},q_{i},q_{i}^{\prime}$, quindi la funzione angolare $T$ dell'
intero sistema dipende da tutte le variabili $p_{i},p_{i}^{\prime},q_{i}%
,q_{i}^{\prime}~i=1,\ldots,n$. Se si vogliono usare le \ref{eqhamjac2},
bisogna prima porre la \ref{T sistema composto} in una forma dove compaiono
dipendenze solo dalle $p_{1},q_{1},p_{n}^{\prime},q_{n}^{\prime}$. Osservando
che lo spazio immagine di $K_{i}$ coincide con lo spazio oggetto di $K_{i+1}$,
si ha
\begin{equation}
p_{i}^{\prime}=p_{i+1}~~q_{i}^{\prime}=q_{i+1}~~i=1,\ldots,n-2,
\end{equation}
la funzione angolare pu\`{o} essere riscritta come
\begin{equation}
T=T_{1}\left(  p_{1},q_{1},p_{1}^{\prime},q_{1}^{\prime}\right)  +T_{2}\left(
p_{1}^{\prime},q_{1}^{\prime},p_{2}^{\prime},q_{2}^{\prime}\right)
+\ldots+T_{n}\left(  p_{n-1}^{\prime},q_{n-1}^{\prime},p_{n}^{\prime}%
,q_{n}^{\prime}\right)  ,
\end{equation}
rimangono da eliminare $2\left(  n-1\right)  $ variabili $p_{1}^{\prime}%
,q_{1}^{\prime},\ldots,p_{n-1}^{\prime},q_{n-1}^{\prime}$. Secondo il
principio di Fermat, la lunghezza del cammino ottico lungo un raggio \`{e}
stazionaria. Allora, assegnando le $p_{1},q_{1},p_{n}^{\prime},q_{n}^{\prime}%
$, le derivate di $T$ rispetto alle altre variabili si annullano%
\begin{align}
\frac{\partial T}{\partial p_{i}^{\prime}}  &  =\frac{\partial T_{i}}{\partial
p_{i}^{\prime}}+\frac{\partial T_{i+1}}{\partial p_{i}^{\prime}}=0,\\
\frac{\partial T}{\partial q_{i}^{\prime}}  &  =\frac{\partial T_{i}}{\partial
q_{i}^{\prime}}+\frac{\partial T_{i+1}}{\partial q_{i}^{\prime}}%
=0,~~i=1,\ldots,n-1.
\end{align}
La \ref{T ottica gauss} ci permette di dare al sistema precedente la forma
\begin{subequations}
\label{relazioni dipendenza coeff T}%
\begin{equation}
2\left(  A_{\xi_{i}^{\prime}}+A_{\xi_{i+1}}\right)  p_{i}^{\prime}%
+A_{\eta_{i+1}}p_{i+1}^{\prime}=0,~~2\left(  A_{\xi_{i}^{\prime}}+A_{\xi
_{i+1}}\right)  q_{i}^{\prime}+A_{\eta_{i+1}}q_{i+1}^{\prime}=0,
\end{equation}
da cui si ricava
\end{subequations}
\begin{subequations}
\label{sistema}%
\begin{equation}
p_{i}^{\prime}=D_{i}p_{i}+D_{i}^{\prime}p_{i+1}^{\prime},~~q_{i}^{\prime
}=D_{i}q_{i}+D_{i}^{\prime}q_{i+1}^{\prime}.
\end{equation}
dove si \`{e} posto
\end{subequations}
\begin{subequations}
\label{sistema coeff}%
\begin{equation}
D_{i}=-\frac{A_{\eta_{i}}}{2\left(  A_{\xi_{i}^{\prime}}+A_{\xi_{i+1}}\right)
},~~D_{i}^{\prime}=-\frac{A_{\eta_{i}+1}}{2\left(  A_{\xi_{i}^{\prime}}%
+A_{\xi_{i+1}}\right)  }.
\end{equation}
Il sistema \ref{sistema} permette di esprimere le incognite $p_{i}^{\prime
},q_{i}^{\prime}$~$i=1,\ldots,n-1$ nelle $p_{1},q_{1},p_{n}^{\prime}%
,q_{n}^{\prime}$.

Il procedimento descritto fin qui verr\`{a} applicato ad una lente di spessore
\ $d$ ed indice di rifrazione $N$. In questo caso avremo due superfici ottiche
$S_{1},~S_{2}$, la cui distanza \emph{lungo l' asse} \emph{ottico} \`{e} $d$.
Per calcolare la funzione angolare intrinseca del sistema $T^{\ast},$ si
fissino i piani base in modo che siano tangenti alle due superfici nei punti
di intersezione con l' asse ottico. Utilizzando la \ref{T ottica gauss} e le
\ref{def coeff T ottica gauss} ed osservando che in queste formule $a^{\prime
}$ \`{e} stato sostituito con $-d$, si ricava l' espressione%
\end{subequations}
\begin{equation}
T^{\ast}=\frac{1}{2P_{1}}\xi_{1}+\frac{1}{2}\left(  \frac{1}{P_{1}}-\frac
{d}{N}\right)  \xi_{1}^{\prime}-\frac{1}{P_{1}}\eta_{1}+\frac{1}{2P_{2}}%
\xi_{2}+\frac{1}{2P_{2}}\xi_{2}^{\prime}-\frac{1}{P_{2}}\eta_{2,}%
\end{equation}
Da cui, tenendo conto delle \ref{relazioni dipendenza coeff T} le
\ref{sistema} si riducono alle
\begin{align}
p_{1}^{\prime}  &  =\frac{1}{P_{1}+P_{2}-\frac{d}{N}P_{1}P_{2}}\left(
P_{2}p_{1}+P_{2}p_{2}^{\prime}\right)  ,\\
q_{1}^{\prime}  &  =\frac{1}{P_{1}+P_{2}-\frac{d}{N}P_{1}P_{2}}\left(
P_{2}q_{1}+P_{2}q_{2}^{\prime}\right)  .
\end{align}
Sostituendo queste relazioni nell' espressioni trovate per $T^{\ast}$
otteniamo
\begin{equation}
T^{\ast}=\frac{1}{2\mu}\left[  \left(  1-\frac{d}{N}P_{2}\right)  \xi
_{1}+\left(  1-\frac{d}{N}P_{1}\right)  \xi_{2}^{\prime}-2\eta\right]  ,
\label{T star 2 sup}%
\end{equation}
dove si \`{e} posto per brevit\`{a}
\begin{equation}
\mu=\frac{1}{P_{1}+P_{2}-\frac{d}{N}P_{1}P_{2}}.
\end{equation}
Possiamo ovviamente identificare i coefficienti nella \ref{T star 2 sup} nel
modo seguente%
\begin{equation}
A_{\xi^{\prime}}^{\ast}=\frac{1}{2\mu}\left(  1-\frac{d}{N}P_{1}\right)
,~~A_{\xi}^{\ast}=\frac{1}{2\mu}\left(  1-\frac{d}{N}P_{2}\right)  ,~~A_{\eta
}^{\ast}=-\frac{1}{\mu}.
\end{equation}

\section{Vignettatura}

Una delle aberrazioni di un sistema ottico centrato, gi\`{a} presente a questo
livello di descrizione, \`{e} la vignettatura. Questa aberrazione consiste in
una caduta di luminosit\`{a} che si pu\`{o} osservare allontanandosi dal
centro del piano immagine ed \`{e} dovuta alla dimensione finita delle lenti
che compongono il sistema ottico. Infatti, non tutti i raggi che entrano nello
strumento raggiungono il piano immagine, ma possono essere bloccati dalle
montature delle lenti. Si consideri un oggetto esteso, illuminato
uniformemente e posto a distanza finita dalla pupilla d' entrata\footnote{Un
diaframma \`{e} un dispositivo di piccole dimensioni longitudinali atto a
limitare la quantit\`{a} di luce che attraversa un sistema ottico. Pu\`{o}
essere posto all' interno dello strumento, in tal caso si definisce pupilla
d'entrata (risp. uscita) l' immagine gaussiana del piano del diaframma
attraverso lo strumento nello spazio oggetto (risp. immagine).}. Si prendano
in considerazione due punti $P_{0},~P_{1},$ il primo sull' asse ottico il
secondo sulla retta ortogonale all' asse ottico passante per il punto $P_{0}$.
Si considerino anche la prima superficie e l'immagine nello spazio oggetto
dell' ultima superficie. Le immagini di queste superfici, proiettate nello
spazio oggetto sono tre cerchi, detti anche aperture. Queste pongono dei
limiti alla quantit\`{a} di luce (e quindi all' energia da essa trasportata)
che pu\`{o} raggiungere lo spazio immagine. Tali limiti variano con la
distanza dall' asse dell' oggetto. Per trovare i raggi luminosi che
attraversano il sistema da un dato punto oggetto si traccino dal punto $P_{0}$
i coni aventi come base le aperture e si proiettino su un piano ortogonale
all' asse ottico. L' intersezione di questi tre cerchi sar\`{a} proporzionale
alla quantit\`{a} di luce proveniente da $P_{0}$ ed \`{e} la massima che si
pu\`{o} ottenere. Tracciando ora i coni da $P_{1}$ e prendendone l'
intersezione si ottiene una una complicata figura anch' essa proporzionale
alla luce proveniente da $P_{1}$ che raggiunge il piano immagine.%

Si assuma che l' oggetto, la prima superficie e l' immagine dell' ultima
superficie giacciono rispettivamente a distanza $s,$ $s_{A},$ e $s_{B}$ dalla
pupilla d' entrata e che l' immagine dell' ultima superficie sono cerchi di
raggi $A,P,B$. La proiezione di questi cerchi sulla pupilla d' entrata da un
punto di coordinate $x=0,$ $y=h,$ nel piano oggetto \`{e} data da
\begin{equation}
x^{2}+y^{2}=P^{2}~~x^{2}+\left(  y-h\right)  ^{2}=P_{A}~~x^{2}+\left(
y-t_{B}\right)  ^{2}=P_{B}%
\end{equation}
con
\begin{equation}
t_{A}=\frac{s_{A}}{s_{A}-s}~~t_{B}=\frac{s_{B}}{s_{B}-s}~~P_{A}=\frac{s_{A}%
}{s_{A}-s}~~P_{B}=\frac{s_{B}}{s_{B}-s}.
\end{equation}
Se un oggetto \`{e} posto all' infinito, un punto assiale \`{e} equivalente ad
un fascio di raggi paralleli all' asse ottico, se il punto fuori asse il punto
\`{e} equivalente ad un fascio di raggi paralleli che formano un angolo
$\alpha$ con l' asse. Si ha per i tre cerchi delle relazioni precedenti
\begin{align}
t_{A}  &  =s_{A}\tan\alpha~~P_{A}=A\\
t_{B}  &  =s_{B}\tan\alpha~~P_{B}=B
\end{align}
Il problema \`{e} stato qui esposto in una forma semplificata, si sono prese
in considerazione solo tre superfici, invece di tutte le superfici che
compongono il sistema. La sostanza del problema non cambia, le relazioni date
vanno applicate per tutte le superfici che compongono il sistema. Va anche
osservato che le sole leggi dell' ottica di Gauss, danno una soluzione
approssiamata al problema della vignettatura. Una sua completa comprensione
\`{e} possibile grazie alle leggi della fotometria.

\section{Sfocatura e profondit\`{a} di campo}

Si \`{e} visto che un sistema ottico centrato, nel limite dell' ottica
Gaussiana, fa convergere tutti i raggi provenienti da un punto del piano
oggetto in un punto del piano immagine, questa situazione si verifica per
tutti e soli i punti del piano oggetto che andranno a formare un immagine "a
fuoco" sul piano immagine. Questa affermazione sembra contraddire l'
osservazione. Infatti \`{e} esperienza comune vedere fotografie in cui oggetti
molto distanti tra loro sono contemporaneamente a fuoco, ad esempio una
persona in primo piano e sullo sfondo un panorama perfettamente nitido.

Si consideri un oggetto puntiforme, situato a distanza $p$ (piano oggetto) da
un obbiettivo focalizzato a distanza $q$ (piano immagine) dall' ultima
superficie ottica. Si consideri un altro punto posto a distanza $p+\Delta p$
dall' obiettivo. I raggi uscenti da questo punto attraverseranno il proprio
punto coniugato prima (o dopo) aver raggiunto il piano immagine. Se il punto
si trova sull' asse ottico i raggi uscenti dal sistema ottico formeranno una
sezione di cono, cos\`{\i} sul piano immagine si former\`{a} il cosidetto
cerchio di confusione, che ha diametro $d$ . Il diametro della pupilla d'
entrata sia $D.$ Utilizzando la relazione di Huygens ed esprimendo le
dimensioni della pupilla d' entrata e del cerchio di confusione in funzione
delle distanze $p,q$ si pu\`{o} ricavare ad esempio se gli indici di
rifrazione degli spazi oggetto ed immagine sono uguali si ricava%
\begin{align}
\frac{1}{p}+\frac{1}{q}  &  =\frac{1}{f},~~\frac{1}{p+\Delta p}+\frac
{1}{q-\Delta q}=\frac{1}{f}\\
\frac{d}{2}  &  \simeq\beta\Delta q,~~\frac{D}{2}\simeq\beta\left(  q-\Delta
q\right)
\end{align}
Ricavando l' espressione del cerchio di confunsione in funzione delle altre
grandezze in gioco si arriva a
\begin{equation}
d=\frac{pq\left(  p+\Delta p\right)  }{q\Delta p+p\Delta p+p^{2}}\simeq
D\frac{q\Delta p}{p^{2}}%
\end{equation}
quindi al crescere dell' apertura del diaframma le dimensioni del circolo di
confusione aumentano. Perci\`{o} fissti $p,q$ e stabilito come tollerabile
(per esempio in relazione con la grana della pellicola fotografica o la
risoluzione dell' occhio umano) un cerchio di confusione di diametro $d$ la
profondit\`{a} di campo $\Delta p$ \`{e} inversamente proporzionale al
diametro $D$ di apertura della lente,%
\begin{equation}
\Delta p={\frac{-\left(  d\,p^{2}\right)  +p^{2}\,q}{d\,p+d\,q-p\,q}}%
\end{equation}
inoltre la profondit\`{a} di campo aumenta quando si focalizza l' obiettivo su
distanze $p$ maggiori. E' proprio sfruttando questa propriet\`{a} di tutti i
sistemi ottici che vengono prodotte macchine fotografiche prive del sistema di
messa a fuoco, in cui l' obbiettivo viene focalizzato sulla distanza che
massimizza $\Delta p,$ detta iperfocale. Sempre sfruttando questa
propriet\`{a} e grazie ad una opportuna scala posta sull' obiettivo, i
fotografi possono scattare in situazioni poco agevoli conoscendo solo l'
intervallo di distanze in cui si trover\`{a} il soggetto, al resto
provveder\`{a} la profondit\`{a} di campo.

\chapter{Le aberrazioni dei sistemi ottici centrati}

\section{Definizioni e prime propriet\`{a}}

Come si \`{e} gi\`{a} osservato, nel limite dell' ottica gaussiana, le
trasformazioni tra piani coniugati sono ortoscopische, ovvero ad un punto
nello spazio oggetto corrisponde un punto nello spazio immagine ed a un piano
corrisponde un piano. Si conservano anche gli angoli tra i vettori nella
trasformazione tra lo spazio oggetto e quello immagine: un' immagine
bidimensionale \`{e} riprodotta fedelmente. Purtroppo l'approssimazione
gaussiana non \`{e} sufficientemente accurata nella maggior parte dei casi e
nella progettazione di un sistema ottico vanno presi in considerazione
contributi di ordine maggiore al primo al fine di minimizzarli. Dopo che un
raggio luminoso ha attraversato un discreto numero di superfici risulta
piuttosto laborioso esprimerlo in funzione dei valori iniziali. Un sistema
ottico centrato $K$, stabilisce una corrispondenza tra i raggi di direzione
$\left(  p,q\right)  $ uscenti dal punto $\left(  x,y\right)  $ appartenenti
allo spazio oggetto e le corrispondenti grandezze dello spazio immagine.
Questa corrispondenza assume la forma
\begin{subequations}
\label{espressione generica raggio}%
\begin{align}
x^{\prime}  &  =x^{\prime}\left(  x,y,p,q\right)  ,~~y^{\prime}=y^{\prime
}\left(  x,y,p,q\right)  ,\\
p^{\prime}  &  =p^{\prime}\left(  x,y,p,q\right)  ,~~q^{\prime}=q^{\prime
}\left(  x,y,p,q\right)  .
\end{align}
Nell' approssimazione di Gauss queste relazioni si riducono a
\end{subequations}
\begin{subequations}
\label{eq moto gauss}%
\begin{gather}
x^{\prime}=Mx,~~y^{\prime}=My,\\
p^{\prime}=\frac{N^{\prime}}{f}x-Mp,~~q^{\prime}=\frac{N^{\prime}}{f}y-Mq.
\end{gather}
Il vettore di aberrazione \`{e} definito come quel vettore le cui componenti
sono espresse dalle differenze
\end{subequations}
\begin{subequations}
\label{vettore aberrazione}%
\begin{align}
\varepsilon_{x}  &  =x^{\prime}-Mx,\\
\varepsilon_{y}  &  =y^{\prime}-My.
\end{align}
Il significato fisico del vettore di aberrazione \`{e} evidente: le sue
componenti si annullano se e solo se il comportamento del sistema ottico
coincide con quello previsto dall'ottica di Gauss. Il problema principale
della teoria delle aberrazioni \`{e} determinare l' espressione analitica
delle \ref{vettore aberrazione} in funzione delle coordinate dell' oggetto,
della direzione dei raggi uscenti e delle caratteristiche del sistema ottico
$K$. Iniziamo a ricavare le prime propriet\`{a} sulle aberrazioni ottiche. E'
noto che lo sviluppo in serie della funzione principale angolare non presenta
potenze dispari delle $p,q,p^{\prime},q^{\prime}$, la sua espressione pu\`{o}
essere scritta come%
\end{subequations}
\begin{equation}
T=\sum_{n}T_{2n}.
\end{equation}
Di conseguenza, poich\'{e} le aberrazioni saranno ricavate dalla derivazione
della funzione angolare presenteranno solo potenze dispari delle $\left(
p,q,p^{\prime},q^{\prime}\right)  $. Tale importante informazione si sarebbe
potuta ricavare anche da un' attenta analisi della forma dell' espressione del
raggio, infatti si pu\`{o} dimostrare che le \ref{espressione generica raggio}
possono essere poste nella forma
\begin{equation}
x_{\alpha}^{\prime}\left(  x,y,p,q\right)  =p_{\alpha}h\left(  u,v,w\right)
+x_{\alpha}k\left(  u,v,w\right)  ~~\alpha=1,2
\label{espressione raggio meno generica}%
\end{equation}
con $u=x^{2}+y^{2},~v=p^{2}+q^{2},~w=px+qy$ e $h$ e $k$ sono funzioni
opportune. Lo sviluppo in serie nelle $\left(  x,y,p,q\right)  $ in un intorno
del punto $\left(  0,0,0,0\right)  $ della
\ref{espressione raggio meno generica} \`{e} equivalente a
\begin{equation}
x^{\prime}=p\sum_{n}h_{n}+x\sum_{n}k_{n}%
\end{equation}
dove $h_{n}$ e $k_{n}$ sono i termini $n$-simi dello sviluppo in serie di
Taylor delle funzioni $h$ e $k$ nelle $\left(  u,v,w\right)  $. Tale analisi
non presenta altri vantaggi, e quindi si torner\`{a} a riferirsi all' utilizzo
della funzione angolare.

I primi due addendi dello sviluppo di $T$ sono responsabili del comportamento
Gaussiano del sistema ottico in esame, il termine successivo, responsabile
delle aberrazioni del terzo ordine, \`{e}%
\begin{equation}
T_{4}=A_{\xi\xi}\xi^{2}+A_{\xi^{\prime}\xi^{\prime}}\xi^{\prime2}+A_{\eta\eta
}\eta^{2}+A_{\xi\eta}\xi\eta+A_{\xi\xi^{\prime}}\xi\xi^{\prime}+A_{\xi
^{\prime}\eta}\xi^{\prime}\eta.
\end{equation}
Il vettore di aberrazione avr\`{a} di conseguenza la seguente espressione%
\begin{align}
\varepsilon_{x}  &  =-\left(  4A_{\xi^{\prime}\xi^{\prime}}p^{\prime}%
\xi+4A_{\eta\eta}p\eta+A_{\xi\eta}\xi p^{\prime}+2A_{\xi\xi^{\prime}}\xi
p^{\prime}+A_{\xi^{\prime}\eta}\left(  2p^{\prime}\eta+\xi^{\prime}p\right)
\right)  ,\\
\varepsilon_{y}  &  =-\left(  4A_{\xi^{\prime}\xi^{\prime}}q^{\prime}%
\xi+4A_{\eta\eta}q\eta+A_{\xi\eta}\xi q^{\prime}+2A_{\xi\xi^{\prime}}\xi
q^{\prime}+A_{\xi^{\prime}\eta}\left(  2q^{\prime}\eta+\xi^{\prime}q\right)
\right)  .
\end{align}
Si pu\`{o} ricavare un' espressione del vettore di aberrazione in funzione
delle sole variabili dello spazio oggetto. Infatti, visto che siamo
interessati a relazioni di terzo grado, si possono sostituire alle $\left(
p^{\prime},q^{\prime}\right)  $ delle relazioni lineari nelle coordinate dello
spazio oggetto senza introdurre errori apprezzabili. Tali relazioni sono
quelle dell' ottica Gaussiana \ref{eq moto gauss}; si ha quindi
\begin{align}
\varepsilon_{x}  &  =p\left[  M\left(  4A_{\xi^{\prime}\xi^{\prime}}%
+4A_{\eta\eta}+A_{\xi\eta}+2A_{\xi\xi^{\prime}}-3A_{\xi^{\prime}\eta}\right)
\xi-A_{\xi^{\prime}\eta}\frac{N^{\prime2}}{f^{2}}u+4\frac{N^{\prime}}%
{f}\left(  MA_{\xi^{\prime}\eta}-A_{\xi\eta}\right)  w\right] \nonumber\\
&  +x\left[  \left(  2A_{\eta\eta}\frac{MN^{\prime}}{f}-4A_{\xi^{\prime}%
\xi^{\prime}}\frac{N^{\prime}}{f}-A_{\xi\xi^{\prime}}\frac{N^{\prime}}%
{f}\right)  \xi+\left(  4A_{\eta\eta}\frac{MN^{\prime}}{f}-A_{\eta\eta}%
\frac{N^{\prime2}}{f^{2}}\right)  w\right]  ,\\
\varepsilon_{y}  &  =q\left[  M\left(  4A_{\xi^{\prime}\xi^{\prime}}%
+4A_{\eta\eta}+A_{\xi\eta}+2A_{\xi\xi^{\prime}}-3A_{\xi^{\prime}\eta}\right)
\xi-A_{\xi^{\prime}\eta}\frac{N^{\prime2}}{f^{2}}u+4\frac{N^{\prime}}%
{f}\left(  MA_{\xi^{\prime}\eta}-A_{\xi\eta}\right)  w\right] \nonumber\\
&  +y\left[  \left(  2A_{\eta\eta}\frac{MN^{\prime}}{f}-4A_{\xi^{\prime}%
\xi^{\prime}}\frac{N^{\prime}}{f}-A_{\xi\xi^{\prime}}\frac{N^{\prime}}%
{f}\right)  \xi+\left(  4A_{\eta\eta}\frac{MN^{\prime}}{f}-A_{\eta\eta}%
\frac{N^{\prime2}}{f^{2}}\right)  w\right]  ,
\end{align}
dove si \`{e} posto $u=x^{2}+y^{2}$ e $w=px+qy$. Si osservi che in questa
espressione del vettore di aberrazione compaiono solo cinque dei sei
coefficienti della funzione angolare, questo si rifletter\`{a} nel fatto che
sono solo cinque i coefficienti di aberrazione indipendenti del terzo ordine
che determinano la qualit\`{a} dell' immagine. La conoscenza del sesto
coefficiente \`{e} necessaria per caratterizzare completamente la variet\`{a}
dei raggi dello spazio immagine in cui il sistema ottico trasforma i raggi
dello spazio oggetto. Ad esempio per determinare le aberrazioni per una
qualunque coppia di piani coniugati e non solo per quella scelta inizialmente.

\section{Le variabili di von Seidel}

Si possono ottenere ulteriori semplificazioni nei calcoli introducendo delle
nuove coordinate che avranno il pregio di \textbf{rimanere costanti} durante
la rifrazione e la riflessione nel limite dell' ottica del primo ordine. Per
valutare gli effetti di un diaframma all' interno di un sistema ottico
centrato conviene introdurre altri piani di riferimento. Sia $\pi_{d}$ il
piano in cui \`{e} posto il diaframma, si indicher\`{a} con $\pi_{e}$ il piano
dello spazio oggetto coniugato nell' ottica di Gauss a $\pi_{d}$, analogamente
il piano coniugato a $\pi_{d}$ nello spazio immagine si indicher\`{a} con
$\pi_{e}^{\prime}$. I due piani $\pi_{e},$ $\pi_{e}^{\prime}$ sono detti
rispettivamente piano della pupilla d' entrata e d' uscita. Se non \`{e}
presente un diaframma nello strumento la pupilla d' entrata coincide con il
piano tangente al vertice della prima superficie dello strumento, quello di
uscita \`{e} la sua immagine nell' ottica di Gauss. Siano $\pi,~\pi_{e}$
rispettivamente il piano oggetto ed il piano della pupilla d'entrata di
equazioni $z=z_{o},~z=z_{e}$ e $\pi^{\prime},~\pi_{e}^{\prime}$ i loro
coniugati nel senso dell' ottica di Gauss (rispettivamente piano immagine e
piano della pupilla d' uscita) identificati da $z=z^{\prime},~z=z_{e}^{\prime
}$, siano $l,~l_{e}$ due lunghezze di riferimento sui primi due piani ed
$l^{\prime},$~$l_{e}^{\prime}$ altre due lunghezze di riferimento sui piani
$\pi^{\prime},~\pi_{e}^{\prime}$ scelte in modo da avere
\begin{equation}
\frac{l^{\prime}}{l}=M,~~\frac{l_{e}^{\prime}}{l_{e}}=M_{e},
\end{equation}
dove $M$ ed $M_{e}$ sono gli ingrandimenti relativi alle coppie dei piani
oggetto-immagine e pupilla d' ingresso-uscita. La quantit\`{a} $J$ definita
come%
\begin{equation}
J=\frac{Nll_{e}}{z_{e}-z_{o}}=\frac{N^{\prime}l^{\prime}l_{e}^{\prime}}%
{z_{e}^{\prime}-z^{\prime}}.
\end{equation}
rimane costante nel limite dell' ottica geometrica. Si possono ora definire le
nuove variabili $X,~X^{\prime}$ come
\begin{align}
X  &  =J\frac{x}{l},~~Y=J\frac{y}{l},\\
X^{\prime}  &  =J\frac{x^{\prime}}{l^{\prime}},~~Y^{\prime}=J\frac{y^{\prime}%
}{l^{\prime}}.
\end{align}
A partire da questa definizione e dalle \ref{eq moto gauss} possiamo
introdurre altre coppie di variabili al posto delle $\left(  p,q\right)  $
$\left(  p^{\prime},q^{\prime}\right)  $. Se con $x_{e},$ $y_{e}$ $\left(
x_{e}^{\prime},y_{e}^{\prime}\right)  $ si indicano le coordinate di un punto
sul piano della pupilla d' ingresso (d' uscita) abbiamo, nel limite dell'
ottica di Gauss%
\begin{align}
X_{e}  &  =\frac{x_{e}}{l_{e}}+\frac{\left(  z_{e}-z_{o}\right)  }{l_{e}}%
\frac{p}{N},~~X_{e}^{\prime}=\frac{x_{e}^{\prime}}{l_{e}^{\prime}}%
+\frac{\left(  z_{e}^{\prime}-z^{\prime}\right)  }{l_{e}^{\prime}}%
\frac{p^{\prime}}{N^{\prime}},\\
Y_{e}  &  =\frac{y_{e}}{l_{e}}+\frac{\left(  z_{e}-z_{o}\right)  }{l_{e}}%
\frac{q}{N},~~Y_{e}^{\prime}=\frac{y_{e}^{\prime}}{l_{e}^{\prime}}%
+\frac{\left(  z_{e}^{\prime}-z^{\prime}\right)  }{l_{e}^{\prime}}%
\frac{q^{\prime}}{N^{\prime}}.
\end{align}
Si osservi che il secondo insieme di variabili a differenza del primo \`{e} adimensionale.

Le $X,~X^{\prime},~Y,$~$Y^{\prime},~X_{e},~X_{e,}^{\prime}~Y_{e,}%
~Y_{e}^{\prime}$ vengono dette variabili di Seidel. L' utilit\`{a} delle
variabili di Seidel si riveler\`{a} nel prossimo paragrafo quando sar\`{a}
dimostrato il teorema di addizione delle aberrazione del terzo ordine, che
permetter\`{a} di comporre agevolmente le aberrazioni del terzo ordine di
sistemi ottici complessi sfruttando la costanza delle nuove coordinate nel
passaggio attraverso il sistema ottico.

\section{L' iconale di Schwarzschild}

\begin{definition}
Se si scrive la funzione angolare in funzione delle nuove coordinate test\`{e}
ricavate, e si aggiunge ad essa una nuova funzione $\psi$ in modo che si
abbia
\begin{subequations}
\label{eq hamjac iconale}%
\begin{align}
\frac{\partial}{\partial X_{e}^{^{\prime}}}\left(  T+\psi\right)   &
=X^{\prime}-X,~~\frac{\partial}{\partial Y_{e}^{^{\prime}}}\left(
T+\psi\right)  =Y^{\prime}-Y,\\
\frac{\partial}{\partial X^{\prime}}\left(  T+\psi\right)   &  =X_{e}^{\prime
}-X_{e},~~\frac{\partial}{\partial Y^{^{\prime}}}\left(  T+\psi\right)
=Y_{e}^{\prime}-Y_{e}%
\end{align}
si ottiene una una funzione $S=T+\psi$ detta \textbf{iconale di Schwarzschild}%
; l' espressione di $\psi$ \`{e}%
\end{subequations}
\begin{equation}
\psi=\frac{z_{e}-z_{o}}{2Nl_{e}^{2}}\left(  X^{2}+Y^{2}\right)  -\frac
{z_{e}^{\prime}-z^{\prime}}{2N^{\prime}l_{e}^{\prime2}}\left(  X^{^{\prime}%
2}+Y^{^{\prime}2}\right)  +\left(  X_{e}^{\prime}-X_{e}\right)  X+\left(
Y_{e}^{\prime}-Y_{e}\right)  Y.
\end{equation}

\end{definition}

Si scriva lo sviluppo in serie di $S$ fino al quarto ordine, che, come le
altre funzioni principali, conterr\`{a} solo potenze pari delle $\left(
x,y\right)  $ e delle $\left(  p,q\right)  $. Tenendo presente la costanza
delle variabili di Seidel e le \ref{eq hamjac iconale}, si evince che il
termine di ordine si annulla. E' ora possibile scrivere a quest' ordine di
approssimazione l' iconale come%
\begin{equation}
S=S_{0}+AR^{4}+BR_{e}^{4}+Ck^{2}+DR^{2}R_{e}^{2}+ER^{2}k+FR_{e}^{2}k
\label{iconale terzo ordine}%
\end{equation}
con $R^{2}=X^{2}+Y^{2},~R_{e}^{2}=X_{e}^{2}+Y_{e}^{2},~k=XX_{e}+YY_{e}.$

Applicando le \ref{eq hamjac iconale} si ricava il vettore di aberrazione
nelle nuove coordinate%
\begin{gather}
\alpha\varepsilon_{x}=\left(  2Ck^{2}+ER^{2}+FR_{e}^{\prime2}\right)
X+\left(  BR_{e}^{\prime2}+DR^{2}+2Fk^{2}\right)  X_{e}^{\prime},\\
\alpha\varepsilon_{y}=\left(  2Ck^{2}+ER^{2}+FR_{e}^{\prime2}\right)
Y+\left(  BR_{e}^{\prime2}+DR^{2}+2Fk^{2}\right)  Y_{e}^{\prime}.
\end{gather}
con $\alpha=\frac{N^{\prime}l^{\prime}}{z_{e}^{\prime}}$. Poich\'{e} $\psi$
\`{e} una funzione quadratica delle variabili di Seidel, il termine del quarto
ordine dell' iconale coincide con il termine del quarto ordine della funzione angolare.

\section{Il teorema di addizione delle aberrazioni del terzo ordine}

Sar\`{a} ora dimostrato il teorema di addizione delle aberrazione del terzo
ordine utilizzando l' iconale di Schwarzschild. Risulteranno fondamentali le
propriet\`{a} delle variabili di Seidel.

Sia $K$ un sistema ottico composto da due superfici $K_{1},~K_{2},$ e siano
$S_{1},~S_{2},$ le rispettive iconali di equazione%
\begin{align}
S_{1}  &  =T_{1}+\frac{z_{e1}}{2N_{1}l_{e1}^{2}}\left(  X_{1}^{2}+Y_{1}%
^{2}\right)  -\frac{z_{e1}^{\prime}}{N_{e1}^{\prime}l_{e1}^{\prime2}}\left(
X_{1}^{^{\prime}2}+Y_{1}^{\prime2}\right)  +\left(  X_{e1}^{\prime}%
-X_{e1}\right)  X_{1}+\left(  Y_{e1}^{\prime}-Y_{e1}\right)  Y_{1},\\
S_{2}  &  =T_{2}+\frac{z_{e2}}{2N_{2}l_{e2}^{2}}\left(  X_{2}^{2}+Y_{2}%
^{2}\right)  -\frac{z_{e2}^{\prime}}{N_{e2}^{\prime}l_{e2}^{\prime2}}\left(
X_{2}^{^{\prime}2}+Y_{2}^{\prime2}\right)  +\left(  X_{e2}^{\prime}%
-X_{e2}\right)  X_{2}+\left(  Y_{e2}^{\prime}-Y_{e2}\right)  Y_{2}.
\end{align}
L' iconale del sistema completo sar\`{a}%
\begin{equation}
S_{1}=T_{1}+\frac{z_{e1}}{2N_{1}l_{e1}^{2}}\left(  X_{1}^{2}+Y_{1}^{2}\right)
-\frac{z_{e2}^{\prime}}{N_{e2}^{\prime}l_{e2}^{\prime2}}\left(  X_{2}%
^{^{\prime}2}+Y_{2}^{\prime2}\right)  +\left(  X_{e2}^{\prime}-X_{e1}\right)
X_{1}+\left(  Y_{e2}^{\prime}-Y_{e1}\right)  Y_{1}.
\end{equation}
Ricordando che $z_{e1}^{\prime}=z_{e2},~l_{e1}^{\prime}=l_{e2}$ e che le
variabili di Seidel non variano, entro termini di Gauss nel passaggio tra i
vari spazi oggetto ed immagine, si ottiene%
\begin{equation}
S=S_{1}+S_{2}+\left(  X_{1}-X_{1}^{\prime}\right)  \left(  X_{e2}^{\prime
}-X_{e2}\right)  +\left(  Y_{1}-Y_{1}^{\prime}\right)  \left(  Y_{e2}^{\prime
}-Y_{e2}\right)
\end{equation}
ed usando le \ref{eq hamjac iconale}%
\begin{equation}
S=S_{1}+S_{2}+\frac{\partial S}{\partial X_{e1}^{\prime}}\frac{\partial
S}{\partial X_{2}}+\frac{\partial S}{\partial Y_{e1}^{\prime}}\frac{\partial
S}{\partial Y_{2}}%
\end{equation}
Si sviluppi ora in serie di Taylor. Dall' assenza di termini del secondo
ordine e ricordando le \ref{eq hamjac iconale} si ottiene che
\begin{equation}
S=S_{1}^{\left(  0\right)  }+S_{2}^{\left(  0\right)  }+S_{1}^{\left(
4\right)  }+S_{2}^{\left(  4\right)  }%
\end{equation}
dove%
\begin{equation}
S_{i}^{\left(  4\right)  }=A_{i}R_{i}^{4}+B_{i}R_{ei}^{4}+C_{i}k_{i}^{2}%
+D_{i}R_{i}^{2}R_{ie}^{2}+E_{i}R_{i}^{2}k_{i}+F_{i}R_{e}^{2}k_{i}~~i=1,2
\end{equation}
con ovvio significato della notazione usata. La costanza delle variabili di
Seidel ci permette di scrivere%
\begin{align}
S^{\left(  4\right)  }  &  =\left(  A_{1}+A_{2}\right)  R^{4}+\left(
B_{1}+B_{2}\right)  R_{e}^{4}+\left(  C_{1}+C_{2}\right)  k^{2}\nonumber\\
&  +\left(  D_{1}+D_{2}\right)  R^{2}R_{e}^{2}+\left(  E_{1}+E_{2}\right)
R^{2}k+\left(  F_{1}+F_{2}\right)  R_{e}^{2}k.
\end{align}
Si pu\`{o} ora enunciare il seguente

\begin{theorem}
Ogni coefficiente di aberrazione di un sistema ottico centrato \`{e} la somma
dei corrispondenti coefficienti delle singole superfici che compongono il sistema.
\end{theorem}

\section{Sviluppo al quarto ordine della funzione angolare}

Si proceda ora allo sviluppo della funzione angolare, che permetter\`{a} di
esprimere i coefficienti di aberrazione in funzione delle caratteristiche
fisiche del sistema. Si seguir\`{a} la stessa via utilizzata nel capitolo
precedente, quando sono stati calcolati i primi termini dello sviluppo. Si
parta dall' espressione generale della funzione angolare
\begin{equation}
T=\left(  p-p^{\prime}\right)  x+\left(  q-q^{\prime}\right)  y+\left(
r-r^{\prime}\right)  z+r^{\prime}a^{\prime}-ra,
\label{definizione T per calcolo 3 ordine}%
\end{equation}
dove \`{e} stato indicato con $P=\left(  x,y,z\right)  $ il punto d'
intersezione tra $S$ e il raggio $\mathbf{p}$. Le coordinate del punto di
intersezione devono estremizzare la lunghezza del cammino ottico, ossia la
funzione
\begin{equation}
T\left(  x,y,z\right)  +\lambda F\left(  x,y,z\right)  ,
\end{equation}
dove $\lambda$ \`{e} un moltiplicatore di Lagrange e $F\left(  x,y,z\right)
=z-z\left(  x^{2}+y^{2}\right)  =0$ \`{e} l' equazione della superficie $S$.
La condizione necessaria per un estremale \`{e}
\begin{equation}
\frac{\partial T}{\partial x}+\lambda\frac{\partial F}{\partial x}%
=0,~~\frac{\partial T}{\partial y}+\lambda\frac{\partial F}{\partial
y}=0,~~\frac{\partial T}{\partial z}+\lambda\frac{\partial F}{\partial z}=0,
\end{equation}
o esplicitamente
\begin{equation}
p-p^{\prime}=\lambda\frac{\partial z}{\partial x},~~q-q^{\prime}=\lambda
\frac{\partial z}{\partial x},~~r-r^{\prime}=-\lambda.
\end{equation}
Sostituendo la terza relazione nelle altre si ricava%
\begin{equation}
\frac{p-p^{\prime}}{r-r^{\prime}}=-\frac{\partial z}{\partial x}%
,~~\frac{q-q^{\prime}}{r-r^{\prime}}=-\frac{\partial z}{\partial y}.
\end{equation}
Si espliciti l' equazione della superficie $S$ entro termini del quarto ordine%
\begin{equation}
z=\frac{1}{2R_{c}}\left(  x^{2}+y^{2}\right)  +\frac{\left(  1+K\right)
}{8R_{c}^{3}}\left(  x^{2}+y^{2}\right)  ^{2}.
\end{equation}
Sempre entro termini del quarto ordine,%
\begin{equation}
r=N-\frac{1}{2N}\left(  p^{2}+q^{2}\right)  +O\left(  4\right)  ,~~r^{\prime
}=N^{\prime}-\frac{1}{2N^{\prime}}\left(  p^{\prime2}+q^{\prime2}\right)
+O\left(  4\right)
\end{equation}
e
\begin{equation}
\frac{1}{r-r^{\prime}}=\frac{1}{N^{\prime}-N}\left[  1-\frac{1}{2N\left(
N^{\prime}-N\right)  }\left(  p^{2}+q^{2}\right)  +\frac{1}{2N^{\prime}\left(
N-N^{\prime}\right)  }\left(  p^{\prime2}+q^{\prime2}\right)  \right]  .
\end{equation}
Inserendo gli sviluppi in serie sin qui trovati nalla
\ref{definizione T per calcolo 3 ordine} dopo semplici, ma noiosi calcoli, si
arriva all' espressione di $T$ al quarto ordine
\begin{subequations}
\label{T4 schw}%
\begin{align}
T_{4}  &  =\frac{1}{8Nz}\left\{  \frac{NR_{c}}{\left(  N^{\prime}-N\right)
^{2}}\left[  \left(  p-p^{\prime}\right)  ^{2}+\left(  q-q^{\prime}\right)
^{2}\right]  -\frac{z}{N}\left(  p^{2}+q^{2}\right)  \right\} \nonumber\\
&  -\frac{1}{8N^{\prime}z^{\prime}}\left\{  \frac{N^{\prime}R_{c}}{\left(
N^{\prime}-N\right)  ^{2}}\left[  \left(  p-p^{\prime}\right)  ^{2}+\left(
q-q^{\prime}\right)  ^{2}\right]  -\frac{z^{\prime}}{N^{\prime}}\left(
p^{\prime2}+q^{\prime2}\right)  \right\} \\
&  -\frac{KR_{c}}{8\left(  N^{\prime}-N\right)  ^{3}}\left[  \left(
p-p^{\prime}\right)  ^{2}+\left(  q-q^{\prime}\right)  ^{2}\right]  .\nonumber
\end{align}

\section{I coefficienti di aberrazione \newline del terzo ordine}

Per avere l' espressione esplicita dei coefficienti di aberrazione \`{e}
necessario esprimere $T_{4}$ in funzione delle variabili di Seidel ed
applicare il teorema di addizione delle aberrazioni del terzo ordine. Si
identifichino ora i punti assiali $z=a$ e $z\prime=a$ con i punti oggetto ed
immagine dell' ottica Gaussiana e si introducano le seguenti notazioni%
\end{subequations}
\begin{equation}
z=a,~~z^{\prime}=a^{\prime},~~w_{e}=a+z_{e},~~w_{e}^{\prime}=a^{\prime}%
+z_{e}^{\prime},
\end{equation}
\begin{subequations}
\label{invarianti Abbe}%
\begin{align}
Q  &  =N\left(  \frac{1}{R_{c}}-\frac{1}{z}\right)  =N^{\prime}\left(
\frac{1}{R_{c}}-\frac{1}{z^{\prime}}\right)  ,\\
L  &  =N\left(  \frac{1}{R_{c}}-\frac{1}{w_{e}}\right)  =N^{\prime}\left(
\frac{1}{R_{c}}-\frac{1}{w_{e}^{\prime}}\right)  ,
\end{align}
dove $Q$ e $L$ sono gli invarianti di Abbe della superficie $S$ relativi alle
coppie dei piani oggetto-immagine ed alle pupille d'
entrata-uscita\footnote{Gli invarianti di Abbe discendono direttamente dalla
relazione di Huygens scritta per una superficie. Infatti da \ $-\frac{N}%
{z}+\frac{N^{\prime}}{z^{\prime}}=\frac{N^{\prime}-N}{R_{c}}$ e dalla medesima
relazione scritta per i pianidelle pupille, dopo un leggero maquillage, si
perviene alle espressione per $Q$ $,L$.}. Si verifica che%
\end{subequations}
\begin{equation}
\frac{l^{\prime}}{l}=\frac{R_{c}-z^{\prime}}{R_{c}-z},~~\frac{l_{e}^{\prime}%
}{l_{e}}=\frac{R_{c}-z_{e}^{\prime}}{R_{c}-z_{e}}.
\end{equation}
e, tenendo conto delle \ref{invarianti Abbe}, si pu\`{o} scrivere
\begin{equation}
\frac{l^{\prime}}{l}=\frac{Nz^{\prime}}{N^{\prime}z},~~\frac{l_{e}^{\prime}%
}{l_{e}}=\frac{Nz_{e}^{\prime}}{N^{\prime}z_{e}}.
\end{equation}
Introdotte le notazioni
\begin{equation}
h=\frac{l_{e}z}{a}=\frac{l_{e}^{\prime}z^{\prime}}{a^{\prime}}%
\end{equation}
le vecchie variabili, espresse in funzione di quelle di Seidel, assumono la
seguente forma%
\begin{align}
p  &  =N\left(  \frac{h}{z}X_{e}^{\prime}-\frac{h_{e}}{w_{e}}X\right)
,~~q=N\left(  \frac{h}{z}Y_{e}^{\prime}-\frac{h_{e}}{w_{e}}Y\right)  ,\\
p^{\prime}  &  =N^{\prime}\left(  \frac{h}{z^{\prime}}X_{e}^{\prime}%
-\frac{h_{e}}{w_{e}^{\prime}}X\right)  ,~~q^{\prime}=N^{\prime}\left(
\frac{h}{z}Y_{e}^{\prime}-\frac{h_{e}}{w_{e}}Y\right)  .
\end{align}
Sostituendo queste relazioni nella\ \ref{T4 schw} ed applicando il teorema di
addizione delle aberrazione del terzo ordine, dopo lunghi e tediosi passaggi
algebrici, si ottengono i coefficienti di aberrazione cercati
\begin{align}
A  &  =\frac{1}{2}\sum_{i}h_{e}^{4}\left[  \left(  N_{i}-N_{i-1}\right)
\frac{K_{i}}{R_{ci}^{3}}+L_{i}^{2}\left(  \frac{1}{N_{i}z_{i}^{\prime}}%
-\frac{1}{N_{i-1}z_{i}}\right)  \right. \nonumber\\
&  +\left(  Q_{i}-L_{i}\right)  ^{2}\left(  \frac{z_{i}^{\prime}}{N_{i}%
w_{ei}^{\prime2}}-\frac{z_{i}}{N_{i-1}w_{ei}^{2}}\right)  \left.
-2L_{i}\left(  Q_{i}-L_{i}\right)  \left(  \frac{1}{N_{i}w_{ei}^{\prime}%
}-\frac{1}{N_{i-1}w_{ei}}\right)  \right]  ,\\
B  &  =\frac{1}{2}\sum_{i}h_{i}^{4}\left[  \left(  N_{i}-N_{i-1}\right)
\frac{K_{i}}{R_{ci}^{3}}+Q_{i}^{2}\left(  \frac{1}{N_{i}z_{i}^{\prime}}%
-\frac{1}{N_{i-1}z_{i}}\right)  \right]  ,\\
C  &  =\frac{1}{2}\sum_{i}h_{i}^{2}h_{ei}^{2}\left[  \left(  N_{i}%
-N_{i-1}\right)  \frac{K_{i}}{R_{ci}^{3}}+Q_{i}^{2}\left(  \frac{1}{N_{i}%
z_{i}^{\prime}}-\frac{1}{N_{i-1}z_{i}}\right)  \right]  ,\\
D  &  =\frac{1}{2}\sum_{i}h_{i}^{2}h_{ei}^{2}\left[  \left(  N_{i}%
-N_{i-1}\right)  \frac{K_{i}}{R_{ci}^{3}}+Q_{i}L_{i}\left(  \frac{1}%
{N_{i}z_{i}^{\prime}}-\frac{1}{N_{i-1}z_{i}^{\prime}}\right)  \right.
\nonumber\\
&  \left.  -Q_{i}\left(  Q_{i}-L_{i}\right)  \left(  \frac{1}{N_{i}%
w_{ei}^{\prime}}-\frac{1}{N_{i-1}w_{ei}^{\prime}}\right)  \right]  ,
\end{align}

\begin{align}
E  &  =\frac{1}{2}\sum_{i}h_{i}h_{ei}^{3}\left[  \left(  N_{i}-N_{i-1}\right)
\frac{K_{i}}{R_{ci}^{3}}+L_{i}^{2}\left(  \frac{1}{N_{i}z_{i}^{\prime}}%
-\frac{1}{N_{i-1}z_{i}^{\prime}}\right)  \right. \nonumber\\
&  \left.  -L_{i}\left(  Q_{i}-L_{i}\right)  \left(  \frac{1}{N_{i}%
w_{ei}^{\prime}}-\frac{1}{N_{i-1}w_{ei}^{\prime}}\right)  \right]  ,\\
F  &  =\frac{1}{2}\sum_{i}h_{i}^{3}h_{ei}\left[  \left(  N_{i}-N_{i-1}\right)
\frac{K_{i}}{R_{ci}^{3}}+Q_{i}L_{i}\left(  \frac{1}{N_{i}z_{i}^{\prime}}%
-\frac{1}{N_{i-1}z_{i}^{\prime}}\right)  \right]  .
\end{align}

\section{Il significato dei coefficienti di aberrazione}

Sar\`{a} ora proposta un' interpretazione geometrica dei coefficienti di
aberrazione del terzo ordine e, conseguentemente, del loro effetto sulla
qualit\`{a} finale dell' immagine prodotta dal sistema in considerazione.
Analizzando la \ref{iconale terzo ordine}, \`{e} banale verificare che il
coefficiente $A$ non ha influenza sulla qualit\`{a} finale dell' immagine in
quanto $X^{\prime}=\frac{\partial S}{\partial X_{e}}$,$Y^{\prime}%
=\frac{\partial S}{\partial Y_{e}}.$ L' analisi va quindi effettuata sugli
altri coefficienti.Conviene esaminare singolarmente i coefficienti ponendo di
volta in volta tutti i coefficienti uguali a zero tranne uno. Inoltre, grazie
alla simmetria assiale del sistema, si pu\`{o} porre, senza perdit\`{a} di
generalit\`{a} $X=0,$ e $X_{e}=R_{e}\cos\alpha$, $Y_{e}=R_{e}\sin\alpha$
.Conviene introdurre anche le curve di aberrazione, tagliate sul piano
immagine dai fasci della superficie conica che ha il vertice nel punto
oggetto, e che dopo la rifrazione hanno come base un cerchio di raggio
$R_{e}^{^{\prime}}$ sul piano $z_{e}^{\prime}.$ L'intersezione dei fasci con
la pupilla di entrata \`{e} approssimativamente un cerchio.Il coefficiente $B$
\`{e} responsabile della \textbf{aberrazione sferica}, infatti si ha
\begin{equation}
\varepsilon_{x}=BR_{e}^{3}\cos\alpha,~~\varepsilon_{y}=BR_{e}^{3}\sin\alpha.
\end{equation}
Le curve di aberrazione formano dei cerchi concentrici con centro nel punto
immagine Gaussiano; il loro raggio cresce in base alla terza potenza della
apertura dello strumento e sono indipendenti dalla posizione del punto
oggetto. La \textbf{distorsione}\emph{ }\`{e} invece causata dal coefficiente
$E$ che compare nell' espressione del vettore di aberrazione nel modo seguente%
\begin{equation}
\varepsilon_{x}=0,~~\varepsilon_{y}=ER^{2}Y.
\end{equation}
Siccome $X_{e,}~$ed $Y_{e}$ non compaiono nelle formule, l'immagine risultante
\`{e} puntiforme, ma le distanze non sono precisamente proporzionali a quelle
dei punti dello spazio oggetto.A seconda del segno di $E$ si parla di
distorsione a cuscinetto o a barilotto. Il termine in $F$ produce la
\textbf{coma}%
\begin{equation}
\varepsilon_{x}=-FYR_{e}^{2}\sin2\alpha,~~\varepsilon_{y}=-FYR_{e}^{2}\left(
2+2\cos\alpha\right)
\end{equation}
Le curve di aberrazione che, fissato $Y$, si formano per vari coni dei raggi
d' ingresso (con $R_{e}$ variabile) sono delle circonferenze tangenti a due
rette uscenti dal punto immagine di Gauss, che formano un angolo di $30%
{{}^\circ}%
$ con l' asse delle $Y.$ Questo aberrazione \`{e} detta coma per via dell'
aspetto asimmetrico che d\`{a} alle immagini. Tale effetto trasforma un
oggetto puntiforme in uno di forma allungata. I primi astronomi confondevano a
causa della coma le stelle con delle comete da cui il nome dell' aberrazione.
Per l' analisi delle altre aberrazioni conviene considerare contemporaneamente
i coefficienti $C$ e $D$ che generano l'\textbf{astigmatismo }e la
\textbf{curvatura di campo. }Il fascio di raggi entrante, di piccola apertura,
ha due linee focali una in direzione saggittale all' asse dello strumento, l'
altra tangente ad una circonferenza che ha il centro sull' asse ottico ed
appartiene ad un piano verticale all' asse stesso. Le due superfici contenenti
le linee focali, al variare della posizione dell' oggetto, si chiamano
rispettivamente superficie immagine saggittale e tangenziale. Nella prima
approssimazione si possono sostituire le due superfici con delle sfere
tangenti all' asse, che che abbiano raggi $\rho_{s}$ $\rho_{t}.$ Si fissi come
positivo il valore di $\rho_{s}$ e $\rho_{t}$ quando il centro della sfera cui
afferiscono si trova davanti al piano immagine, assumendo come verso positivo
quello di propagazione della luce. Le immagini si formano nitide sulle sfere
test\`{e} introdotte, proiettando queste sul piano immagine si ottengono i
seguenti valori del vettore di aberrazione%
\begin{equation}
\varepsilon_{x}=\frac{Y^{2}R_{e}\cos\theta}{2\rho_{t}N}~~\varepsilon_{y}%
=\frac{Y^{2}R_{e}\sin\theta}{2\rho_{s}N}%
\end{equation}%
\begin{equation}
\frac{1}{\rho_{t}}=2N^{\prime}\left(  2C+D\right)  ,~~\frac{1}{\rho_{s}%
}=2N^{\prime}D.
\end{equation}
Le quantit\`{a} $2C+D$ e $D$ si chiamano convessit\`{a} d' immagine
tangenziale e saggittale. Si parla di astigmatismo per indicare la met\`{a}
della differenza, mentre la met\`{a} della somma viene detta curvatura di
campo. Mandando a zero tutte la altre aberrazioni si forma un' immagine nitida
sulla sfera di raggio $\rho$ tangente al piano immagine sull' asse ottico.%
\begin{gather}
\frac{1}{2}\left(  \frac{1}{\rho_{t}}-\frac{1}{\rho_{s}}\right)  =2N^{\prime
}C,\\
\frac{1}{\rho}=\frac{1}{2}\left(  \frac{1}{\rho_{s}}+\frac{1}{\rho_{t}%
}\right)  =2N^{\prime}\left(  C+D\right)
\end{gather}
%

\begin{center}
\includegraphics[
trim=0.000000in 0.000000in -0.001250in -0.004166in,
natheight=2.603900in,
natwidth=3.125400in,
height=6.7305cm,
width=8.0572cm
]%
{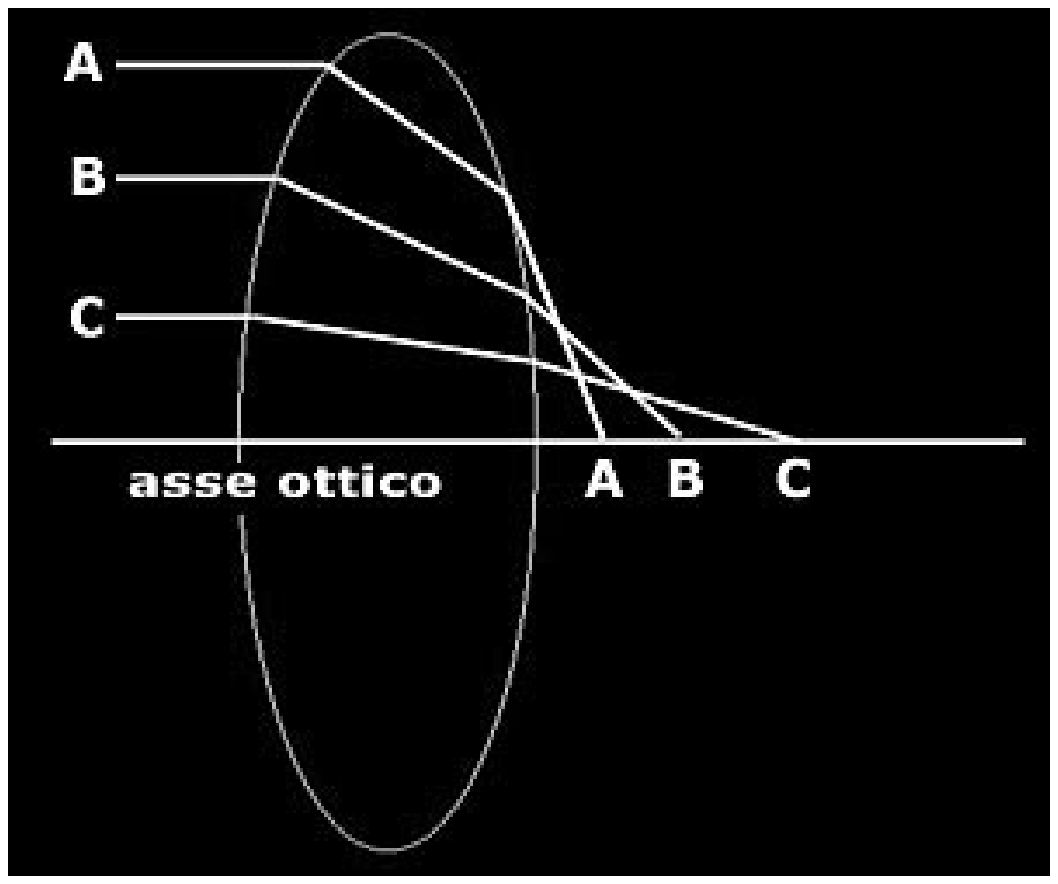}%
\\
{}Aberrazione sferica.
\end{center}
\begin{center}
\includegraphics[
trim=0.000000in 0.000000in 0.000761in -0.004166in,
natheight=2.603900in,
natwidth=1.521200in,
height=6.7305cm,
width=3.9517cm
]%
{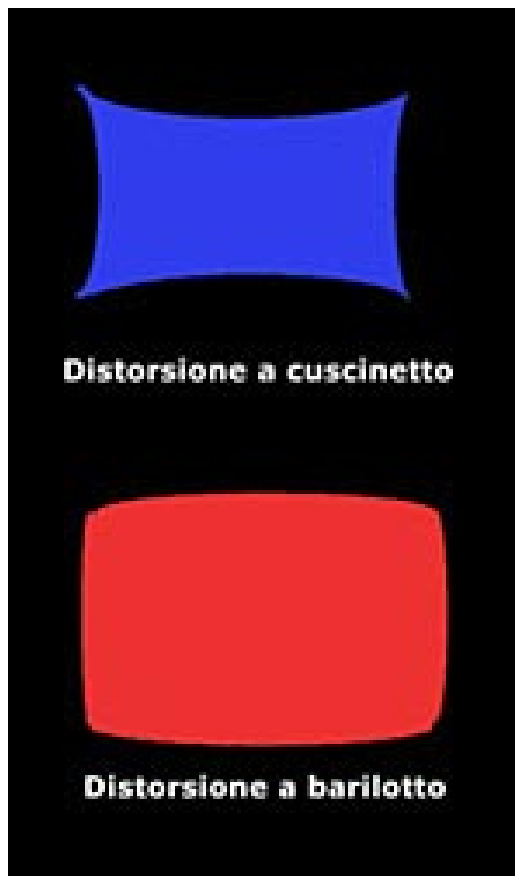}%
\\
Distorsioni prodotte su un soggetto di forma rettangolare.
\end{center}
\begin{center}
\includegraphics[
natheight=2.603900in,
natwidth=4.166700in,
height=2.6455in,
width=4.2168in
]%
{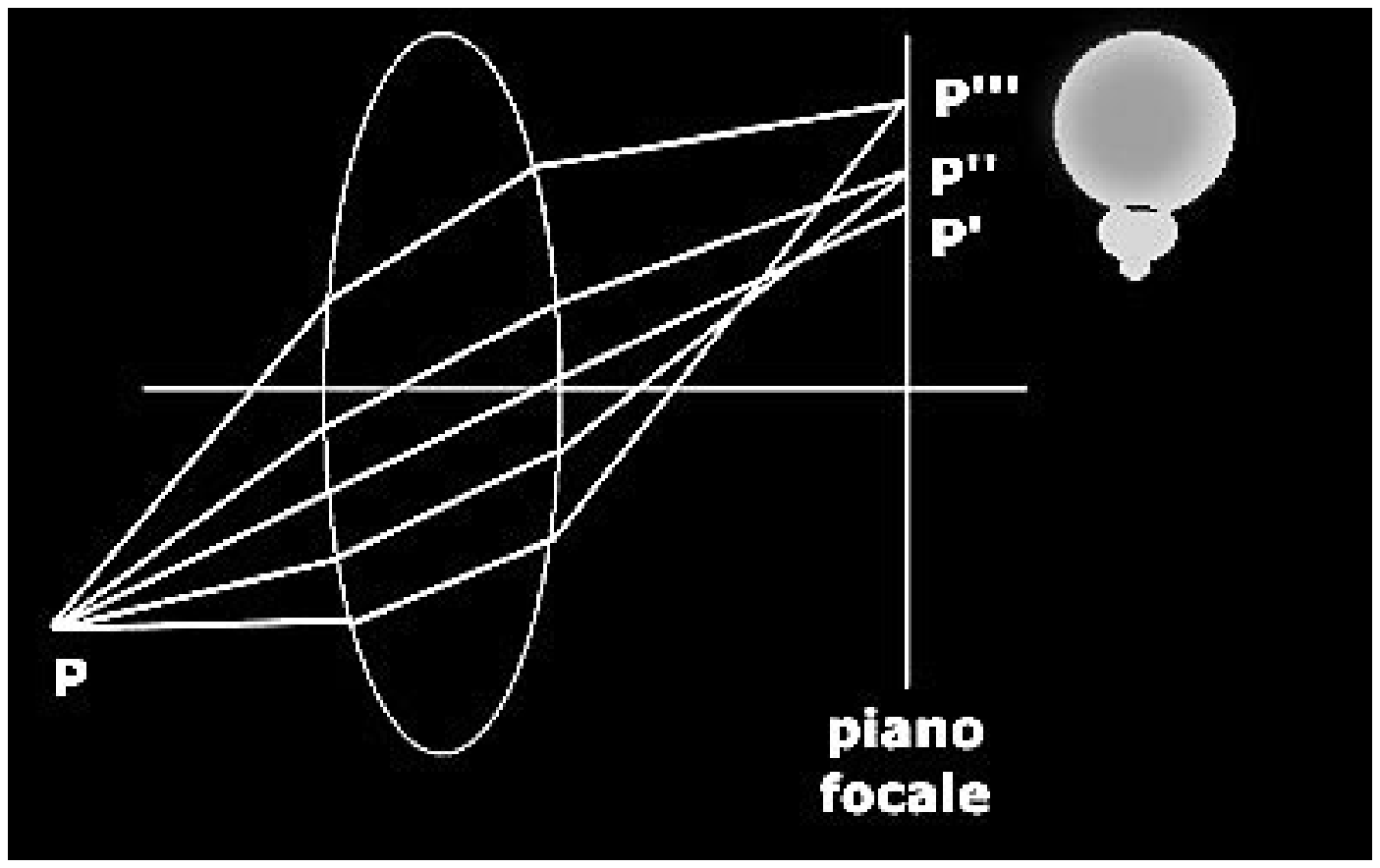}%
\\
Coma
\end{center}
\begin{center}
\includegraphics[
natheight=1.740000in,
natwidth=3.479100in,
height=1.7775in,
width=3.5259in
]%
{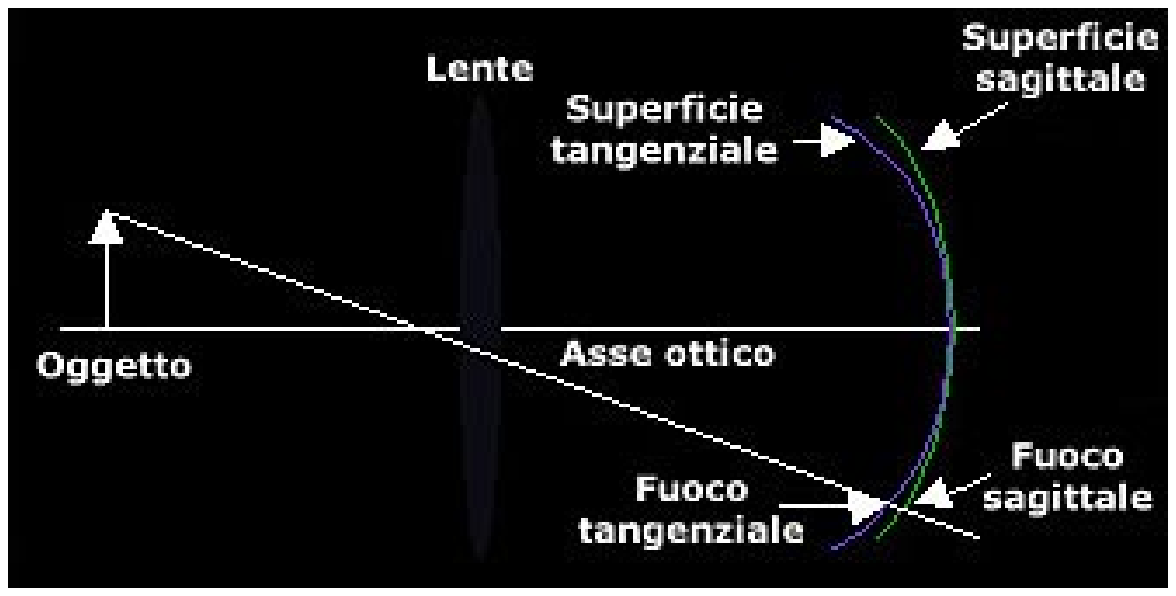}%
\\
Astigmatismo
\end{center}
\begin{center}
\includegraphics[
natheight=2.593600in,
natwidth=3.646100in,
height=2.6351in,
width=3.6936in
]%
{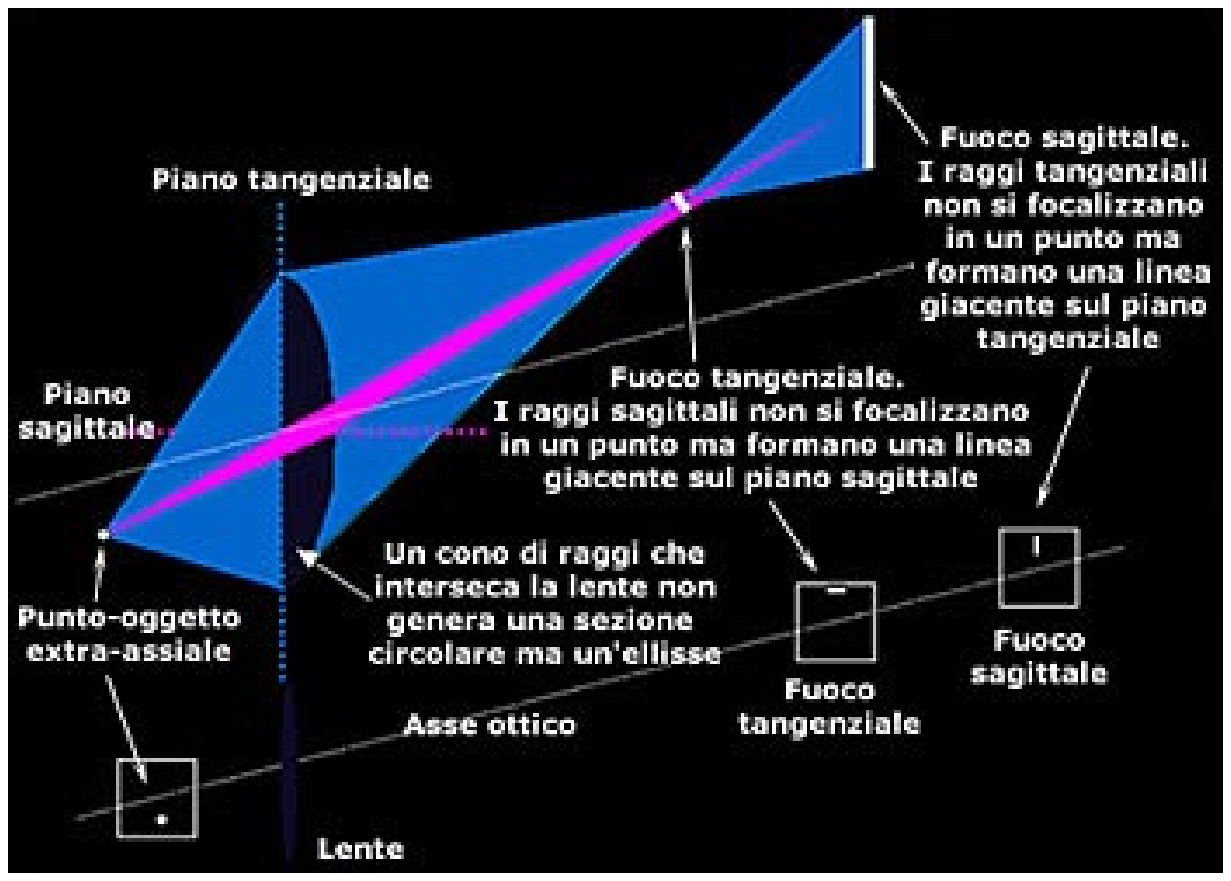}%
\\
Curvatura di campo
\end{center}

\section{Le aberrazioni del quinto ordine}

Per ricavare le aberrazioni del quinto ordine i ragionamenti sono
assolutamente analoghi a quelli fatti fin' ora. Si sviluppi
l\textquotedblright\ iconale ad un ordine successivo di approssimazione%
\begin{gather}
S_{6}=S_{1}R^{6}+S_{2}R_{0}^{4}R_{e}^{2}+S_{3}R^{4}w^{2}+S_{4}R^{2}R_{e}%
^{4}+S_{5}R^{2}R_{e}^{2}w^{2}\\
+S_{6}RR_{e}w^{2}+S_{7}R_{E}^{6}+S_{8}R_{e}^{2}w^{2}+S_{9}R_{E}w^{2}%
+S_{10}w^{6},
\end{gather}
dove le $S_{i}\ $si considerano costanti arbitrarie. Ponendo anche in questo
caso $Y_{0}=0$ dalle \ref{eq hamjac iconale} si ricava%
\begin{align}
-Y^{\prime}  &  =\frac{\partial S^{\left(  4\right)  }}{\partial Y_{e}}%
+2Y_{e}\left(  S_{2}Y^{4}+2S_{4}Y^{4}R_{e}+S_{5}Y^{3}Y_{e}+3S_{7}R_{e}%
^{2}+2S_{8}YY_{e}R_{e}^{2}+S_{9}Y^{2}Y_{e}^{2}\right) \\
X-X^{\prime}  &  =\frac{\partial S^{\left(  4\right)  }}{\partial X_{e}%
}+2X_{e}\left(  S_{2}X^{4}+2S_{4}X^{4}R_{e}+S_{5}X^{3}X_{e}+3S_{7}R_{e}%
^{2}+2S_{8}XX_{e}R_{e}^{2}+S_{9}X^{2}X_{e}^{2}\right)  +\\
&  X\left(  S_{3}X^{4}+S_{5}X^{2}R_{e}^{2}+2S_{6}X^{3}X_{e}+S_{8}R_{e}%
^{4}+2S_{9}XX_{e}R_{e}^{2}+3S_{10}X^{2}X_{e}^{2}\right)  .\nonumber
\end{align}
Siccome $S_{1}$ non compare in queste espressioni, rimangono nove coefficienti
indipendenti del quinto ordine. Questi si possono isolare uno per volta e
considerare le rispettive curve di aberrazione. Si pone al solito $X_{e}%
=R_{e}\cos\theta,Y_{e}=R_{e}\sin\theta$ e si indichino le variazioni cos\`{\i}
ottenute con $\Delta X,\Delta Y$. La nomenclatura utilizzata \`{e} quella
proposta da Schwarzschild, in letteratura esistono altre classificazioni.

\begin{itemize}
\item $S_{7}$ - aberrazione sferica di secondo grado: queste curve di
aberrazione sono delle circonferenze, i raggi delle quali sono indipendenti
dalla posizione dell' oggetto e crescono con la quinta potenza dell' apertura
dello strumento.%
\begin{align}
\Delta X  &  =-6S_{7}R_{e}^{5}\cos\theta\\
\Delta Y  &  =-6S_{7}R_{e}^{5}\sin\theta.
\end{align}

\item $S_{8}$ - coma circolare: le curve di aberrazione sono delle
circonferenze di raggio $2S_{8}Y_{0}R_{e}^{4}$ tangenti ad un piano che forma
un angolo di $41%
{{}^\circ}%
8^{\prime}$ ($\sin41%
{{}^\circ}%
8^{\prime}=2/3$) con l' asse delle $Y$%
\begin{align}
\Delta X  &  =-S_{8}XR_{e}^{4}\cos\theta,\\
\Delta Y  &  =-S_{8}XR_{e}^{4}\sin\theta.
\end{align}

\item $S_{4}$ - aberrazione sferica sagittale: le curve di aberrazione sono
delle circonferenze, i raggi delle quali crescono con il quadrato della
distanza assiale e la terza potenza dell' apertura%
\begin{align}
\Delta X  &  =-4S_{4}X^{2}R_{e}^{3}\cos\theta,\\
\Delta Y  &  =-4S_{4}X^{2}R_{e}^{3}\sin\theta.
\end{align}

\item $S_{9}$ - ali :in questo caso le curve di aberrazioni sono del sesto
ordine ed inalcuni casi assumono la suggestiva forma di ali. Tutte le curve
generate da raggi uscenti dallo stesso punto oggetto sono centrate nel punto
immagine coniugato al punto oggetto%
\begin{align}
\Delta X  &  =-2S_{9}X^{2}R_{e}^{3}\cos\theta\left(  1+\cos^{2}\theta\right)
\\
\Delta Y  &  =-2S_{9}X^{2}R_{e}^{3}\cos^{2}\theta\sin\theta.
\end{align}

\item $S_{10}$ - freccia: la curva di aberrazione consiste in una retta, che
si stende dall' immagine di Gauss verso il bordo dell' immagine.%
\begin{align}
\Delta X  &  =-3S_{10}X^{3}R_{e}^{2}\cos^{2}\theta,\\
\Delta Y  &  =0
\end{align}

\item $S_{5}$ - coma laterale: le curve di aberrazione hanno la stessa forma
della coma consueta, si distinguono soltanto nel fatto che crescono con la
terza potenza della distanza assiale.%
\begin{align}
\Delta X  &  =-S_{5}X^{3}R_{e}^{2}\left(  1+2\cos^{2}\theta\right)  ,\\
\Delta Y  &  =-S_{5}X^{3}R_{e}^{2}2\cos\theta\sin\theta.
\end{align}

\item $S_{2}$ $S_{6}$ - astigmatismo e curvatura di campo sagittali: anche in
questo caso conviene considerare due aberrazioni contemporaneamente. Le curve
di aberrazione sono delle ellissi. Si chiamer\`{a} la met\`{a} della
differenza astigmatismo e la met\`{a} della somma convessit\`{a} laterale d'
immagine.%
\begin{align}
\Delta X  &  =-2\left(  S_{2}+S_{6}\right)  X^{4}R_{e}\cos\theta\\
\Delta Y  &  =-2S_{2}X^{4}R_{e}\sin\theta
\end{align}

\item $S_{3}$ - distorsione sagittale: analogamente a quella del terzo ordine
questa aberrazione non influenza le immagini puntiformi, ma contribuisce
ulteriormente, insieme a quella del terz' ordine, alla deformazione di oggetti
estesi%
\begin{align}
\Delta X  &  =-S_{3}X^{5},\\
\Delta Y  &  =0.
\end{align}

\end{itemize}

\chapter{La teoria delle aberrazioni per mezzo delle \newline equazioni di
Hamilton}

\section{L' ottica al primo ordine per un sistema ottico centrato}

Si consideri l'Hamiltoniana
\begin{equation}
H=-\sqrt{N^{2}\left(  X^{2}+Y^{2},z\right)  -\left(  P^{2}+Q^{2}\right)
}=-\sqrt{N^{2}\left(  U,z\right)  -V},common used approach to the aberration theory is based on the Hamilton's principal functions. There is another approach, more intuitive, based on the Hamilton's equations. This is the so called
\end{equation}
in cui l' indice di rifrazione \`{e} continuo ed a simmetria assiale, e
$X,Y,P,Q$ indicano le soluzioni non approssimate delle equazioni di Hamilton e
si sono fatte le posizioni $X^{2}+Y^{2}=U,P^{2}+Q^{2}=V.$ Sviluppando in serie
di Taylor rispetto alle $U,V$ e troncando lo sviluppo al primo ordine si ha
\begin{equation}
H\simeq-N\left(  z\right)  +\frac{1}{2}\left.  \frac{\partial N}{\partial
U}\right\vert _{0}U+\frac{1}{2N\left(  z\right)  }V,
\end{equation}
dove si \`{e} posto $N\left(  0,z\right)  =N\left(  z\right)  .$ Le equazioni
di Hamilton per la nuova Hamiltoniana sono dunque
\begin{subequations}
\label{eq ham lun 1}%
\begin{align}
\dot{x}  &  =\frac{1}{N\left(  0,z\right)  }p,~~\dot{p}=-Dx,\\
\dot{y}  &  =\frac{1}{N\left(  0,z\right)  }q,~~\dot{q}=-Dy,
\end{align}
avendo posto $D=\left.  \frac{\partial N}{\partial U}\right\vert _{0}.$ La
situazione che per\`{o} presenta il maggior interesse dal punto di vista
fisico \`{e} quella in cui l' indice di rifrazione \`{e} costante a tratti,
come si verifica nei consueti strumenti ottici da osservazione e negli
obiettivi fotografici. Questa condizione sar\`{a} ottenuta come limite del
caso continuo. Prima di proseguire, conviene introdurre una opportuna
notazione per il caso discreto.

Saranno indicate con indice pari tutte quelle quantit\`{a} che si intendono
valutate sui piani tangenti ai vertici delle superfici ottiche, mentre per il
piano immagine verranno adoperate le consuete notazioni con apice. Siano quindi
\end{subequations}
\begin{itemize}
\item $z_{0},z_{2,}\cdots z_{2n},z^{\prime}$ ascisse dei piani tangenti alle
superfici ottiche. In particolare $z_{0}$ \`{e} il piano oggetto, $z^{\prime}$
il piano immagine;

\item $x_{0},x_{2},\cdots x_{2n},x^{\prime}$ $y_{0},y_{2},\cdots
y_{2n},y^{\prime}$ coordinate del raggio sui piani $z_{0},z_{2},\cdots
z_{2n},z^{\prime}.$

\item $R_{2},R_{4},\cdots R_{2n}$ i raggi di curvatura delle superfici ottiche.
\end{itemize}

Si utilizzeranno indici dispari per tutte le quantit\`{a} che ha senso
definire solo nelle regioni tra due piani tangenti

\begin{itemize}
\item $t_{1},t_{3}\cdots t_{2n+1}$ distanze assiali tra i piani tangenti
$z_{2n}$ e $z_{2n+2};$

\item $N_{1},N_{3}\cdots N_{2n+1}$ indice di rifrazione tra i piani tangenti
$z_{2n}$ e $z_{2n+2};$

\item $p_{1},p_{3},\cdots p_{2n+1}$ $q_{1},q_{3},\cdots q_{2n+1}$ momenti
cinetici tra i piani tangenti $z_{2n}$ e $z_{2n+2}.$
\end{itemize}

Nel caso in cui l' indice di rifrazione \`{e} costante a tratti, i raggi
luminosi altro non sono che dei tratti di retta connessi continuamente. Le
loro espressioni altro sono
\begin{equation}
x\left(  z\right)  =\frac{p}{N}z+c_{x},~~y\left(  z\right)  =\frac{q}%
{N}z+c_{y},
\end{equation}
con $c_{x},$ $c_{y}$ costanti. Si pu\`{o} scrivere la soluzione alle prime due
equazione di Hamilton come
\begin{equation}
x_{2n+2}-x_{2n}=\frac{t_{2n+1}}{N_{2n+1}}p_{2n+1},~~y_{2n+2}-y_{2n}%
=\frac{t_{2n+1}}{N_{2n+1}}q_{2n+1}.
\end{equation}
Per ottenere una soluzione analoga per le altre due equazioni conviene
supporre l' indice di rifrazione continuo in una piccola regione di ampiezza
$2\delta$ centrata nel punto $z_{2n}$ e successivamente passare al limite per
$\delta\rightarrow0$.%
\begin{equation}
\int_{z_{2n}-\delta}^{z_{2n}-\delta}pdz=-\int_{z_{2n}-\delta}^{z_{2n}-\delta
}\frac{N^{\prime}\left(  z\right)  }{R}x\left(  z\right)  =-\int
_{z_{2n}-\delta}^{z_{2n}-\delta}\frac{\Delta N}{\Delta z}\frac{1}{R_{c}%
}x\left(  z\right)  dz
\end{equation}
il passaggio al limite conduce a%
\begin{equation}
p_{2n+1}-p_{2n-1}=-x_{2n}\frac{\Delta N}{R_{2n}},
\end{equation}
ed a
\begin{equation}
q_{2n+1}-q_{2n-1}=-y_{2n}\frac{\Delta N}{R_{2n}}.
\end{equation}
Queste ultime relazioni altro non sono che la formula di Huygens, infatti
effettuando le sostituzioni
\begin{equation}
\frac{a}{n_{2n-1}}=\frac{-x_{2n}}{p_{2n-1}},~~\frac{a^{\prime}}{n_{2n+1}%
}=\frac{x_{2n}}{p_{2n+1}},
\end{equation}%
\begin{center}
\includegraphics[
natheight=2.099600in,
natwidth=2.059800in,
height=1.682in,
width=1.6496in
]%
{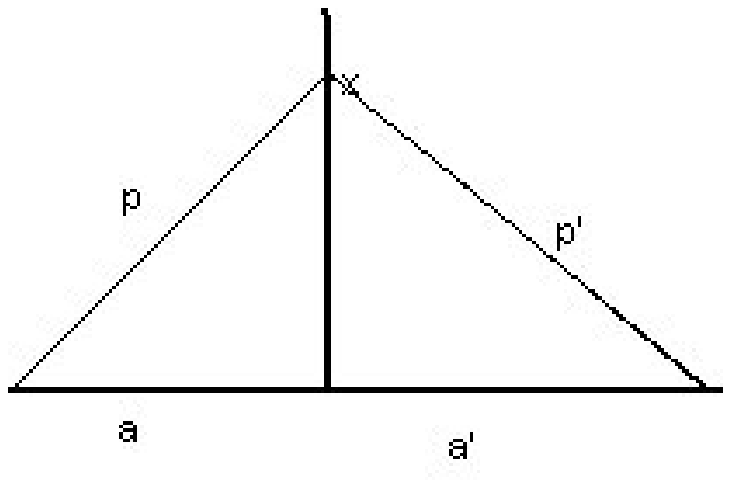}%
\end{center}
si ottiene
\begin{equation}
\frac{N_{2n-1}}{a}+\frac{N_{2n+1}}{a^{\prime}}=\frac{N_{2n+1}-N_{2n-1}}{R_{c}%
}.
\end{equation}
La presenza di un diaframma nello strumento va trattata con alcuni
accorgimenti. Sia $z_{e}$ l' ascissa del diaframma sull' asse ottico, conviene
descrivere il raggio luminoso tramite due soluzioni particolari delle
\ref{eq ham lun 1}, che saranno indicate con $h\left(  z\right)  ,k\left(
z\right)  ,g\left(  z\right)  ,f\left(  z\right)  $ tali che
\begin{align}
x\left(  z\right)   &  =x_{0}h\left(  z\right)  +x_{e}k\left(  z\right)
,~~y\left(  z\right)  =y_{0}h\left(  z\right)  +y_{e}k\left(  z\right)  ,\\
p\left(  z\right)   &  =x_{0}g\left(  z\right)  +x_{e}f\left(  z\right)
,~~q\left(  z\right)  =y_{0}g\left(  z\right)  +y_{e}f\left(  z\right)  .
\end{align}
con $f\left(  z\right)  =N\left(  z\right)  k^{\prime}\left(  z\right)
,~g\left(  z\right)  =N\left(  z\right)  h^{\prime}\left(  z\right)  .$ Le
soluzioni $h\left(  z\right)  ,$ $k\left(  z\right)  $ soddisfano le seguenti
condizioni al contorno
\begin{subequations}
\label{condizioni al cont1}%
\begin{align}
k\left(  z_{0}\right)   &  =0,~~k\left(  z^{\prime}\right)  =0~~k\left(
z_{e}\right)  =1,\\
h\left(  z_{0}\right)   &  =1,~~h\left(  z^{\prime}\right)  =M,~~h\left(
z_{e}\right)  =0,
\end{align}
$M$ indicando l' ingrandimento gaussiano del sistema. Le due soluzioni
$h\left(  z\right)  ,k\left(  z\right)  $ sono dette rispettivamente raggio
assiale e raggio di campo.

\section{L'invariante di Lagrange per i raggi assiale e di campo}

Due coppie di soluzioni $\left(  x,p\right)  ,\left(  \xi,\pi\right)  $ delle
equazioni di Hamilton per lo stesso sistema ottico centrato possono essere
messe in relazione grazie ad un invariante. Si osservi che valgono le seguenti
relazioni%
\end{subequations}
\begin{align}
\delta_{2n+1}  &  =\frac{x_{2n+2}-x_{2n}}{p_{2n+1}}=\frac{\xi_{2n+2}-\xi_{2n}%
}{\pi_{2n+1}},\\
-D_{2n}  &  =\frac{p_{2n+1}-p_{2n-1}}{x_{2n}}=\frac{\pi_{2n+1}-\pi_{2n-1}}%
{\xi_{2n}}.
\end{align}
queste relazioni possono essere messe nella forma
\begin{equation}
\det\left(
\begin{array}
[c]{cc}%
\xi_{2n+2} & \pi_{2n+1}\\
x_{2n+2} & p_{2n+1}%
\end{array}
\right)  =\det\left(
\begin{array}
[c]{cc}%
\xi_{2n} & \pi_{2n+1}\\
x_{2n} & p_{2n+1}%
\end{array}
\right)  ,
\end{equation}
e
\begin{equation}
\det\left(
\begin{array}
[c]{cc}%
\xi_{2n} & \pi_{2n-1}\\
x_{2n} & p_{2n-1}%
\end{array}
\right)  =\det\left(
\begin{array}
[c]{cc}%
\xi_{2n+2} & \pi_{2n-1}\\
x_{2n+2} & p_{2n-1}%
\end{array}
\right)  .
\end{equation}
Al variare dell' indice $n$ questi determinanti sono sempre uguali, e verranno
indicati con $\Gamma.$ Si pu\`{o} esprimere questo invariante in funzione dei
raggi assiale e di campo%
\begin{equation}
\det\left(
\begin{array}
[c]{cc}%
h_{2n} & g_{2n+1}\\
k_{2n} & f_{2n+1}%
\end{array}
\right)  =\det\left(
\begin{array}
[c]{cc}%
h_{2n} & g_{2n-1}\\
k_{2n} & f_{2n-1}%
\end{array}
\right)  .
\end{equation}

\section{Le aberrazioni del terzo ordine. Vantaggi dell' uso delle equazioni
di Hamilton}

Proseguendo nello sviluppo in serie dell' Hamiltoniana si otterr\`{a} la
teoria delle abberrazioni del terzo ordine. I vantaggi di questo approccio
sono svariati. In primo luogo tale approccio \`{e} concettualmente pi\`{u}
semplice ed intuitivo di quello basato sull' uso delle funzioni principali di
Hamilton. Inoltre, quando si utilizzano le funzioni principali, avendo ognuna
di queste delle condizioni di non applicabilit\`{a} (ad esempio i sistemi
telescopici per la $T,$ funzione angolare), bisogna aver chiaro il tipo di
sistema ottico di cui si intendono minimizzare le aberrazioni. E' ovvio che l'
utilizzo delle equazioni di Hamilton, avendo queste valenza globale, non
presenta questo tipo di limitazioni. Le espressioni trovate per i coefficienti
di aberrazione sono molto pi\`{u} manegevoli rispetto a quelle trovate con
altri approcci. Inoltre, non sono formulate in funzione di variabili
adimensionali, ma direttamente nelle coordinate sui piani della pupilla d'
entrata e dell' oggetto. Partire da un mezzo con indice di rifrazione continuo
permette di applicare direttamente le relazioni alle moderne lenti che
presentano questa caratteristica. Il teorema di addizione delle aberrazioni
del terzo ordine discende naturalmente dalla $\sigma-$addittivit\`{a} dell'
integrale, tanto che non si avverte neanche la necessit\`{a} di enunciarlo.

\section{Le aberrazioni del terzo ordine. Derivazione}

Si sviluppi l' Hamiltoniana al second' ordine in $U,V$%
\begin{equation}
H=H_{0}+H_{U}U+H_{V}V+\frac{1}{2}\left(  H_{U^{2}}U^{2}+H_{UV}UV+H_{V^{2}%
}V^{2}\right)  +\cdots
\end{equation}
e si consideri lo sviluppo delle soluzioni nei parametri $x_{o},y_{o}%
,x_{e},y_{e}$
\begin{subequations}
\label{sviluppi soluzioni eq hamilton lunemburg}%
\begin{align}
X  &  =X_{1}+X_{3}+\cdots\\
Y  &  =Y_{1}+Y_{3}+\cdots\\
P  &  =P_{1}+P_{3}+\cdots\\
Q  &  =Q_{1}+Q_{3}+\cdots
\end{align}
dove il pedice indica il grado del polinomio nei parametri $x_{o},y_{o}%
,x_{e},y_{e}.$ I coefficienti dei polinomi sono delle funzioni di $z$. D'ora
in avanti, per brevit\`{a}, si considereranno solo le equazioni per le
variabili $\left(  x,p\right)  $ essendo quelle per le $\left(  y,q\right)  $
identiche a meno di fattori. Le equazioni di Hamilton sono
\end{subequations}
\begin{equation}
\dot{X}_{1}+\dot{X}_{3}+\ldots=2\left(  H_{V}+H_{UV}U+H_{V^{2}}V+\ldots
\right)  \left(  P_{1}+P_{3}+\ldots\right)  ,
\end{equation}
e posto $U_{2}=X_{1}^{2}+Y_{1}^{2}$ e $V_{2}=P_{1}^{2}+Q_{1}^{2}$ diventano%
\begin{equation}
\dot{X}_{1}+\dot{X}_{3}+\ldots=2H_{V}P_{1}+2H_{V}P_{3}+2\left(  H_{UV}%
U_{2}+H_{V^{2}}V_{2}\right)  P_{1}+\ldots
\end{equation}
dove sono stati omessi termini di grado maggiore al terzo. Uguagliando i
polinomi del medesimo grado segue
\begin{subequations}
\label{eq ham lun 1ord}%
\begin{align}
\dot{X}_{1}-2H_{V}P_{1}  &  =0,\\
\dot{P}_{1}+2H_{U}X_{1}  &  =0,
\end{align}
\end{subequations}
\begin{subequations}
\label{eq ham lun 3}%
\begin{align}
\dot{X}_{3}-2H_{V}P_{3}  &  =2\left(  H_{UV}U_{2}+H_{V^{2}}V_{2}\right)
P_{1},\\
\dot{P}_{3}+2H_{U}X_{3}  &  =-2\left(  H_{U^{2}}U_{2}+H_{UV}V_{2}\right)
X_{1}.
\end{align}
Le equazioni del terzo ordine sono le soluzioni di equazioni non omogenee. I
termini non omogenei sono funzioni note delle soluzioni parassiali e gli
operatori differenziali sulla sinistra sono gli stessi che compaiono nelle
equazioni omogenee per l' ottica di Gauss. Il problema era determinare le
soluzioni delle equazioni canoniche con le seguenti condizioni al contorno%
\end{subequations}
\begin{equation}
X\left(  z_{0}\right)  =x_{0},~~X\left(  z_{e}\right)  =x_{e}.
\label{cond cont eq ham lun}%
\end{equation}
Una volta determinate le soluzioni parassiali possiamo scaricare le condizioni
al contorno su quest' ultime. Le aberrazioni del terzo ordine dovranno quindi
soddisfare le seguenti condizioni al contorno%
\begin{equation}
X_{3}\left(  z_{0}\right)  =X_{3}\left(  z_{e}\right)  =0. \label{cond cont3}%
\end{equation}
Possiamo quindi riformulare il problema iniziale nel modo seguente:

\begin{problem}
Trovare una soluzione delle equazioni \ref{eq ham lun 3} che soddisfino le
condizioni al contorno \ref{cond cont3}.
\end{problem}

Il problema si risolve prendendo in considerazione le equazioni
\begin{gather}
\dot{X}_{3}-\frac{1}{N}P_{3}=2\left(  H_{UV}U_{2}+H_{V^{2}}V_{2}\right)
P_{1};\\
\dot{k}\left(  z\right)  -\frac{1}{N}g\left(  z\right)  =0.
\end{gather}
Da queste, segue
\begin{equation}
\dot{X}_{3}g-\dot{k}P_{3}=-2\left(  H_{U^{2}}U_{2}+H_{V^{2}}V_{2}\right)
P_{1}f\left(  z\right)  . \label{rif primo}%
\end{equation}
Analogamente, dalle relazioni%
\begin{gather}
\dot{P}_{3}+DX_{3}=-2\left(  H_{U^{2}}U_{2}+H_{UV}V_{2}\right)  \dot{X}_{1},\\
\dot{f}+Dk=0
\end{gather}
si ottiene
\begin{equation}
X_{3}\dot{f}-k\dot{P}_{3}=2\left(  H_{U^{2}}U_{2}+H_{UV}V_{2}\right)  X_{1}k.
\label{rif secondo}%
\end{equation}
Sommando la \ref{rif primo} e la \ref{rif secondo} si ha
\begin{equation}
\frac{d}{dz}\left(  X_{3}f-kP_{3}\right)  =2\left(  H_{U^{2}}U_{2}+H_{UV}%
V_{2}\right)  X_{1}k+2\left(  H_{U^{2}}U_{2}+H_{UV}V_{2}\right)  X_{1}k.
\end{equation}
Integrando l' equazione, facendo uso delle condizioni al contorno e dell'
espressione dell' invariante ottico di Lagrange al primo ordine sul piano
immagine%
\begin{equation}
\Gamma=Mf\left(  z^{\prime}\right)
\end{equation}
si ha%
\begin{equation}
X_{3}\left(  z^{\prime}\right)  =\frac{2M}{\Gamma}\int_{z_{o}}^{z^{\prime}%
}\left[  \left(  H_{UV}U_{2}+H_{V^{2}}V_{2}\right)  P_{1}f+\left(  H_{U^{2}%
}U_{2}+H_{UV}V_{2}\right)  X_{1}k\right]  dz, \label{ab x lun}%
\end{equation}
analogamente
\begin{equation}
Y_{3}\left(  z^{\prime}\right)  =\frac{2M}{\Gamma}\int_{z_{o}}^{z^{\prime}%
}\left[  \left(  H_{UV}U_{2}+H_{V^{2}}V_{2}\right)  Q_{1}f+\left(  H_{U^{2}%
}U_{2}+H_{UV}V_{2}\right)  Y_{1}k\right]  dz. \label{ab y lun}%
\end{equation}
Gli integrali \ref{ab x lun}, \ref{ab y lun} rappresentano le aberrazioni del
terzo ordine. Si possono identificare i coefficienti di aberrazioni
sostituendo tutte le relazioni fin qui trovate nelle \ref{ab x lun},
\ref{ab y lun} ed eguagliando quest'ultime all' espressione del vettore di
aberrazione trovata grazie all' ausilio delle funzioni principali. Si hanno le
seguenti espressioni
\begin{subequations}
\label{ab3 lun}%
\begin{align}
A  &  =\frac{2}{\Gamma}\int_{z_{0}}^{z^{\prime}}H_{UU}k^{4}+2H_{UV}k^{2}%
f^{2}+H_{VV}f^{4}dz,\\
B  &  =\frac{6}{\Gamma}\int_{z_{0}}^{z^{\prime}}H_{UU}k^{3}h+H_{UV}kf\left(
hg+kf\right)  +H_{VV}f^{3}gdz,\\
C  &  =\frac{6}{\Gamma}\int_{z_{0}}^{z^{\prime}}H_{UU}k^{2}h^{2}%
+2H_{UV}kfhg+H_{VV}f^{2}g^{2}dz+2\Gamma\int_{z_{0}}^{z^{\prime}}H_{UV}dz,\\
D  &  =\frac{2}{\Gamma}\int_{z_{0}}^{z^{\prime}}H_{UU}k^{2}h^{2}%
+2H_{UV}\left(  h^{2}g^{2}+k^{2}f^{2}\right)  +H_{VV}f^{2}g^{2}dz,\\
E  &  =\frac{6}{\Gamma}\int_{z_{0}}^{z^{\prime}}H_{UU}kh^{3}+H_{UV}hg\left(
hg+kf\right)  +H_{VV}fg^{3}dz,
\end{align}
dove si \`{e} preferito omettere le dipendenze da $z$ per non appesantire la
notazione. Da un' attenta analisi delle relazioni precedenti si ricava%
\end{subequations}
\begin{equation}
3D-C=\Gamma\int_{z_{0}}^{z^{\prime}}H_{UV}dz.
\end{equation}
Tale importante risultato va sotto il nome di Teorema di Petzval, da cui
discendono due conseguenze. In primis, uno strumento ottico, per cui la somma
di Petzval
\begin{equation}
\int_{z_{0}}^{z^{\prime}}\frac{1}{R_{c}}d\left(  \frac{1}{N}\right)
\end{equation}
\`{e} nulla, risulta corretto per l' astigmatismo e la curvatura di campo. In
secondo luogo, si pu\`{o} semplificare l' analisi fin qui condotta
restringendo l' attenzione ai soli raggi meridionali. In tal modo si possono
ricavare direttamente quattro dei cinque coefficienti $\left(  A,B,C,E\right)
$, l' ultimo potr\`{a} essere ricavato dal teorema di Petzval. Al fin di
trovare la forma esplicita dei coefficienti dello sviluppo dell' Hamiltoniana,
si prendano in considerazione le caratteristiche geometriche del mezzo. Posto,
per semplicit\`{a} di notazione,%
\begin{equation}
N\left(  0,z\right)  =N\left(  z\right)  =N,~~N_{U}\left(  0,z\right)
=N_{U}\left(  z\right)  =N_{U}~~N_{UU}\left(  0,z\right)  =N_{UU}\left(
z\right)  =N_{UU}%
\end{equation}
si ha
\begin{equation}
H_{UU}=-N_{UU},~~H_{UV}=-\frac{N_{U}}{2N^{2}},~~H_{VV}=\frac{1}{4N^{3}}.
\label{dH}%
\end{equation}
Si esprimano $N_{U}$ ed $N_{UU}$ in funzione delle caratteristiche geometriche
delle superfici $N\left(  U,z\right)  =$ costante. Sia $Z=G\left(  U\right)  $
l' equazione della superficie ottica che passa per il punto $z=z_{c}$ dell'
asse ottico, si ha
\begin{equation}
N\left(  U,G\left(  U\right)  \right)  =N\left(  0,z\right)  ,
\end{equation}
derivando rispetto ad $U$ e valutando le derivate sull' asse ottico si
ottiene
\begin{subequations}
\label{n in funz di r}%
\begin{gather}
N_{U}+N^{\prime}G^{\prime}\left(  0\right)  =0,\\
N_{UU}N^{\prime}G^{\prime\prime}\left(  0\right)  +N^{\prime\prime}(G^{\prime
}\left(  0\right)  )^{2}=0.
\end{gather}
Si sviluppi in serie $G\left(  U\right)  $ in un intorno dell' asse ottico%
\end{subequations}
\begin{equation}
G\left(  U\right)  =z+G^{\prime}\left(  0\right)  U+G^{\prime\prime}\left(
0\right)  U^{2}=z+\frac{1}{2R_{c}}U+\frac{a}{2}U^{2},
\end{equation}
usando le \ref{n in funz di r} e le \ref{dH} si trova%
\begin{align}
H_{UU}  &  =aN^{\prime}-\frac{1}{4}\frac{d}{dz}\left(  \frac{N^{\prime}}%
{R_{c}^{2}}\right)  ,\\
H_{UV}  &  =-\frac{1}{4R_{c}}\frac{d}{dz}\left(  \frac{1}{N}\right)  ,\\
H_{VV}  &  =\frac{1}{4N^{3}.}%
\end{align}
Sostituendo le relazioni fin qui trovate nelle \ref{ab3 lun}, dopo un' attenta
manipolazione degli integrandi, ed introducendo le seguenti notazioni
\begin{equation}
S=k\frac{\left(  \frac{d}{dz}\frac{f\left(  z\right)  }{N}\right)  ^{2}\left(
\frac{d}{dz}\frac{f\left(  z\right)  }{N^{2}}\right)  }{\left(  \frac{d}%
{dz}\frac{1}{N\left(  z\right)  }\right)  ^{2}},~~A_{s}=K\frac{k^{4}}%
{R_{c}^{3}}\frac{dN\left(  z\right)  }{dz},~~P=\frac{1}{R}\frac{d}{dz}\left(
\frac{1}{N\left(  z\right)  }\right)  ,~~\omega=\frac{\left(  \frac{d}%
{dz}\frac{g\left(  z\right)  }{N}\right)  }{\left(  \frac{d}{dz}\frac{f\left(
z\right)  }{N}\right)  },
\end{equation}
si possono esprimere i coefficienti del polinomio
\begin{equation}
\varepsilon_{x}=Ax_{e}^{3}+Bx_{e}^{2}x_{o}+Cx_{e}x_{o}^{2}+Ex_{o}^{3}%
\end{equation}
tramite le formule
\begin{subequations}
\label{coeff abb lun}%
\begin{align}
A  &  =-\frac{1}{2\Gamma}\int_{z_{o}}^{z^{\prime}}S+A_{s}k^{4}dz,\\
B  &  =-\frac{3}{2\Gamma}\int_{z_{o}}^{z^{\prime}}S\omega+A_{s}k^{3}hdz,\\
C  &  =-\frac{3}{2\Gamma}\int_{z_{o}}^{z^{\prime}}S\omega^{2}+A_{s}k^{2}%
h^{2}dz-\frac{\Gamma}{2}\int_{z_{o}}^{z^{\prime}}Pdz,\\
E  &  =-\frac{1}{2\Gamma}\int_{z_{o}}^{z^{\prime}}S\omega^{3}+A_{s}%
kh^{3}dz-\frac{\Gamma}{2}\int_{z_{o}}^{z^{\prime}}P\omega dz
\end{align}
Per i dettagli del calcolo dei coefficienti di aberrazione si rimanda al
paragrafo successivo. Le relazioni fin qui ricavate valgono per un sistema
ottico con indice di rifrazione continuo. Il passaggio al limite discreto
permette di determinare, per ogni superficie del sistema ottico, le seguenti
quantit\`{a}%
\end{subequations}
\begin{equation}
S_{i}=k_{i}\frac{\left(  \Delta_{i}\frac{f\left(  z\right)  }{N}\right)
^{2}\left(  \Delta_{i}\frac{f\left(  z\right)  }{N^{2}}\right)  }{\left(
\Delta_{i}\frac{1}{N}\right)  ^{2}},~~A_{si}=\frac{K_{i}}{2R_{ci}^{3}}\Delta
N,~~P_{i}=\frac{1}{R_{ci}}\left(  \Delta_{i}\frac{1}{N_{i}}\right)
,~~\omega_{i}=\frac{\left(  \Delta_{i}\frac{g\left(  z\right)  }{N}\right)
}{\left(  \Delta_{i}\frac{f\left(  z\right)  }{N}\right)  }.
\end{equation}
Si possono esprimere, in funzione di queste, i coefficienti di aberrazione
\begin{subequations}
\label{coeff lun}%
\begin{align}
A  &  =-\frac{1}{2\Gamma}\sum_{i}\left(  S_{i}+A_{si}k_{i}^{4}\right)  ,\\
B  &  =-\frac{3}{2\Gamma}\sum_{i}\left(  S_{i}\omega_{i}+A_{si}k_{i}^{3}%
h_{i}\right)  ,\\
C  &  =-\frac{3}{2\Gamma}\sum_{i}\left(  S_{i}\omega_{i}^{2}+A_{si}k_{i}%
^{2}h_{i}^{2}\right)  -\frac{\Gamma}{2}\sum_{i}P_{i},\\
E  &  =-\frac{1}{2\Gamma}\sum_{i}\left(  S_{i}\omega_{i}^{3}+A_{si}k_{i}%
h_{i}^{3}\right)  -\frac{\Gamma}{2}\sum_{i}P_{i}\omega_{i}.
\end{align}
Dove i parametri che compaiono si possono ricavare grazie alle seguenti
relazioni di ricorrenza.%
\end{subequations}
\begin{subequations}
\begin{align}
k_{2n+2}-k_{2n}  &  =\frac{t_{2n+1}}{N_{2n+1}}f_{2n+1},~~h_{2n+2}-h_{2n}%
=\frac{t_{2n+1}}{N_{2n+1}}g_{2n+1},\\
f_{2n+1}-f_{2n-1}  &  =-k_{2n}\frac{\Delta N}{R_{2n}},~~g_{2n+1}%
-g_{2n-1}=-h_{2n}\frac{\Delta N}{R_{2n}}.
\end{align}
Va osservato che le uniche restrizioni realmente importanti nelle condizioni
al contorno imposte sui raggi assiale e di campo sono $k\left(  z_{0}\right)
=0$ e $h\left(  z_{e}\right)  =0$. Esse richiedono che i due raggi passino
rispettivamente attraverso i punti oggetto e il centro della pupilla d'
entrata. Se $h\left(  z_{0}\right)  \not =1,$e $k\left(  z_{e}\right)
\not =1$ attraverso le stesse formule si ottengono i coefficienti del
polinomio%
\end{subequations}
\begin{equation}
A\tilde{x}_{e}^{3}+B\tilde{x}_{o}\tilde{x}_{e}^{2}+C\tilde{x}_{0}^{2}\tilde
{x}+E\tilde{x}_{0}^{3}%
\end{equation}
dove
\begin{equation}
\tilde{x}_{0}=\frac{x_{0}}{h\left(  z_{0}\right)  },~~\tilde{x}_{e}%
=\frac{x_{e}}{k\left(  z_{e}\right)  }.
\end{equation}
Queste condizioni sono equivalenti all' utilizzo di un sistema di unit\`{a} di
misura "naturali" sui due piani.

\section{Calcolo dei coefficienti del terzo ordine}

Verr\`{a} ora illustrato in dettaglio il procedimento che porta alle
\ref{coeff lun}. In primo luogo dalle \ref{eq ham lun 1ord} si possono
ricavare le seguenti relazioni che esprimono il raggio di curvatura in
funzione dei raggi parassiali%
\begin{equation}
\frac{1}{R_{c}}=-\frac{f^{\prime}}{N^{\prime}k}=-\frac{g^{\prime}}{N^{\prime
}h}. \label{R(Nfk)}%
\end{equation}
Si osservi che la funzione $k\left(  z\right)  $ si annulla sugli estremi di
integrazione (si veda la \ref{condizioni al cont1}). Siano $L\left(  z\right)
$ $F\left(  z\right)  $ due funzioni opportunamente regolari, integrando per
parti ne discende
\begin{equation}\label{integrazione parti}
\int_{z_{0}}^{z^{\prime}}L\left(  z\right)  k\left(  z\right)  F^{\prime
}\left(  z\right)  dz=-\int_{z_{0}}^{z^{\prime}}\left[  F\left(  z\right)
\frac{d}{dz}\left(  L\left(  z\right)  k\left(  z\right)  \right)  \right]  dz
\end{equation}
Ci si concentri sul coefficiente $A$ che esprime l' aberrazione sferica.
Sostituendo le \ref{R(Nfk)} nella \ref{coeff lun}a, portando $R_{c}$ all'
interno dell' operatore di derivazione che compare nell' espressione del
coefficiente $H_{UU}$ dello sviluppo in serie dell' Hamiltoniana si ha
\begin{align}
A  &  =\int_{z_{0}}^{z^{\prime}}\frac{f{(}z{)}^{4}}{2\,N{(}z{)}^{3}}%
-\frac{f(z)^{2}\,k(z)\,f^{\prime}(z)}{N(z)^{2}}-\frac{k(z)\,f{^{\prime}%
(}z)^{3}}{2\,N^{\prime}(z)^{2}}-\frac{K\,k(z)\,f^{\prime}(z)^{3}}%
{2\,N^{\prime}(z{)}^{2}}+\nonumber\\
&  \frac{k(z)\,f^{\prime}(z)^{2}\,k^{\prime}(z)}{N^{\prime}(z)}-\frac
{k(z{)}^{2}\,f^{\prime}(z)\,f^{\prime\prime}(z)}{N^{\prime}(z)}+\frac
{k(z)^{2}\,f^{\prime}(z)^{2}\,N^{\prime\prime}(z)}{2\,N^{\prime}(z{)}^{2}}dz,
\end{align}
integrando per parti gli ultimi due addendi grazie alla
\ref{integrazione parti}, si ha
\begin{equation}
A=\int_{z_{0}}^{z^{\prime}}\frac{{f(z)}^{4}}{2\,{N(z)}^{3}}-\frac{{f(z)}%
^{2}\,k(z)\,f^{\prime}(z)}{{N(z)}^{2}}-\frac{k(z)\,{f^{\prime}(z)}^{3}%
}{2\,{N^{\prime}(z)}^{2}}-\frac{K\,k(z)\,{f^{\prime}(z)}^{3}}{2\,{N^{\prime
}(z)}^{2}}+\frac{2\,k(z)\,{f^{\prime}(z)}^{2}\,k^{\prime}(z)}{N^{\prime}%
(z)}dz.
\end{equation}
Si aggiunga e sottragga la quantit\`{a}
\begin{equation}
\frac{f^{2}\left(  z\right)  k\left(  z\right)  f^{\prime}\left(  z\right)
}{N^{2}\left(  z\right)  }.
\end{equation}
Avendo l' accortezza di integrare per parti prima di sommare, si ottiene%
\begin{align}
A  &  =\int_{z_{0}}^{z^{\prime}}\frac{-5\,f{(z)}^{2}\,k(z)\,f^{\prime}%
(z)}{2\,N{(}z{)}^{2}}-\frac{k(z)\,f{^{\prime}(}z{)}^{3}}{2\,N^{\prime}%
(z{)}^{2}}-\frac{K\,k(z)\,f^{\prime}(z)^{3}}{2\,N^{\prime}(z)^{2}}\nonumber\\
&  +\frac{2\,f(z)\,k(z)\,f^{\prime}(z)^{2}}{N(z)\,N^{\prime}(z)}%
+\frac{f(z)^{3}\,k(z)\,N^{\prime}(z)}{N(z)^{3}}dz.
\end{align}
Evidenziando per parti si perviene infine alla forma definitiva del
coefficiente di aberrazione%
\begin{equation}
A=\int_{z_{0}}^{z^{\prime}}\left[  k\left(  z\right)  \frac{\left(  \frac
{d}{dz}\frac{g\left(  z\right)  }{N\left(  z\right)  }\right)  ^{2}\left(
\frac{d}{dz}\frac{g\left(  z\right)  }{N^{2}\left(  z\right)  }\right)
}{\left(  \frac{d}{dz}\frac{1}{N\left(  z\right)  }\right)  ^{2}}%
-K\frac{k\left(  z\right)  ^{4}}{R_{c}^{3}}\frac{d}{dz}N\left(  z\right)
\right]  dz.
\end{equation}
L' integrando \`{e} quasi ovunque nullo, l' unico punto dell' asse con
contributo diverso da zero si trova sulla superficie ottica. Si pu\`{o} quindi
restringere il dominio ad una piccola regione di integrazione centrata intorno
a $z_{c}$ $\left[  z_{c}-\delta,z_{c}+\delta\right]  $. Le quantit\`{a} che
compaiono sotto il segno di derivata nell' integrando hanno valori ben
definiti a destra e sinistra della superficie ottica. Possiamo quindi supporre
che varino linearmente nell' intervallo chiuso $\left[  z_{c}-\delta
,z_{c}+\delta\right]  $ e sostituire alle derivate i rapporti incrementali per
poi passare al limite per $\delta\rightarrow0.$ Si ha
\begin{gather}
A=\lim_{\delta\rightarrow0}\int_{z_{c}-\delta}^{z_{c}+\delta}\left[
k\frac{\left(  \Delta\frac{g\left(  z\right)  }{N}\right)  ^{2}\left(
\Delta\frac{g\left(  z\right)  }{N^{2}}\right)  }{\left(  \Delta\frac{1}%
{N}\right)  ^{2}}2\delta-2K\frac{k^{4}}{R_{c}^{3}}\Delta N\delta\right]
dz=\nonumber\\
k\frac{\left(  \Delta\frac{g}{N}\right)  ^{2}\left(  \Delta\frac{g}{N^{2}%
}\right)  }{\left(  \Delta\frac{1}{N}\right)  ^{2}}-K\frac{k^{4}}{R_{c}^{3}%
}\Delta N.
\end{gather}
Per precedere al calcolo degli altri coefficienti di aberrazione \`{e}
necessario dimostrare le seguenti relazioni
\begin{equation}
\frac{h^{\prime}}{k^{\prime}}=\frac{g^{\prime}}{f}=\frac{h^{\prime\prime}%
}{k^{\prime\prime}}. \label{mia dim}%
\end{equation}
Si osservi che le equazioni di Hamilton si disaccoppiano fornendo due sistemi
di equazioni differenziali identici per i raggi assiale e di campo, con l'
unica differenza delle condizioni al contorno. Sia $\left(  l\left(  z\right)
,N\left(  z\right)  l^{\prime}\left(  z\right)  \right)  $ la soluzione
generale di questo sistema. Imposte le condizioni al contorno ai raggi assiale
e di campo si ha
\begin{align}
k\left(  z\right)   &  =\frac{l\left(  z\right)  -l\left(  z_{e}\right)
}{l\left(  z_{0}\right)  -l\left(  z_{e}\right)  }=\alpha l\left(  z\right)
+\beta,\\
h\left(  z\right)   &  =1-\frac{l\left(  z\right)  -l\left(  z_{e}\right)
}{l\left(  z_{0}\right)  -l\left(  z_{e}\right)  }=-\alpha l\left(  z\right)
+\gamma,
\end{align}
con $\alpha,\beta,\gamma$ costanti. E' banale verificare che
\begin{equation}
\frac{h^{\prime}\left(  z\right)  }{k^{\prime}\left(  z\right)  }%
=\frac{h^{\prime\prime}\left(  z\right)  }{k^{\prime\prime}\left(  z\right)
}.
\end{equation}
Per definizione $f^{\prime}\left(  z\right)  =N\left(  z\right)  k^{\prime
}\left(  z\right)  $ e $g^{\prime}\left(  z\right)  =N\left(  z\right)
h^{\prime}\left(  z\right)  $ da cui si arriva alla \ref{mia dim}. Sostitueno
le relazioni appena trovate nelle formule che esprimono i coefficienti di
aberazione si arriva alle \ref{coeff abb lun}.

\section{Le aberrazioni del quinto ordine}

Si pu\`{o} estendere il procedimento utilizzato nel calcolo delle aberrazioni
del terzo ordine per calcolare le aberrazioni del quinto ordine. In primo
luogo tutti gli sviluppi in serie devono essere estesi all' ordine
immediatamente successivo, poi andranno riscritti ordinandoli nelle potenze
dei dati iniziali. Il primo passo \`{e} dunque
\begin{subequations}
\label{sviluppi soluzioni eq hamilton lunemburg  5 ord}%
\begin{align}
X  &  =X_{1}+X_{3}+X_{5}\ldots,\\
Y  &  =Y_{1}+Y_{3}+Y_{5}\ldots,\\
P  &  =P_{1}+P_{3}+P_{5}\ldots,\\
Q  &  =Q_{1}+Q_{3}+Q_{5}\ldots
\end{align}%
\end{subequations}
\begin{align}
H\left(  U,V,z\right)   &  =H_{0}+H_{U}U+H_{V}V+\frac{1}{2}\left(  H_{U^{2}%
}U^{2}+H_{UV}UV+H_{V^{2}}V^{2}\right)  +\\
&  +\frac{1}{6}\left(  H_{U^{3}}U^{3}+H_{U^{2}V}U^{2}V+H_{UV^{2}}%
UV^{2}+H_{V^{3}}V^{3}\right)  \ldots
\end{align}
Sostituendo queste relazioni nelle equazioni di Hamilton, posto $U_{4}%
=X_{1}X_{3}+Y_{1}Y_{3},~V_{4}=P_{1}P_{3}+Q_{1}Q_{3}$ e raggruppando i polinomi
di grado uguale si ha
\begin{align}
\dot{X}_{1}  &  =2H_{V}P_{1},\\
\dot{X}_{3}  &  =2H_{V}P_{3}+2\left(  H_{UV}U_{2}+H_{V^{2}}V_{2}\right)
P_{1},\\
\dot{X}_{5}  &  =2H_{V}P_{5}+\left(  H_{UV}U_{2}+H_{V^{2}}V_{2}\right)
P_{3}+\nonumber\\
&  2\left(  H_{UV}U_{4}+H_{V^{2}}V_{4}+\frac{1}{2}H_{U^{2}V}U_{2}%
^{2}+H_{UV^{2}}U_{2}V_{2}+\frac{1}{2}H_{V^{3}}V_{2}^{2}\right)  P_{1},
\end{align}
per i momenti si ha invece%
\begin{subequations}
\begin{align}
\dot{P}_{1}  &  =-2H_{U}X_{1},\\
\dot{P}_{3}  &  =-2H_{U}X_{3}-2\left(  H_{U^{2}}U_{2}+H_{UV}V_{2}\right)
X_{1},\\
\dot{P}_{5}  &  =-2H_{U}X_{5}-2\left(  H_{UV}U_{2}+H_{V^{2}}V_{2}\right)
X_{3}-\nonumber\\
&  2\left(  H_{UV}V_{4}+H_{U^{2}}U_{4}+H_{U^{2}V}U_{2}V+\frac{1}{2}H_{UV^{2}%
}V_{2}^{2}+\frac{1}{2}H_{U^{3}}U_{2}^{2}\right)  X_{1}.
\end{align}
\ \ Le prime due coppie di relazioni sono uguali a quelle trovate per il
calcolo delle aberrazioni del terzo ordine, le loro soluzioni sono quindi
note. Va ora affrontato il compito di calcolare quelle del quinto ordine.
Dalle \ref{cond cont eq ham lun} e dalle equazioni per i raggi parassiali,
imponendo che anche le aberrazioni del quinto ordine soddisfino condizioni al
contorno simili alle \ref{cond cont3}, ossia
\end{subequations}
\begin{equation}
X_{5}\left(  z_{o}\right)  =Y_{5}\left(  z_{o}\right)  =X_{5}\left(
z_{e}\right)  =Y_{5}\left(  z_{e}\right)  =0,
\end{equation}
si pu\`{o} ricavare, cos\`{\i} come \`{e} stato fatto per le aberrazioni del
terzo ordine,
\begin{gather}
\frac{d}{dz}\left(  X_{5}g\left(  z\right)  -k\left(  z\right)  P_{5}\right)
=2\left(  H_{UV}U_{2}+H_{V^{2}}V_{2}\right)  P_{3}f\left(  z\right)
+\nonumber\\
2\left(  H_{UV}U_{4}+H_{V^{2}}V_{4}+\frac{1}{2}H_{U^{2}V}U_{2}^{2}+H_{UV^{2}%
}U_{2}V_{2}+\frac{1}{2}H_{V^{3}}V_{2}^{2}\right)  P_{1}f\left(  z\right)
+\nonumber\\
2\left(  H_{UV}U_{2}+H_{V^{2}}V_{2}\right)  X_{3}k\left(  z\right)
+\nonumber\\
2\left(  H_{UV}V_{4}+H_{U^{2}}U_{4}+H_{U^{2}V}U_{2}V_{2}+\frac{1}{2}H_{UV^{2}%
}V_{2}^{2}+\frac{1}{2}H_{U^{3}}U_{2}^{2}\right)  X_{1}k\left(  z\right)  ,
\end{gather}%
\begin{align}
X_{5}  &  =\frac{2M}{\Gamma}\int_{z_{o}}^{z^{\prime}}\left(  H_{UV}%
U_{2}+H_{V^{2}}V_{2}\right)  P_{3}f\left(  z\right)  +\nonumber\\
&  \left(  H_{UV}U_{4}+H_{V^{2}}V_{4}+\frac{1}{2}H_{U^{2}V}U_{2}^{2}%
+H_{UV^{2}}U_{2}V_{2}+\frac{1}{2}H_{V^{3}}V_{2}^{2}\right)  P_{1}f\left(
z\right)  +\nonumber\\
&  \left(  H_{UV}U_{2}+H_{V^{2}}V_{2}\right)  X_{3}k\left(  z\right)
+\nonumber\\
&  \left(  H_{UV}V_{4}+H_{U^{2}}U_{4}+H_{U^{2}V}U_{2}V_{2}+\frac{1}%
{2}H_{UV^{2}}V_{2}^{2}+\frac{1}{2}H_{U^{3}}U_{2}^{2}\right)  X_{1}k\left(
z\right)  dz.
\end{align}
Si evince chiaramente che questo procedimento pu\`{o} essere esteso a
qualunque ordine.

\chapter{Esempi di applicazioni della teoria}

\section{Il telescopio Newtoniano}

Il primo esemplare di questo tipo di telescopio fu costruito da Newton nel
1668 (un altro esemplare, tuttora conservato, fu presentato alla Royal Society
nel 1672). Il telescopio Newtoniano \`{e} composto da un grosso specchio posto
sul fondo dello strumento detto primario, e da uno specchio secondario, molto
pi\`{u} piccolo posto davanti al primario \ Lo schema \`{e} riportato in
figura.%
\begin{figure}
[h]
\begin{center}
\includegraphics[
natheight=2.135200in,
natwidth=1.937200in,
height=2.1741in,
width=1.9761in
]%
{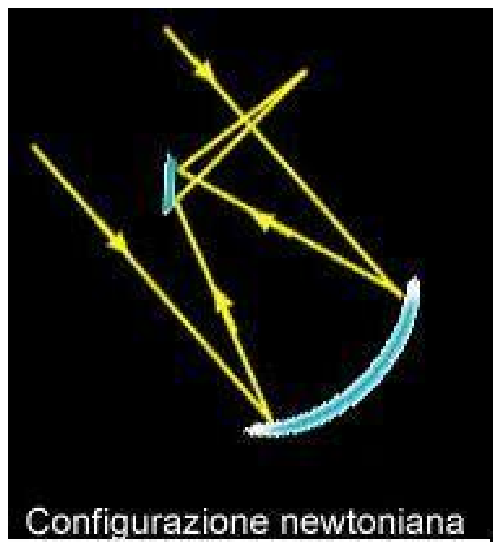}%
\end{center}
\end{figure}
La grande semplicit\`{a} di questo schema, con una sola superficie ottica, e
la sua veneranda et\`{a}, non devono far pensare che sia una soluzione
obsoleta. Si tratta del telescopio preferito dagli astrofili, anche per motivi
non strettamente legati alla resa ottica: l' ecomicit\`{a} e la manegevolezza.
Come si pu\`{o} osservare dalla figura il compito dello specchio secondario
\`{e} quello di deviare i raggi emergenti dallo specchio primario in modo da
permettere un' agevole osservazione delle immagini. Lo specchio secondario
\`{e} quindi uno specchio piano, inclinato di $45%
{{}^\circ}%
$ rispetto all'asse ottico del primario e di piccole dimensioni. Le prime due
caratteristiche del secondario servono a non introdurre differenze di cammino
ottico tra i diversi raggi che provengono dal primario e che convergerebbero
nel fuoco dello strumento, le piccole dimensioni sono necessarie per non
oscurare in maniera significativa il campo visivo dello specchio primario. Per
avere un' immagine stigmatica la forma del primario deve essere parabolica.
Infatti i raggi paralleli all' asse ottico, che coincide con l' asse del
paraboloide, verrano tutti deviati verso il fuoco della paraboloide ove si
forma l' immagine. Il fatto che le immagini extrassiali formate da uno
specchio parabolico siano molte soggette a coma, insieme alle grandi aperture
relative impiegate, rende il campo di utilizzo di questo strumento decisamente
limitato. Sfruttando la teoria delle aberrazioni fin qui sviluppata si
verificher\`{a} che forma parabolica dello specchio \`{e} quella che offre la
migliore configurazione. Si fissi un sistema di riferimento levogiro, con l'
origine nel vertice della superficie ottica e l' asse $z$ coincidente con l'
asse della superficie, lo specchio rivolge la concavit\`{a} nel verso negativo
delle $z,$ e la luce viaggia da sinistra verso destra. L' oggetto che si vuole
osservare \`{e} molto distante dal vertice della superficie, e sia $\delta
_{1}$ questa distanza in modo da trattare in seguito questo termine infinito
con un passaggio al limite. L' immagine si forma nel fuoco dello strumento che
ha coordinate $\left(  0,0,-R_{c}/2\right)  ,$ dove $-R_{c}$ \`{e} il raggio
di curvatura della superficie nel vertice. Invece delle condizioni al contorno
consuete conviene utilizzare le seguenti%
\begin{equation}
h_{0}=\delta_{1}/\mu,~~h_{2}=0,~~k_{0}=0,~~k_{2}=1.
\end{equation}
Si \`{e} visto nel paragrafo 5.4 che questa scelta non influenza i
coefficienti di aberrazione, ma si scarica sulle potenze di $x_{0},x_{e}$ che
sono la base del polinomio di aberrazione, l' introduzione di $\mu$ serve solo
per motivi dimensionali, ad esempio se tutte le lunghezze sono espresse in
metri si pu\`{o} porre $\mu=1m$. Si pu\`{o} ora procedere al calcolo degli
$h_{i},k_{i},f_{i},g_{i}.$ Utilizzando le formule ricorsive si ha
\begin{equation}%
\begin{tabular}
[c]{llll}%
$h_{0}=\delta_{1}/\mu$ & $g_{1}=-1/\mu$ & $h_{2}=0$ & $g_{3}=-1/\mu$\\
$k_{0}=0$ & $f_{1}=1/\delta_{1}$ & $k_{2}=1$ & $f_{3}=1/\delta_{1}-\frac
{2}{R_{c}}$%
\end{tabular}
\end{equation}
e si ottiene $\Gamma=1/\mu.$ Procedendo ora col calcolo dei coefficienti e
andando al limite per $\delta_{1}\rightarrow\infty,$ si ottiene%
\begin{gather}
S=\frac{2}{R_{c}^{3}},~~A_{s}=2\frac{K}{R_{c}^{3}},\\
\omega=\frac{R_{c}}{\mu},~~P=-\frac{2}{R_{c}}%
\end{gather}
e le aberrazioni sono date da
\begin{align}
A  &  =-\frac{\left(  1+K\right)  }{R_{c}^{3}}x_{e}^{3},\\
B  &  =-\frac{3}{R_{c}^{2}}\tan\frac{\alpha}{2}x_{e}^{2}\mu\\
C  &  =\left(  -\frac{3}{R_{c}}+\frac{1}{R_{c}}\right)  \tilde{x}_{0}%
^{2}xe=-\frac{2}{R_{c}}\tan^{2}\frac{\alpha}{2}x_{e}\mu^{2}\\
D  &  =-C\\
E  &  =0
\end{align}
dove $\alpha$ \`{e} l' angolo sotto il quale si vede l' oggetto e $\tan
\frac{\alpha}{2}=\frac{x_{0}}{\delta_{1}}$. Quindi l' unico coefficiente che
mostra una dipendenza, attraverso $K$, dalla forma del primario \`{e} l'
aberrazione sferica. Basta porre $K=-1$ per renderlo nullo e dedurre che la
forma migliore per lo specchio primario \`{e} un paraboloide di rotazione.
Purtroppo non si possono minimizzare le altre aberrazioni se non facendo
crescere a dismisura il raggio della superficie, ma il risultato cos\`{\i}
ottenuto \`{e} gi\`{a} degno di rilievo tenendo presente che si sta
utilizzando una sola superficie ottica!

\section{La camera di Schmidt}

Attorno al 1930 allo schema Newtoniano furono apportate importanti modifiche
per permettere la produzione di immagini fotografiche con grossi sistemi
riflettori. A differenza di uno specchio parabolico che d\`{a} un' immagine
perfetta per i raggi assiali e immagini afflitte da coma a brevi distanze
dall' asse, uno specchio sferico dar\`{a} un' immagine uniforme su un' ampia
superficie sferica concentrica ad esso, ma afflitta da una grossa aberrazione
sferica. Schmidt introdusse all' ingresso dello strumento una sottile lastra,
detta lastra correttrice, posta nel centro di curvatura dello specchio, con
una faccia piana ed l' altra lavorata con un profilo asferico del quarto
ordine.%
\begin{figure}
[h]
\begin{center}
\includegraphics[
natheight=2.156000in,
natwidth=4.187400in,
height=2.1949in,
width=4.2376in
]%
{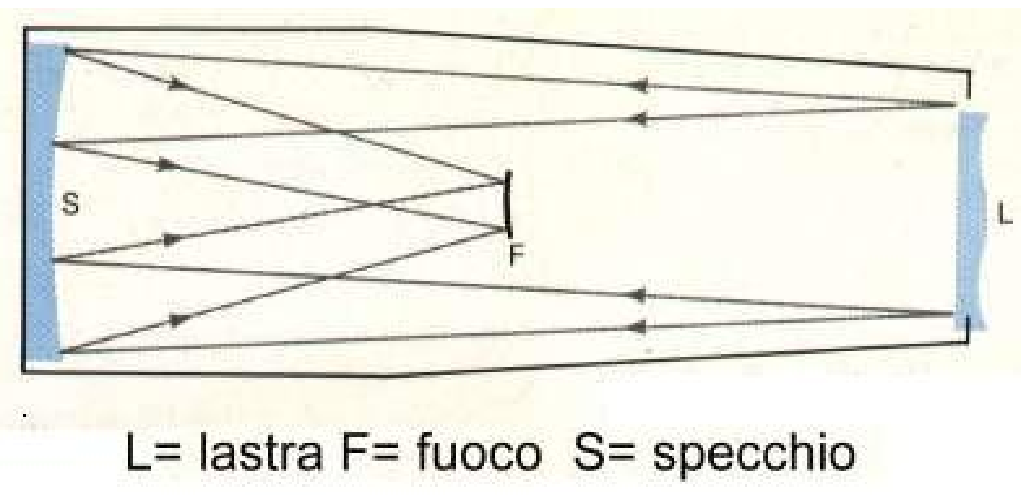}%
\end{center}
\end{figure}
Lo scopo della lastra \`{e} correggere il fascio di raggi entranti in modo da
compensare esattamente l' aberrazione sferica introdotta dallo specchio
sferico. Si pu\`{o} facilmente derivare un' espressione per il profilo della
lastra correttrice della camera di Schmidt. Si calcoleranno le aberrazioni
prodotte da una configurazione di questo tipo nell' approssimazione in cui lo
spessore della lastra sia trascurabile rispetto alle lunghezze in gioco, in
modo da determinare i coefficienti che caratterizzano il profilo asferico e
quindi la forma della lastra. Sia
\begin{equation}
g\left(  z\right)  =R_{c}-\frac{\beta}{4}\left(  x^{2}+y^{2}\right)  ^{2}%
\end{equation}
l' equazione del profilo della lastra, dove $R_{c}$ \`{e} il raggio di
curvatura dello specchio e $\beta$ il parametro incognito che si vuole
determinare. Si \`{e} supposto che il raggio di curvatura $\rho$ della lastra
sia infinito sull' asse ottico. I dati dei raggi assiale e di campo, necessari
per condurre l' analisi, sono%
\begin{gather}%
\begin{tabular}
[c]{cccc}%
$h_{0}=\delta_{1}/\mu$ & $g_{1}=-1/\mu$ & $h_{2}=0$ & $g_{3}=-1/\mu$\\
$k_{0}=0$ & $f_{1}=1/\mu$ & $k_{2}=1$ & $f_{3}=1/\delta_{1}$%
\end{tabular}
\\%
\begin{tabular}
[c]{cccc}%
$h_{4}=0$ & $g_{5}=-1/\mu$ & $h_{6}=-\frac{R_{c}}{\mu}$ & $g_{7}=1/\mu$\\
$k_{4}=1$ & $f_{5}=\frac{1}{\delta_{1}}-\frac{(N-1)}{\rho}$ & $k_{6}%
=1+f_{5}R_{c}$ & $f_{7}=-\frac{2}{R_{c}}-f_{5},$%
\end{tabular}
\end{gather}
dove $N$ \`{e} l' indice di rifrazione della lastra, $\delta_{1}$ \`{e} la
distanza dello spazio oggetto dalla pupilla d' entrata e $\mu$ \`{e} stato
introdotto per ragioni dimensionali, anche in questo caso si pu\`{o} porre, se
le lunghezze sono espresse in $m$, $\mu=1m$. Procedendo al calcolo dei
coefficienti di aberrazione sferica per la lastra corretrice e per lo specchio
sferico si ha
\begin{align}
A_{2}  &  =\beta\left(  N-1\right)  ~\text{(lastra correttrice),}\\
A_{6}  &  =\frac{-2}{R_{c}^{3}}~\ \ \ \ \ \ \ \ \text{(specchio sferico).}%
\end{align}
Sommando i due coefficienti, ed imponendo che l' aberrazione sferica
risultante sia nulla si ricava per $\beta$ il seguente valore%
\begin{equation}
\beta=\frac{2}{R_{c}^{3}\left(  N-1\right)  }.
\end{equation}

\chapter{Conclusioni}

I risultati ottenuti in questo lavoro possono aprire alcuni scenari e campi d' indagine.

I due nuovi invarianti ottici $L_{1},$ $L_{2},$ insieme all' invariante di
Lagrange, potrebbero semplificare l'analisi di sistemi ottici composti
esclusivamente da superfici sferiche (la maggioranza assoluta degli obiettivi
fotografici presenta proprio questa caratteristica). In particolare potrebbe
essere possibile scrivere una delle funzioni principali in funzione di
$\mathbf{L}^{2}=\left(  L_{1}^{2}+L_{2}^{2}+L_{3}^{2}\right)  $ e di uno solo
degli invarianti di rotazione. Se tale circostanza fosse verificata, alcune
delle aberrazioni potrebbero essere dipendenti tra loro.

Nel capitolo cinque sono date le formule di partenza per estendere l'
approccio di Luneburg al quinto ordine. La speranza \`{e} che le formule
ricavate da questo metodo risultino pi\`{u} manegevoli di quelle disponibili.
Inoltre, questo metodo non necessita di complicate trasformazioni di
variabili, n\`{e} di un teorema di addizione.

Infine, grazie alle formule di trasformazione introdotte alla fine del secondo
capitolo si pu\`{o} dimostrare che il raggio uscente da un sistema ottico
\`{e} una funzione continua del dato iniziale. La teoria delle aberrazioni
potrebbe essere interpretata come uno sviluppo in serie in un intorno di una
posizione di equilibrio stabile per le equazioni di Hamilton, l' asse ottico.


\begin{thebibliography}{99}                                                                                               %


\bibitem {1}A. Romano Meccanica Razionale, Liguori Editore 1991

\bibitem {2}R.K. Luneburg Mathematical Theory of Optics, University of
California Press 1965

\bibitem {3}K. Schwarzschild Untersuchungen zur geometrischen, Optik I 1905

\bibitem {4}K. Schwarzschild Untersuchungen zur geometrischen, Optik II 1905

\bibitem {5}M. Born, E.Wolf Principles of Optics, 1970 Pergamon Press

\bibitem {6}A. Romano Geometrical Optics, Theory and Applications in corso di redazione

\bibitem {7}H. A. Buchdahl An Introduction to Hamiltonian Optics, Cambridge
University Press

\bibitem {8}W. T. Welford Geometrical Optics, North Holland Publishing Company Amsterdam

\bibitem {9}Herzberger Modern Geometrical Optics, Interscience Publishers Inc
New York

\bibitem {10}Onofri Destri Istituzioni di Fisica Teorica, Nuova Italia Scientifica

\bibitem {11}Giulio Starita Esercizi di Meccanica Razionale, Liguori Editore 1991

\bibitem {12}C. Miranda Lezioni di Analisi Matematica, Liguori Editore 1965

\bibitem {13}D. Lynden Bell Exact Optics: A unification of optical telescope
design arXiv:phisics%
$\backslash$%
0203082v1 2002

\bibitem {14}Warren J. Smith, Modern optical engineering, 3rd ed.,
McGraw-Hill, 2000
\end{thebibliography}
\end{document}